\documentclass[a4paper,11pt]{article}
\usepackage{jcappub}
\usepackage{graphicx, xcolor}
\usepackage[normalem]{ulem}
\usepackage{amsmath}
\usepackage{subcaption}
\usepackage{float}
\usepackage{etoolbox}
\usepackage{dblfloatfix}
\usepackage{booktabs,array}
\title{\boldmath Kinetic Theory of the EFTofLSS: A Halo-based Approach}

\author[a,1]{Om Kishor Hebbar\note{Corresponding author.},} 
\author[b]{Caio B. de S. Nascimento} 
\author[a]{and Arka Banerjee} 
\affiliation[a]{Department of Physics, Indian Institute of Science Education and Research, Homi Bhabha Road, Pashan, Pune 411008, India} \affiliation[b]{Perimeter Institute for Theoretical Physics, Waterloo, ON, Canada} 
\emailAdd{om.kishorhebbar@students.iiserpune.ac.in}

\abstract{The Effective Field Theory of Large Scale Structure (EFTofLSS) provides a coarse-grained description of the non-linear gravitational clustering of dark matter, in which the effects of short-scale physics are encoded in an effective stress tensor and a set of counterterms. In this work, we develop a framework to investigate the physical origin of these counterterms by relating the effective stress tensor to the dynamics of individual dark matter halos in cosmological N-body simulations. We find that halo accretion contributes to the trace of the halo-based effective stress tensor, contributing to the background pressure and effective sound speed, while anisotropic accretion generates the traceless part, consequently contributing to the shear viscosity. For the halo population with \(M_\mathrm{vir}>10^{12} M_\odot/h\), we find that halos account for \(\sim\!95\%\), \(\sim\!85\%\), and \(\sim\!70\%\) of the full box background pressure, effective sound speed, and shear viscosity, respectively, depending on the EFT cutoff scale. The contributions to the counterterms are dominated by cluster-scale halos, \(M_\mathrm{vir} \sim 10^{14} - 10^{15} M_\odot/h\) and halos with larger inflow rates, supporting an interpretation in terms of accretion-driven, out-of-equilibrium dynamics. Our results provide a short-scale, kinetic-theory interpretation of the EFTofLSS counterterms and establish a framework for connecting the properties and dynamics of short-scale structures to physics-informed priors for large-scale cosmological observables.}

\begin{document}
\maketitle
\flushbottom

\section{Introduction}

The Effective Field Theory (EFT) of Large Scale Structure (LSS) \cite{carrasco2012effective,baumann2012cosmological,pajer2013renormalization,carrasco2014twoloopIR,carrasco2014twoloops,porto2014lagrangian,senatore2015irresummed,carroll2014consistent,ivanov2022eftlss} describes dark matter on large scales as an effective fluid \cite{baumann2012cosmological,carrasco2012effective,pietroni2012coarse,hertzberg2014effective,mercolli2014velocity,manzotti2014coarse,floerchinger2017rg}, in which short-scale physics enters through counterterms. It has since been extended to higher loops \cite{carrasco2014twoloopIR,carrasco2014twoloops,konstandin2019threeloops} and higher-point statistics \cite{angulo2015bispectrum,baldauf2015bispectrum,bertolini2016trispectrum,baldauf2021twoloopbispec}, and validated against N-body simulations \cite{foreman2016precision,baldauf2015apparent,baldauf2016density}. It was originally proposed to formulate a well-defined and controlled alternative to conventional cosmological perturbation theory \cite{bernardeau2002large,scoccimarro1996loop,crocce2006renormalized,carlson2009critical,goroff1986coupling,makino1992analytic,jain1994second}\footnote{Other frameworks attempt to go beyond the single-stream fluid approximation of standard perturbation theory by working directly with the phase-space dynamics of dark matter. These include Kinetic Field Theory \cite{bartelmann2016microscopic,Bartelmann_2019}, perturbative solutions of the Vlasov-Poisson system in full phase space \cite{nascimento2024phase,heisler2026phase}, Vlasov Perturbation Theory, which evolves the hierarchy of cumulants of the phase-space distribution function \cite{garny2023vptlinear,garny2023vptnonlinear,garny2025vptcumulants,garny2025vptlcdm}, and the BBGKY hierarchy initialised with the Zel'dovich approximation \cite{tassev2026dominated}.}. The EFTofLSS approach begins by coarse-graining the standard cosmological fluid equations \cite{baumann2012cosmological,carrasco2012effective}, which introduces an effective stress tensor encoding the back-reaction of short-wavelength fluctuations, through both their kinetic and gravitational contributions, on the long-wavelength fields.

The effective stress tensor can be written both as an expansion in terms of long-wavelength fields \cite{carrasco2012effective,porto2014lagrangian,carrasco2014twoloops,abolhasani2016systematic}, as well as in terms of short-wavelength fields \cite{baumann2012cosmological,nascimento2025semi} (see also \cite{Ansari:2024efj,Ansari:2025nsf} for time non-locality in this expansion). On large scales, the symmetries of the theory constrain the terms in the long-wavelength expansion obtained by averaging over short-wavelength modes \cite{carrasco2012effective,baumann2012cosmological,abolhasani2016systematic,pajer2013renormalization}. The coefficients of these terms, known as counterterms, are not fixed by the symmetries and remain undetermined \cite{carrasco2012effective,abolhasani2016systematic}. They are generally measured through cosmological N-body simulations \cite{carrasco2012effective,carrasco2014twoloops,foreman2016precision,baldauf2015apparent,cataneo2017efficient,mcquinn2016cosmological,Karandikar_2024,nascimento2025semi}, or fitted to observed data alongside cosmological parameters \cite{damico2020boss,ivanov2020boss,desi2024v}.

The effective stress tensor is generated by the multi-streaming of particles on small scales. N-body simulations demonstrate that dark matter halos are regions of extensive multi-streaming \cite{zeldovich1970gravitational,shandarin1989large,pueblas2009generation,abel2012tracing,buehlmann2019largescale,hahn2015velocity,kugel2024cosmic}. Halos are nearly virialised objects \cite{lyndenbell1967statistical,gunn1972infall,binney2008galactic,navarro1996structure,bryan1998statistical,cole1996structure}, which are stabilised by the velocity dispersion sourced by this multi-streaming. In the EFTofLSS formalism, a perfectly virialised object has a vanishing effective stress tensor \cite{baumann2012cosmological}, hence it would not contribute to the counterterms. However, dark matter halos, especially large, cluster-scale halos, are not perfectly virialised \cite{neto2007statistics,kravtsov2012formation}; they are actively merging and accreting matter \cite{lacey1993merger,wechsler2002concentrations,mcbride2009mass,fakhouri2010merger,correa2015accretion,fillmore1984selfsimilar,bertschinger1985selfsimilar}, leading to large out-of-equilibrium contributions. Thus, physically we expect the effective stress tensor to be overwhelmingly sourced by these bound halos \cite{buehlmann2019largescale,nascimento2025semi,Karandikar_2024}. 

In this paper, we aim to map out the explicit physical origins of these counterterms \cite{nascimento2025semi,Karandikar_2024}, essentially amounting to a Kinetic Theory of the EFTofLSS \cite{chapman1970mathematical,lifshitz1981physical}. In standard applications of the EFTofLSS, the counterterms are treated as free parameters \cite{damico2020boss,ivanov2020boss,colas2020efficient,philcox2020combining,ivanov2020planck,chudaykin2020classpt,damico2021wcdm,nishimichi2020blinded,philcox2022boss,zhang2022bossxi,damico2024bispectrum,ivanov2023bispectrum,philcox2022bispectrum,cabass2022single,simon2023eboss,chen2020velocileptors,pezzotta2024euclid,chudaykin2023eboss,taule2023twoloop,taule2024constraints,aviles2024testing,piga2023constraints} and marginalised over when fitting cosmological observables \cite{desi2024v,desi2024vii,chudaykin2025desi1,chudaykin2025desi2,chudaykin2025desi3,ivanov2026desi4,chudaykin2026desi5,elbers2025neutrino,maus2025velocileptors,desi2025hod}. While marginalisation ensures that unmodelled short-scale physics does not bias cosmological inference \cite{baldauf2016lss}, mapping out the physical processes generating these terms provides a path toward physically motivated priors (see \cite{ivanov2024sbp,ivanov2025sbp,ivanov2025analytic,zhang2025hod,desi2025hod,chudaykin2026desi5} for priors calibrated on halo occupation distribution (HOD) mock catalogues, which target the full set of galaxy EFT parameters; cf.\ footnote~\ref{fn:caveats}), which mitigate prior-volume projection effects \cite{carrilho2023,chudaykin2024priors,reeves2025} and yield tighter constraints on cosmological parameters of interest \cite{braganca2023peeking,ivanov2025sbp,chudaykin2026desi5}. 

In this work, we compute the effective stress tensor in terms of the short-scale fields \cite{baumann2012cosmological}, directly from dark matter halos \cite{Behroozi_2012} inside N-body simulations \cite{springel2005gadget2,Springel_2021}, extending the toy-model `top-down' construction in \cite{Karandikar_2024} and analytic methods in \cite{nascimento2025semi} to the completely numerical cosmological setting. We find that halos account for a major fraction of the total counterterms, with actively accreting cluster-scale halos serving as a primary driver of these contributions. We establish that the trace of the effective stress tensor is sourced by halo accretion, and anisotropic accretion \cite{bond1996filaments,hahn2007properties,libeskind2018tracing} generates the traceless part. By demonstrating that a significant fraction of the counterterm magnitude can be mapped to halo accretion dynamics, we establish a foundation for physics-informed marginalisation schemes \cite{ivanov2025analytic,zhang2025hod,chudaykin2026desi5} that bridge short-scale non-linear halo dynamics with large-scale observational predictions\footnote{\label{fn:caveats}We note some caveats to this statement. For galaxy observables, the bias coefficients are degenerate with the gravity-only EFT counterterms considered here \cite{mcdonald2009clustering,assassi2014renormalized,senatore2015bias,angulo2015biased,mirbabayi2015biased,desjacques2018largescale}. In redshift space, additional counterterms appear \cite{Senatore:2014vja}, and they also absorb the effects of fingers-of-God \cite{ivanov2020boss,chudaykin2020classpt}, which are themselves generated by virial motions within halos. For weak lensing, which directly probes the matter field, baryonic feedback modifies the counterterms \cite{Lewandowski:2014rca,Braganca:2020nhv}, and intrinsic alignments further complicate the modelling \cite{Vlah:2019byq}. Finally, we do not consider stochastic terms \cite{baumann2012cosmological,carrasco2012effective}, nor the additional counterterms appearing at two-loop order \cite{carrasco2014twoloops}.}. We note that halo accretion rate is not merely a simulation diagnostic, but a quantity increasingly targeted by observations, most notably through the splashback radius \cite{diemer2014dependence,adhikari2014splashback,more2015splashback,diemer2017sparta}, detected in the galaxy number density profiles of clusters \cite{more2016detection,baxter2017halo,chang2018splashback,shin2019measurement} and in their weak lensing profiles \cite{chang2018splashback,shin2019measurement}.

In addition to applying our methodology to cosmological N-body simulations, we test our computation on some simple toy models involving Plummer spheres \cite{Plummer1911,aarseth1974comparison,binney2008galactic}. These toy models are chosen such that they illustrate the essential features of the calculation and the results, without introducing the additional complexity associated with cosmological simulations. This allows us to validate our implementation in controlled settings, where the expected behaviour can be easily understood. The isolated Plummer sphere acts as a robustness check of our computation against the EFTofLSS non-renormalization theorem \cite{baumann2012cosmological}, and the two colliding Plummer sphere scenarios capture the intuition underlying the physical process we investigate \cite{Karandikar_2024}.

This paper is structured as follows. In Section~\ref{sec:haloanalysis},  we describe the computation of the effective stress tensor for individual halos in N-body simulations, and extract their contributions to the counterterms, identifying the halo properties that correlate with larger contributions. In Section~\ref{sec:fullbox}, we compute the full box effective stress tensor and counterterms, compare them with the halo-based estimates to quantify the fraction sourced by halos, and study their redshift evolution. The Plummer-sphere toy models discussed above are presented in Appendix~\ref{app:plummer}, while various validation tests and supplementary calculations are collected in Appendices~\ref{app:smallbox}--\ref{app:cee}. 

\vspace{-0.2 cm}

\section{Halo-Based Analysis}\label{sec:haloanalysis}

In the EFTofLSS, the effects of short-scale modes on the long-wavelength fields are captured by the effective stress tensor. The effective stress tensor can be written in terms of an expansion in the long-wavelength fields \cite{carrasco2012effective}: 
\begin{equation}\label{eqn:expansion}
\left\langle [\tau^{ij}]_{\Lambda} \right\rangle_{\delta_l} = \bar{P} \delta^{ij} + \bar{\rho} \Bigg[ c_s^2 \delta_l \delta^{ij} - \frac{c_{bv}^2}{Ha} \delta^{ij} \partial_k v_l^k - \frac{3}{4} \frac{c_{sv}^2}{Ha} \left( \partial^j v_l^i + \partial^i v_l^j - \frac{2}{3} \delta^{ij} \partial_k v_l^k \right) \Bigg] + \Delta \tau^{ij} + \dots\,.
\end{equation}
Here, \(\bar{P}\) represents the background pressure, \(c_s\) denotes the speed of sound, and \(c_{bv}\) and \(c_{sv}\) correspond to the bulk and shear viscosity parameters respectively. Importantly, for our purposes, the stress tensor can also be expressed directly in terms of the short-scale fields \cite{baumann2012cosmological}:
\begin{equation}
\tau_{ij} = \overbrace{\rho_m v_i^s v_j^s}^{\text{Kinetic Term}} + \underbrace{\frac{2 \partial_i\phi^s \partial_j\phi^s - \delta_{ij} \partial_k\phi^s \partial_k\phi^s}{8 \pi G}}_{\text{Potential Term}}\,,
\end{equation}
where $\rho_m$ is the matter density, and $v^s_i$ and $\phi^s$ are the short-scale velocities and gravitational potentials, respectively. The kinetic term captures the momentum flux carried by short-scale velocity dispersion, while the potential term is the corresponding gravitational stress sourced by the short-scale potential. Once smoothed over $R_\Lambda$ and averaged over the short-scale modes in the presence of long-wavelength perturbations, $\tau_{ij}$ is expanded in terms of the long-wavelength fields, with the counterterms appearing as the coefficients of this expansion. Our aim in this section is to evaluate $\tau_{ij}$ for the dark matter halos in a cosmological N-body simulation, and to use it to extract the counterterms and identify the halo properties which drive their magnitude.

We use the \textsc{Gadget-4} N-body simulation code \cite{Springel_2021} and simulate a \(1000\) \(\mathrm{Mpc}/h\) box with \(1024^3\) particles, resulting in a mass resolution of \(8.1 \times 10^{10} M_{\odot}/h\), and use a force resolution of \(24.4\,\mathrm{kpc}/h\). This simulation box probes a large cosmological volume, allowing us to obtain better statistics for massive dark matter halos. We also simulate a \(250\) \(\mathrm{Mpc}/h\) box with \(1024^3\) particles, resulting in a mass resolution of \(1.3 \times 10^{9} M_{\odot}/h\), and use a force resolution of \(6.1\,\mathrm{kpc}/h\). This allows us to extend our analysis to lower halo masses (see Appendix~\ref{app:smallbox1}) and is used in creating the visualisations in Figures~\ref{fig:visual} and~\ref{fig:haloalign}. Dark matter halos are identified using \textsc{Rockstar}, the phase space temporal halo finder \cite{Behroozi_2012}. Unlike the isolated Plummer-sphere toy models, dark matter halos are not isolated systems and coexist with long-wavelength modes in their surroundings. Hence, in the following analysis we introduce the EFT cutoff scale $R_\Lambda$ to split the fields into their short and long modes. Section~\ref{sec:halostress} describes the computation of the halo-based effective stress tensor, taking inspiration from the toy model computation in Appendix~\ref{app:plummer}, and Section~\ref{sec:extract} describes the extraction of the counterterms.

\subsection{Halo-Based Effective Stress Tensor}\label{sec:halostress}

We introduce the EFT cutoff scale, \(R_\Lambda\), in real space. For our purposes, the natural choice of the smoothing window function is the real space Top Hat filter. For each halo, we consider particles within a sphere of radius, \(R_h = \alpha \times R_\mathrm{vir}\), where \(R_\mathrm{{vir}}\) is the virial radius and we vary the radius cut \(\alpha \in \{1.5, 2.5, 3\} \). In all of the following results, we choose \(\alpha= 3 \), and Appendix~\ref{app:cut} discusses the results as \(\alpha\) is varied. We note that, since \(R_h\) extends to \(3\,R_{\mathrm{vir}}\), the halo-based stress tensor already implicitly captures a portion of the near-halo filamentary infall associated with each halo, rather than isolating the halo in a strict sense. Although the splashback radius \cite{diemer2014dependence,adhikari2014splashback,more2015splashback} is a natural choice, it is difficult to find for individual halos, leading to our choice of multiples of the virial radius. In the following subsections, we describe the computations of the kinetic and potential terms.

\subsubsection{The Kinetic Term}

To compute the kinetic term, we must isolate the short-scale velocities. To achieve this, we first estimate the long-wavelength bulk velocity of the halo, which we compute as the mean velocity of all particles within a sphere of radius $R_\Lambda$, centred on the halo. Since $R_\Lambda$ is chosen to be much larger than the halo analysis radius, $R_h$, we approximate this mean velocity as uniform across the halo and identify it with the local long-wavelength bulk velocity.

The short-scale velocity is obtained by subtracting this bulk velocity from the total velocity, for each halo particle. This procedure is equivalent to applying a real space Top Hat filter with smoothing scale $R_{\Lambda}$ to the velocity field and subtracting the resulting long-wavelength component. The equivalence relies on the scale separation \( R_h \ll R_{\Lambda}\), which allows the long-wavelength velocity field to be treated as uniform over the extent of the halo. 

\noindent The kinetic term is then computed by the following sum over all the halo particles : 
\vspace{-0.2cm}

\begin{equation}
K_{ij} = \frac{1}{V_{\mathrm{halo}}}\sum_p m_p v^s_i v^s_j\,.
\end{equation}
Here we have divided by the volume of the halo, \(V_{\mathrm{halo}} =\frac{4}{3} \pi R_h^3\).

\subsubsection{The Potential Term}\label{sec:potterm}

To compute the potential term, we must isolate the spatial derivatives of the short-scale gravitational potential. For this purpose, we place each halo at the centre of an isolated cube of side length $L$\footnote{In Appendix~\ref{app:fullapprox}, we show that all contributions to $\phi^s$ are captured provided that $L > 2(R_h + 2R_{\Lambda})$. Such a large $L$ renders the computation prohibitively expensive; we therefore adopt the approximation $L > 2R_{\Lambda}$, which is demonstrated in the same appendix to be in excellent agreement with the complete calculation.}. This is justified because only the short-scale fields are of interest here, so any mode with a wavelength exceeding the size of the cube does not contribute and may be neglected. The potential is then obtained by depositing the halo particles onto a grid to construct the density field, and solving Poisson's equation in Fourier space.

\vspace{-0.3cm}

\begin{equation}
\nabla^2 \phi = 4 \pi G (\rho - \bar\rho) \implies \phi(k) = - \frac{4 \pi G}{k^2} \,\delta\rho(k)\,, \qquad \delta\rho \equiv \rho - \bar\rho\,.
\end{equation}

We then smooth this potential by multiplying it by the Fourier transform of the real space Top Hat filter, and subtract the smoothed field from the full field, to isolate the short-scale gravitational potential. 

\vspace{-0.1cm}

\begin{equation}
\phi^s(k) = \phi(k) - \left( W_{\Lambda}(k) \cdot \phi(k) \right)= \phi(k) - \phi^l(k)\,.
\end{equation}

The spatial derivatives of the short-scale potential are constructed in Fourier space, by taking products with the respective wavenumber component:

\vspace{-0.2cm}

\begin{equation}
 \phi^s(k) = -\frac{4 \pi G}{k^2} \rho^s(k) \implies ik^j \phi^s(k) = - \frac{4 \pi G \, ik^j}{k^2} \rho^s(k)\,.
\end{equation}

Finally we inverse Fourier transform to isolate the derivatives of the short-scale gravitational potential, in real space. The potential term is then calculated as a sum over contributions from all grid cells in the isolated cube:
\begin{equation}
P_{ij} = \frac{1}{V_{\mathrm{halo}}} \frac{(\Delta x)^3}{8\pi G}\sum_{\mathrm{grid}} (2 \hspace{0.05cm}\partial_i \phi^s \hspace{0.05cm}\partial_j \phi^s - \delta_{ij} \hspace{0.05cm}\partial_k \phi^s \hspace{0.05cm}\partial_k \phi^s )\,,
\end{equation}
where \((\Delta x)^3\) is the volume of a single grid cell.

In the following we choose the EFT cutoff scale, \(R_{\Lambda} \in \{16, 20, 24, 28\} \hspace{0.1 cm} \mathrm{Mpc}/h\), and the side length of the isolated cube, L \( \in \{40, 48, 50, 58\} \hspace{0.1 cm} \mathrm{Mpc}/h\), respectively. Hence the volume averaged, halo-based effective stress tensor is : \(\tau_{ij}^{\mathrm{halo}} = K_{ij} + P_{ij} \), and we complete our discussion on the computation of the effective stress tensor for dark matter halos.

\subsection{Extracting the Halo-Based Counterterms}\label{sec:extract}

Equipped with the halo-based effective stress tensor computation, we perform it for all halos in the \(z = 0\) snapshot of the \(1000\,\mathrm{Mpc}/h\) box with \(M_\mathrm{vir} > 10^{13} M_{\odot}/h\). At this resolution, halos with \(M_\mathrm{vir} \sim 10^{13} M_{\odot}/h\) contain \(\sim 100\) particles. In order to prevent any double counting, if the centre of a smaller halo is contained within the analysis radius \(R_h\) of a larger halo, the stress tensor contribution from the smaller halo is not included.  All error bars in the following plots are estimated using jackknife resampling of the simulation box, divided into 125 sub-volumes. 

\begin{figure}[!t]

    \centering

    \includegraphics[
        width=0.91\linewidth,
        trim=5 5 10 0,
        clip
    ]{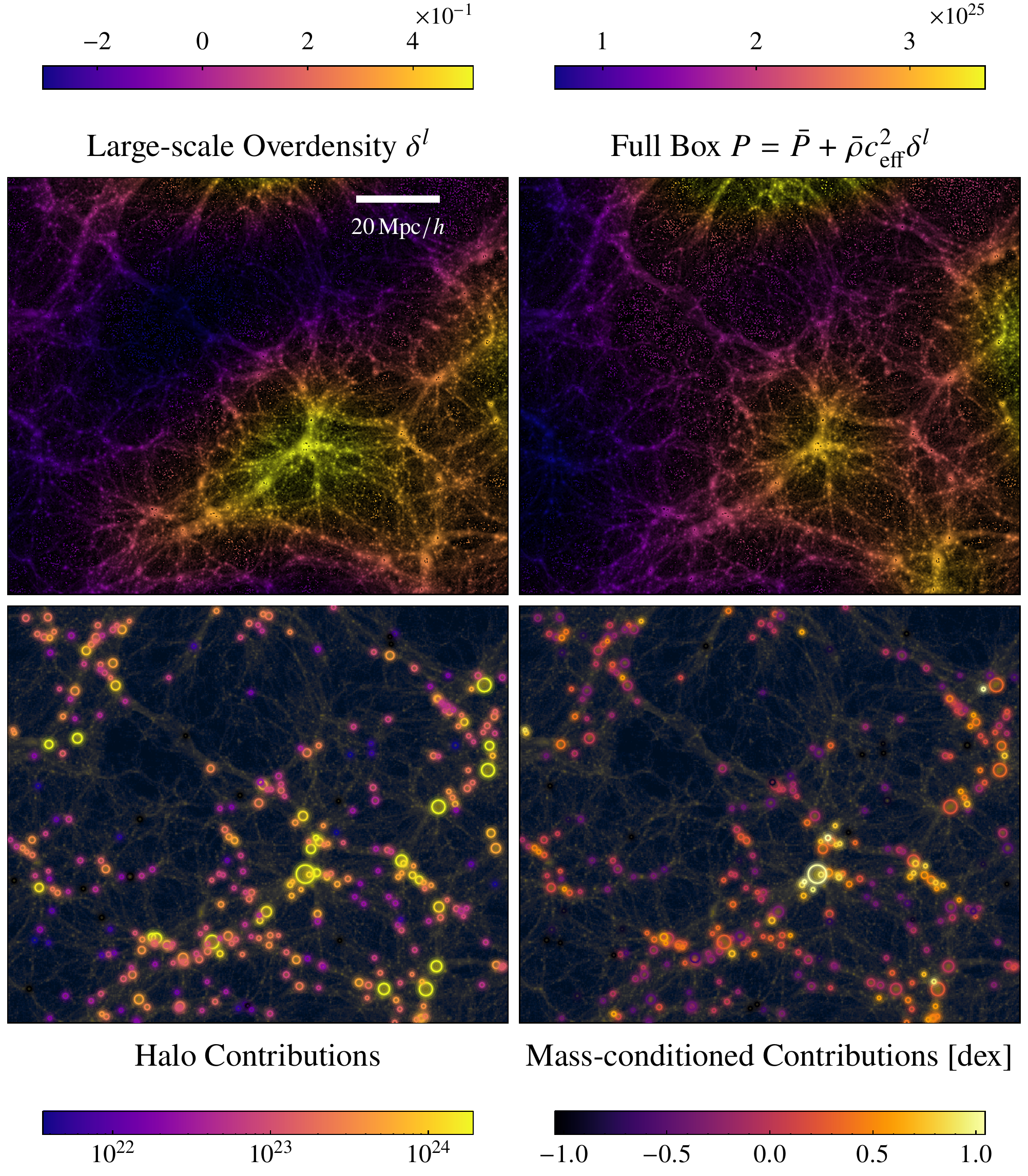}

    \caption{\small A slice of our \(250\hspace{0.1cm}\mathrm{Mpc}/h\) simulation box painting the Cosmic Web by (\textit{top left}) the large-scale overdensity \(\delta^l\) and (\textit{top right}) the full box pressure (in \((\mathrm{km/s})^2\,M_\odot h^2/ \mathrm{Mpc}^3\)). The close visual correspondence between these two panels is a direct reflection of the leading-order EFT relation between the pressure and \(\delta^l\). The bottom left panel shows the pressure sourced by individual halos with \(M_\mathrm{vir} > 10^{12.5} M_{\odot}/h\), with circle radius proportional to virial mass \(M_\mathrm{vir}\) and colour indicating pressure contributions from the halos, defined as one-third of the trace of the halo's effective stress tensor. The spatial correspondence between halo positions and regions of high full box pressure is visually apparent. Moreover, halos of similar mass can source different pressures, suggesting that their coupling to the large-scale modes strongly depends on their local environment. To make this fact stand out over the more dominant trend with mass, we include the bottom right panel.
    We bin halos by mass and colour each halo by to its logarithmic deviation from the median contribution in its mass bin. This foreshadows our finding that the counterterms are significantly sourced by massive, accreting halos.}
    
    \label{fig:visual}

\end{figure}

In the following subsections, we extract these coefficients from the effective stress tensor expansion, Eq.~(\ref{eqn:expansion}). It is important to note, however, that the quantity entering the long-wavelength expansion is the effective stress tensor smoothed over the EFT cutoff scale, \([\tau^{ij}]_{\Lambda}\), whereas our calculations thus far have yielded the effective stress tensor of individual halos, \(\tau^{ij}\). Thus we require an additional smoothing procedure to obtain the long-wavelength stress tensor relevant for the EFT description. This procedure is described in the following subsections. Our broader objective is to understand the physical processes which lead to the generation of the counterterms. Thus, after extracting the counterterms, we focus on identifying properties of halos which correlate with larger contributions to the effective stress tensor, with the aim of identifying the physical characteristics of halos that drive the counterterm contributions. 

Before the extraction of the counterterms, Figure~\ref{fig:visual} illustrates two central points motivating our approach: the close correspondence between the large-scale overdensity and the full box pressure directly reflects the leading-order EFT expansion of the pressure counterterm, while the similarly close correspondence with the halo pressure map shows that this pressure is predominantly sourced by individual halos, motivating the halo-based picture we develop below. The bottom right panel further highlights the variation in pressure contributions among halos of similar mass. We divide the halos into fifteen mass bins and compute the median logarithmic pressure contribution in each bin. Each halo is then coloured according to its logarithmic deviation from the median of its mass bin. The figure suggests that factors beyond halo mass, such as the local environment and accretion activity, influence halo contributions to the counterterms.

\subsubsection{Background Pressure and Effective Sound Speed}\label{sec:counter1}

In this section we focus on computing the counterterms related to the trace of the effective stress tensor. Specifically, we will analyse the physical origins of the background pressure, \(\bar{P}\) and the effective sound speed, \(c_{\mathrm{eff}}^2\), where: 
\begin{equation}
c_{\mathrm{eff}}^2 = c_s^2 - \frac{c_{bv}^2}{Ha} \frac{\langle \theta_l \delta_l \rangle}{\langle \delta_l^2 \rangle}\,,
\end{equation}
where \(\theta_l = \partial_k v^k_l\) is the large-scale velocity divergence.

For the smoothing scales we are interested in, \(\delta_l\) and \(\theta_l\) are highly correlated\footnote{\(\delta_l\) and \(\theta_l\) are not completely degenerate as we are not in the completely linear regime. For example, at a smoothing scale of \(28 \hspace{0.1cm} \mathrm{Mpc}/h\) they have a correlation coefficient r\(\sim 0.9\).}, resulting in noisy individual measurements of \(c_s^2\) and \(c_{bv}^2\). For standard EFTofLSS observables, it is only the particular linear combination of \(c_s^2\) and \(c_{bv}^2\) given above as the effective sound speed, that is measurable at leading order in perturbation theory\footnote{In the EFTofLSS literature, the shear viscosity \(c_{sv}^2\) is typically lumped together with those from the other counterterms, such as \(c_s^2\) and \(c_{bv}^2\), into the combination \(c_{\mathrm{eff}}^2\), since these contributions are degenerate at the level of the matter power spectrum. Here, we instead treat the shear viscosity contribution separately, since we show below that it can be estimated independently of this degeneracy.}. The individual coefficients are not separately observable unless one has additional information or observables that break the degeneracy. Thus, in the following we focus on \(c_{\mathrm{eff}}^2\).
To smooth the effective stress tensor, we use the real space Top Hat filter, consistent with our calculation for the halos. 100,000 spheres of radius \(R_{\Lambda}\) are randomly sampled throughout the simulation box, and their contributions are averaged over. We have checked that our results are not sensitive to the choice of the number of random spheres chosen. Within each sphere the total long-wavelength pressure due to the halos, \(\hat{P}^l\), is computed as: 
\begin{equation}\label{eqn:halopressure}
\hat{P}^l =
\frac{\displaystyle
\sum_{\text{halos} \hspace{0.1cm}\in\hspace{0.1cm}\text{sphere}}
(P^{\mathrm{halo}} \cdot V^{\mathrm{halo}})
}{V_{\mathrm{sphere}}} \hspace{0.5 cm}\mathrm{where,}\hspace{0.2cm} P^{\mathrm{halo}} = \frac{1}{3}Tr(\tau_{ij}^{\mathrm{halo}})\,.
\end{equation} \\
Thus, in a single sphere, \(\hat{P}^l\) is simply the volume-averaged sum over volume weighted pressures of all halos contained in the sphere. We also construct a long-wavelength overdensity field \(\delta^l\), by first constructing the overdensity field \(\delta\) on a \(1024^3\) grid and then smoothing it by a real space Top Hat filter. 

As a result, for each sphere we have a measurement for \((\delta^l, \hat{P}^l)\), and we can use the following relation: 
\begin{equation}
    \hat{P}^l = \bar{P} + c_{\mathrm{eff}}^2 \hspace{0.1 cm}\bar{\rho} \hspace{0.1 cm}\delta^l\,,
\end{equation}
from which the counterterms can be extracted as: 
\begin{equation}\label{eqn:avgpressure}
    {\langle \hat{P}^l \rangle = \bar{P}} \hspace{0.5 cm}\mathrm{and} \hspace{0.5 cm} {c_{\mathrm{eff}}^2 = \frac{\langle \hat{P}^l \delta^l\rangle}{\bar{\rho} \langle \delta^l \delta^l\rangle}}\,,
\end{equation}
where the averaging is performed over the centres of all 100,000 random spheres.

\begin{figure}[!t]
    \centering
    \begin{subfigure}[t]{0.49\textwidth}
        \centering
        \includegraphics[width=\linewidth]{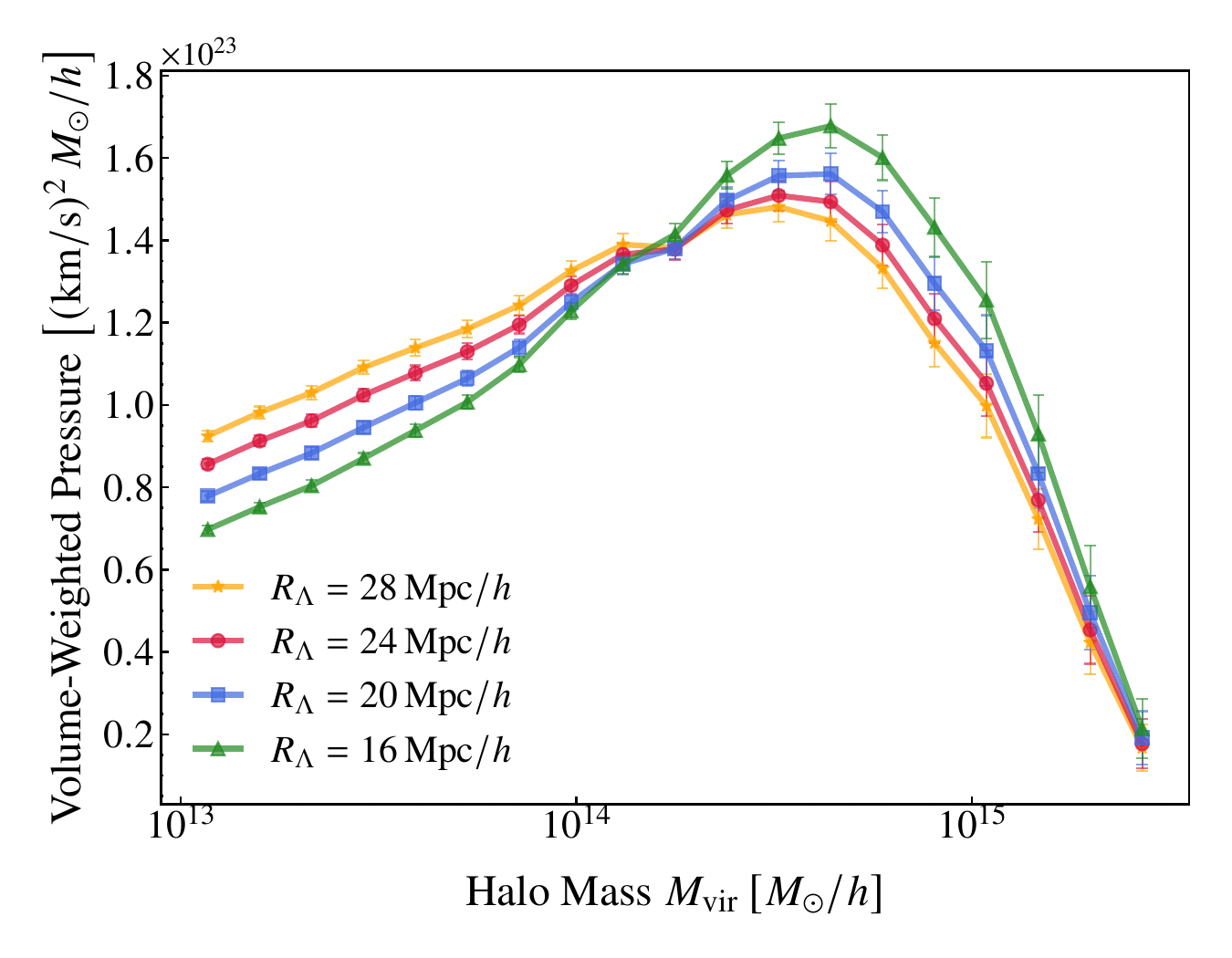}
    \end{subfigure}
    \hfill
    \begin{subfigure}[t]{0.49\textwidth}
        \centering
        \includegraphics[width=\linewidth]{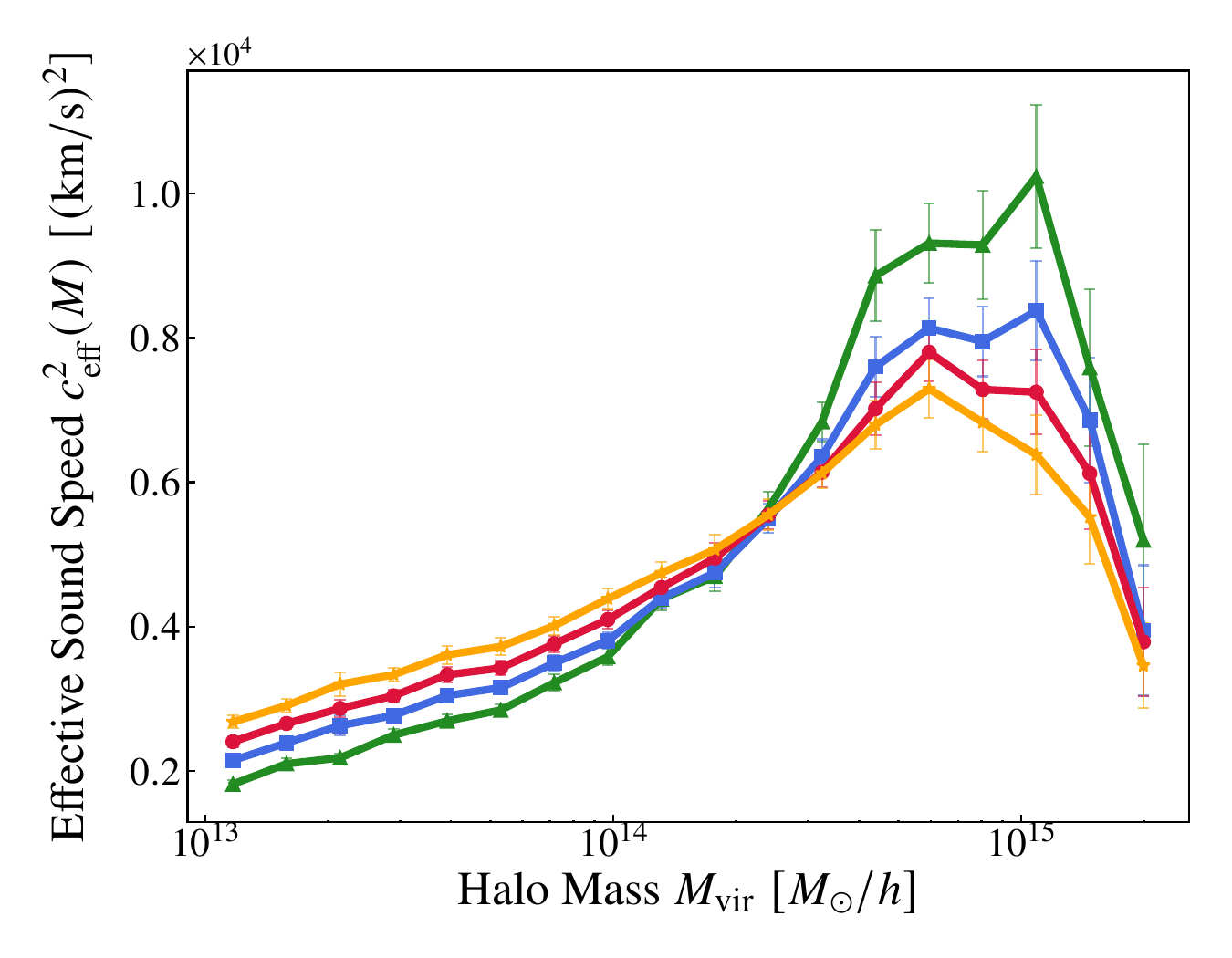}
    \end{subfigure}
    \caption{Contributions to \(\bar{P}\) and \(c_{\mathrm{eff}}^2\) by Halo Mass, for various \(R_{\Lambda}\). We note that the contributions peak in the cluster-scale halo mass range in both cases, due to the combination of halo abundance and the magnitude of the effective stress tensor.}
    \label{fig:halomass1}
\end{figure}

\begin{figure}[!t]
    \centering
    \begin{subfigure}{0.49\textwidth}
        \centering
        \includegraphics[width=\linewidth]{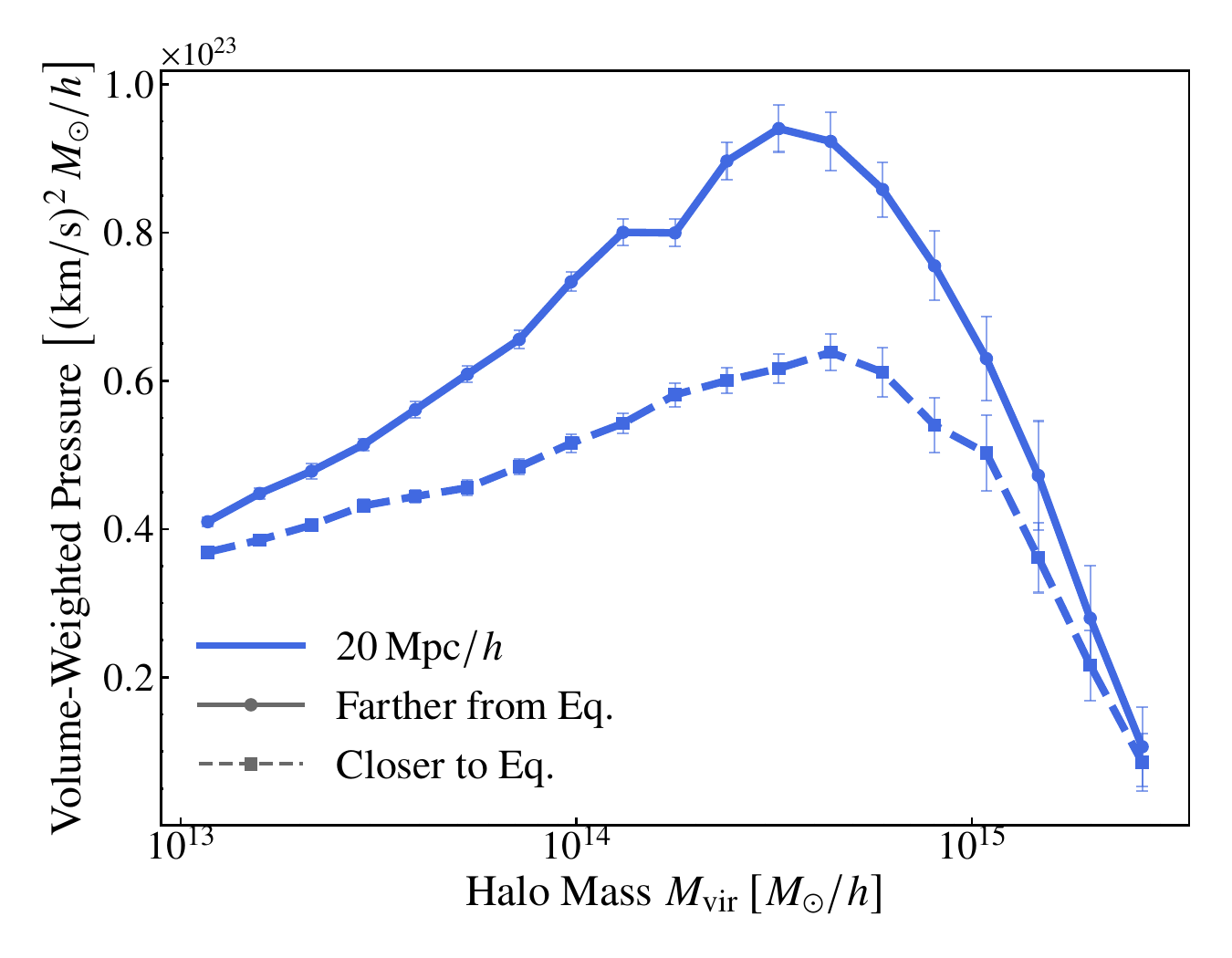}
    \end{subfigure}
    \hfill
    \begin{subfigure}{0.49\textwidth}
        \centering
        \includegraphics[width=\linewidth]{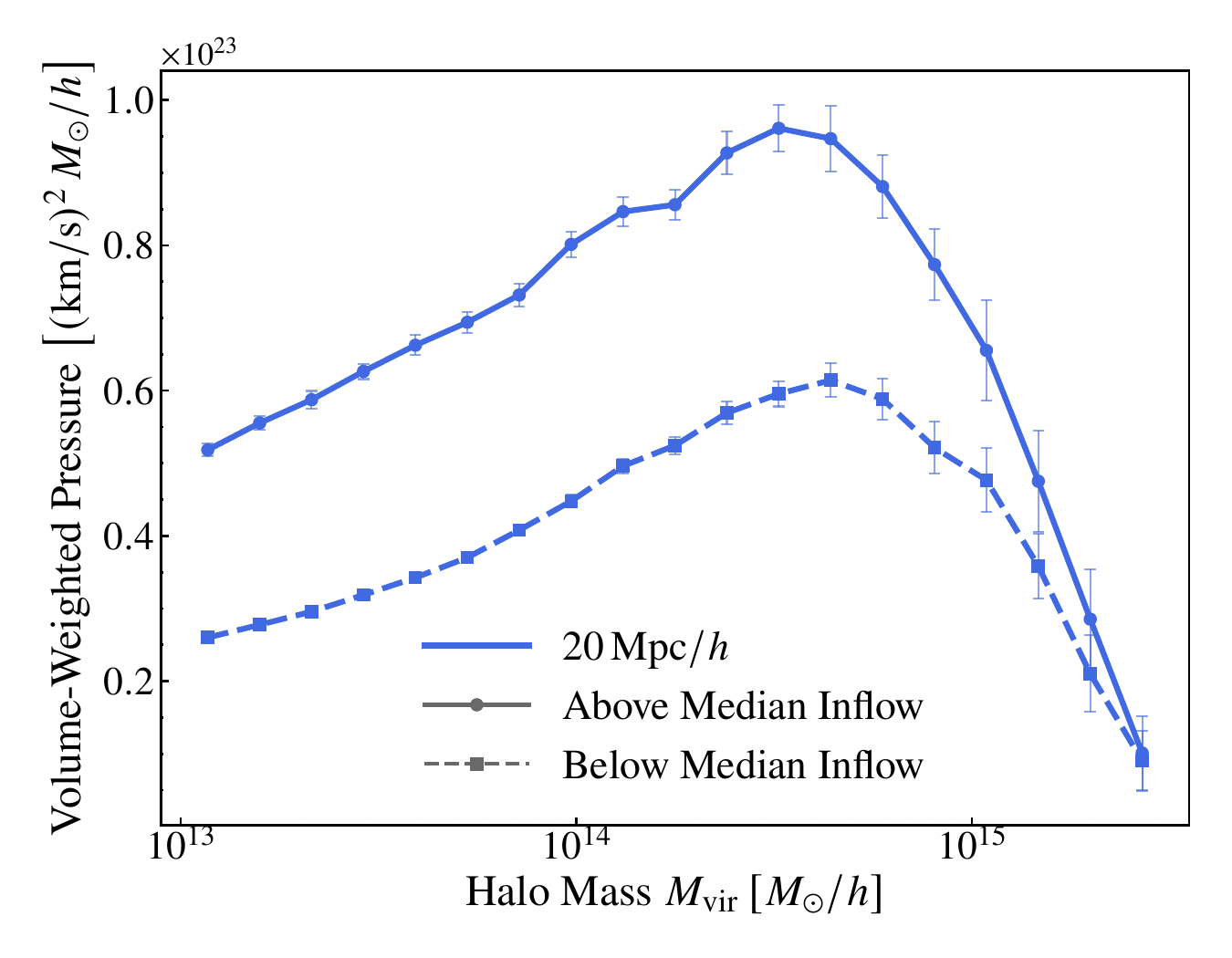}
    \end{subfigure}
    \caption{Contributions to \(\bar{P}\) split by Virial Departure \(|2(T/|U|) - 1|\) (\textit{left}) and total inflow rate (\textit{right}). We show results for \(R_\Lambda=20\,\mathrm{Mpc}/h\) for clarity. The qualitative trends remain stable across smoothing scales. The above median population dominates for a large range of halo masses in both cases, hinting at the underlying physical process.}
    \label{fig:press1}
\end{figure}

We split the halo population into mass bins and perform the above calculation for each mass bin separately. The results are shown in Figure~\ref{fig:halomass1}. The contributions show a clear peak in the $10^{14}$--$10^{15}\,M_{\odot}/h$ halo mass range for all $R_{\Lambda}$, which is the cluster-scale halo mass range. This peak arises from the competition between halo abundance and the magnitude of the halo-based effective stress tensor. At the low-mass end, halos are numerous but each carries a small effective stress tensor, so their total contribution is suppressed. This suppression is investigated further in Appendix~\ref{app:smallbox1}. At the high-mass end, each halo carries a large effective stress tensor, but the halo mass function falls steeply, so very few such halos exist in the box and their total contribution is suppressed again. Cluster-scale halos lie in between these regimes. Even though the number of $\sim 10^{14}\,M_{\odot}/h$ halos is small compared to lower masses, each one contributes enough that together they contribute the most. Since cluster-scale halos are generally less relaxed, as they are actively accreting \cite{neto2007statistics,kravtsov2012formation,fakhouri2010merger}, we expect out-of-equilibrium dynamics to underlie these large contributions.

To identify the underlying physical processes, in the following we select some quantities with simple physical interpretations for each halo. In each mass bin, we then split the contributions to the counterterms by the median value of that quantity. As a result, in each mass bin we obtain above and below median populations. If we observe that either population dominates the contribution for a large mass range, we gain insight into the underlying physics. In most of the following, for clarity we show results for \(R_\Lambda=20\,\mathrm{Mpc}/h\) and note that the qualitative trends remain stable across smoothing scales.

\begin{figure}[!t]
    \centering
    \begin{subfigure}{0.49\textwidth}
        \centering
        \includegraphics[width=\linewidth]{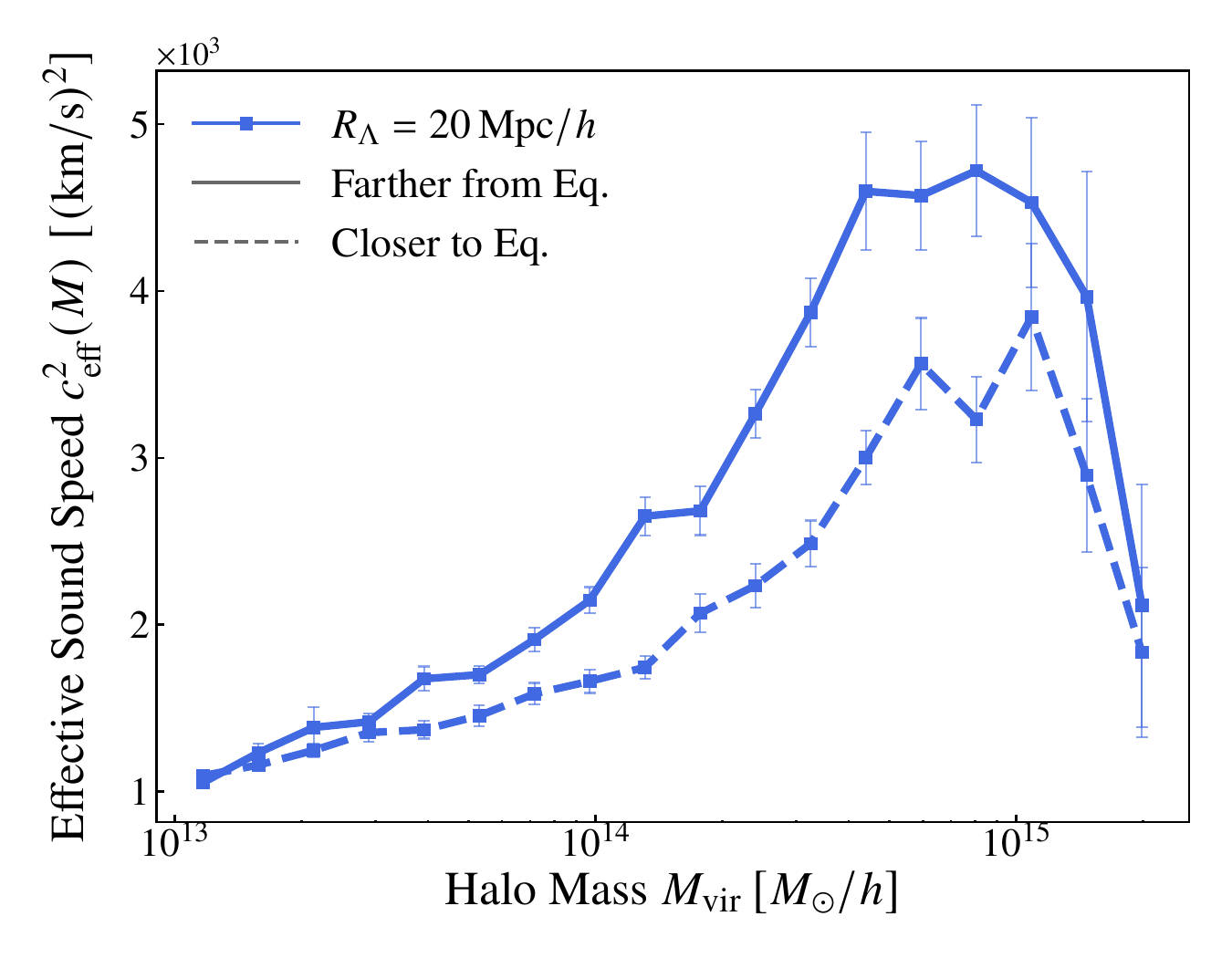}
    \end{subfigure}
    \hfill
    \begin{subfigure}{0.49\textwidth}
        \centering
        \includegraphics[width=\linewidth]{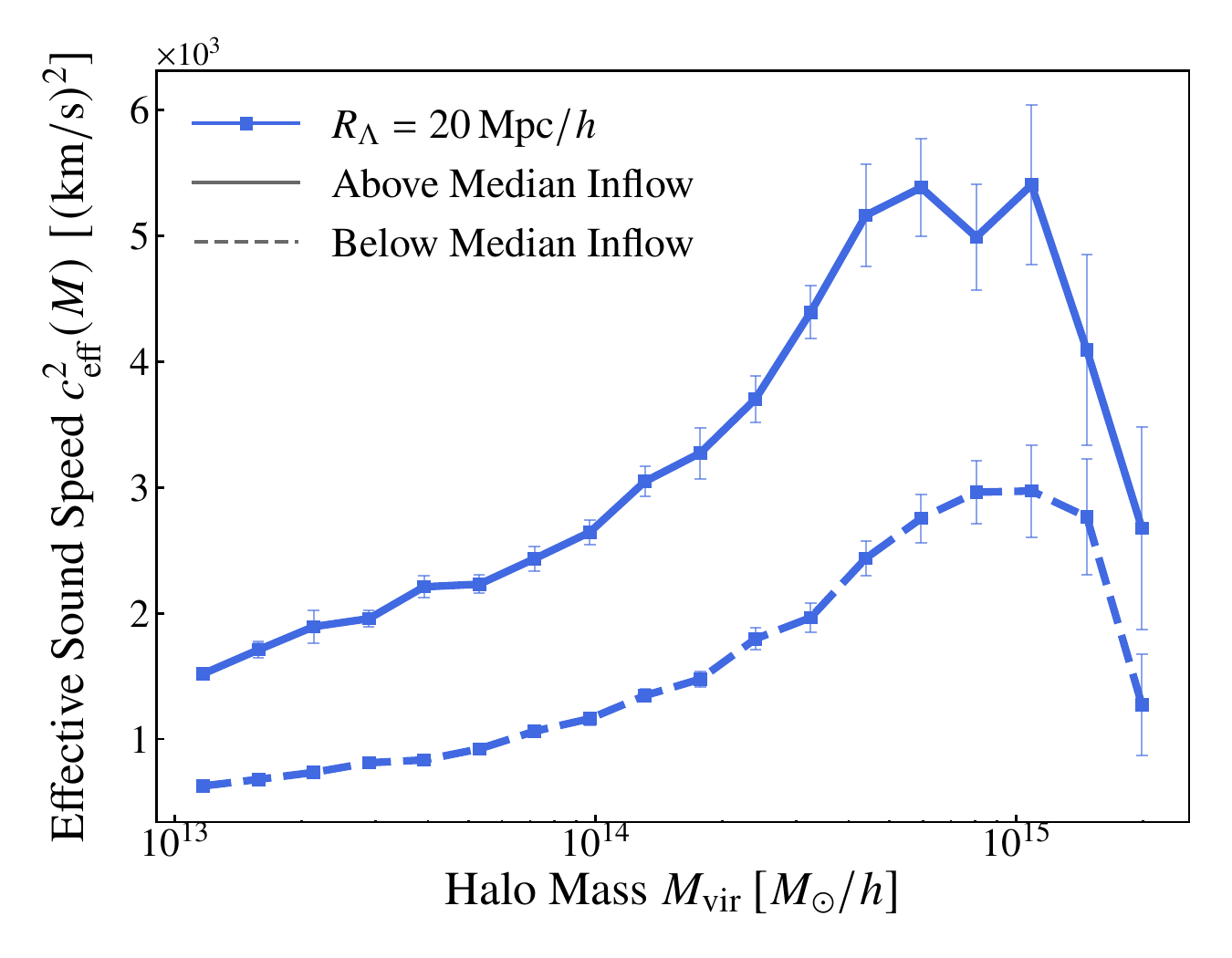}
    \end{subfigure}
    \caption{Contributions to \(c_{\mathrm{eff}}^2\) split by Virial Departure \(|2(T/|U|) - 1|\) (\textit{left}) and total inflow rate (\textit{right}) for \(R_\Lambda=20\,\mathrm{Mpc}/h\). We observe clear separation between the two populations, leading us to expect that out-of-equilibrium halos dominate and allowing us to identify the physical process as halo accretion.}
    \label{fig:ceff2virsplit}
\end{figure}

In the left panel of Figure~\ref{fig:press1} and Figure~\ref{fig:ceff2virsplit}, we choose to split by the ratio of Kinetic and Potential energies, \(T/|U|\), provided by \textsc{Rockstar}. More specifically, we split by the virial departure, \(|2(T/|U|) - 1|\) \cite{neto2007statistics}. As the name suggests, halos with low \(|2(T/|U|) - 1|\) are nearly virialised whereas, halos with high \(|2(T/|U|) - 1|\) are far from virial equilibrium. As seen clearly in Figures~\ref{fig:press1} and~\ref{fig:ceff2virsplit}, the above median population dominates the contribution for nearly the entire halo mass range under consideration. Although this doesn't immediately provide us a physical explanation, it reaffirms the expectation that halos far from equilibrium have larger contributions to the counterterms. Next, we choose to split by the ``total inflow rate'', defined as the total radially infalling momentum in the region between \(R_\mathrm{vir}\) and \(3 \times R_\mathrm{vir}\) for each halo. This quantity can be understood as a proxy for halo mass accretion rate \cite{mcbride2009mass, fakhouri2010merger} and is related to the \(\Gamma\) parameter introduced in \cite{diemer2014dependence}. The details of the computation of total inflow rate can be found in Appendix~\ref{app:aniso}. In the right panels of Figure~\ref{fig:press1} and Figure~\ref{fig:ceff2virsplit}, we observe that total inflow rate provides even more separation between the above and below median population. From this, we expect that \textit{accretion by dark matter halos} is an essential physical process underlying the generation of the background pressure and the effective sound speed. This is reinforced by the intuition gained in Appendix~\ref{app:collide}, where we show that colliding Plummer spheres generate the diagonal elements of the stress tensor. 

\begin{figure}[!t]
    \centering
    \includegraphics[width=0.55\textwidth]{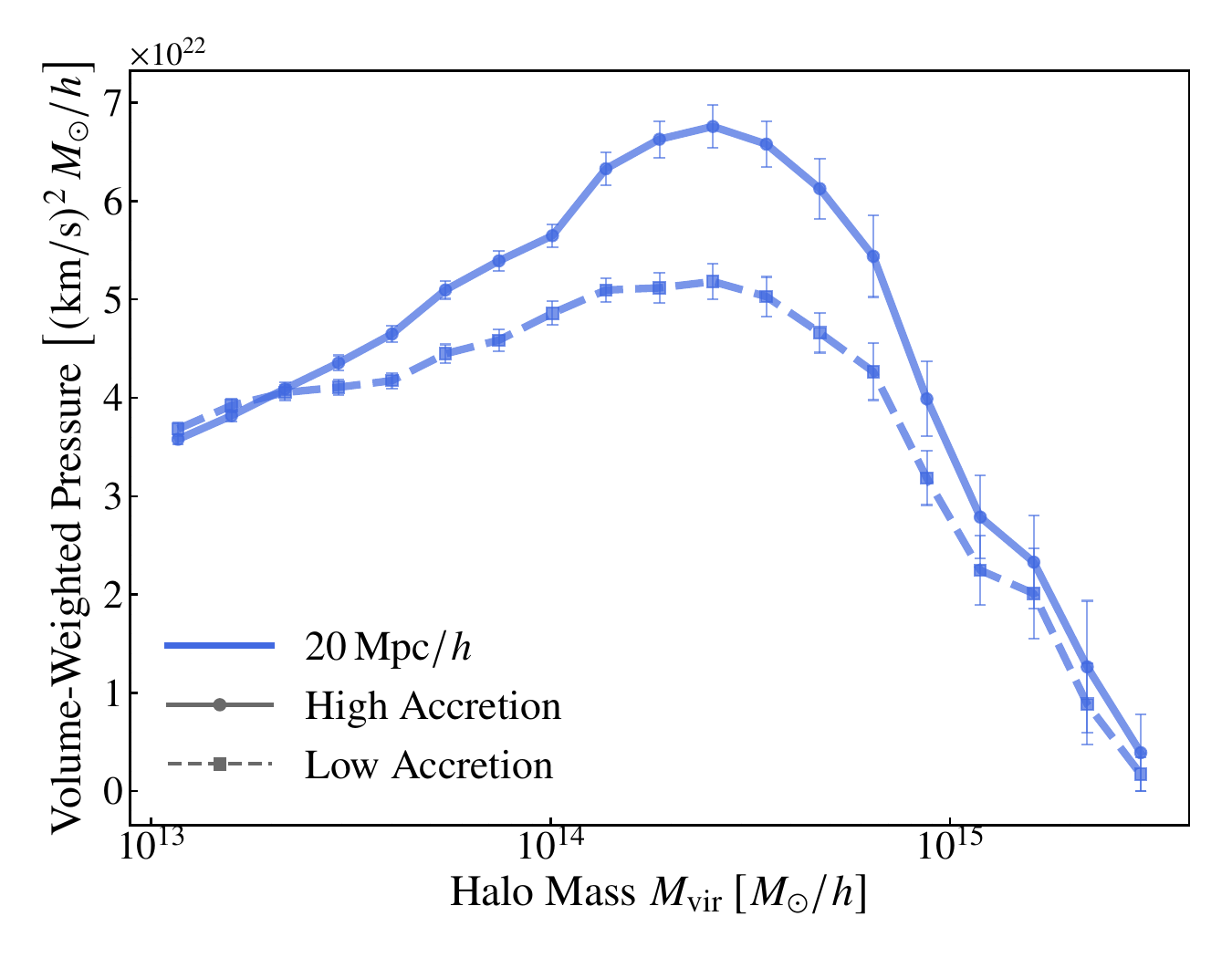}
    \caption{Contributions to \(\bar{P}\) split by \textsc{MDPL2}'s Accretion Rate. We show results for \(R_\Lambda=20\,\mathrm{Mpc}/h\) for clarity. The qualitative trends remain stable across smoothing scales. The above median population dominates, especially near the peak in the contribution. By recovering the same results on a different simulation, cosmology and accretion rate definition, we demonstrate the robustness of our calculation.}
    \label{fig:pressureaccrate}
\end{figure}

This connection can be understood in terms of accretion driving halos away from virial equilibrium. In a virialised system, the kinetic and gravitational contributions to the effective stress tensor cancel, whereas accretion introduces mass and momentum fluxes which can sustain departures from such an equilibrium. The extent of this departure depends on the competition between the accretion timescale, \(t_{\rm acc}\equiv M_{\rm vir}/\dot{M}_{\rm vir}\), and the dynamical timescale, \(t_{\rm dyn}\sim R_{\rm vir}/v_{\rm vir}\), which characterises the timescale for gravitational relaxation.  When accretion occurs on a timescale comparable to or shorter than the dynamical time, the halo may not relax completely before further material is accreted. This provides a physical interpretation for the association between accretion and the halo-based stress tensor.

To demonstrate robustness, we repeat our halo-based analysis on the $z = 0$ snapshot of the \textsc{MDPL2} simulation \cite{Klypin_2016}. \textsc{MDPL2} is a completely independent simulation, run with a slightly different cosmology from our own, allowing us to demonstrate that our results do not depend on the particular simulation or cosmological parameters we have chosen. The different cosmology leads to a different halo mass function, so the overall quantitative results are not identical. However, as we show in Appendix~\ref{app:mdpl2}, the stress tensor associated with halos of a given mass is consistent between the two analyses, and the qualitative features remain the same. Moreover, the \textsc{Rockstar} halo catalogues provided with \textsc{MDPL2} include the instantaneous halo mass accretion rate, computed as the change in halo mass between consecutive simulation snapshots divided by the time interval. Splitting the halo population by this quantity, rather than by the total inflow rate defined from the particle data, demonstrates our expectation with an independently defined measure of accretion.

\begin{figure}[!t]
    \centering

    \begin{subfigure}{0.49\textwidth}
        \centering
        \includegraphics[width=\linewidth]{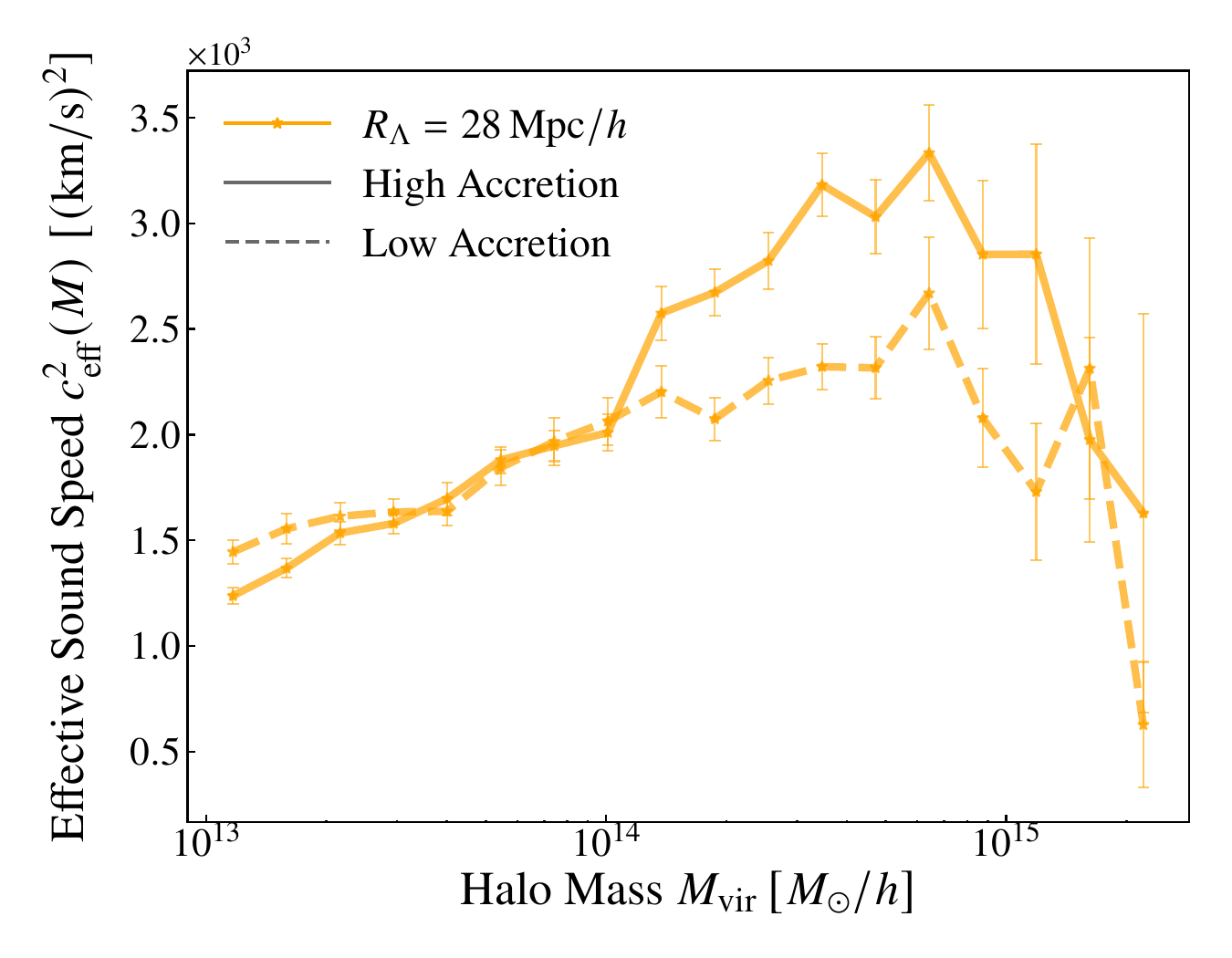}
        \label{fig:ceff2accrate1a}
    \end{subfigure}
    \hfill
    \begin{subfigure}{0.49\textwidth}
        \centering
        \includegraphics[width=\linewidth]{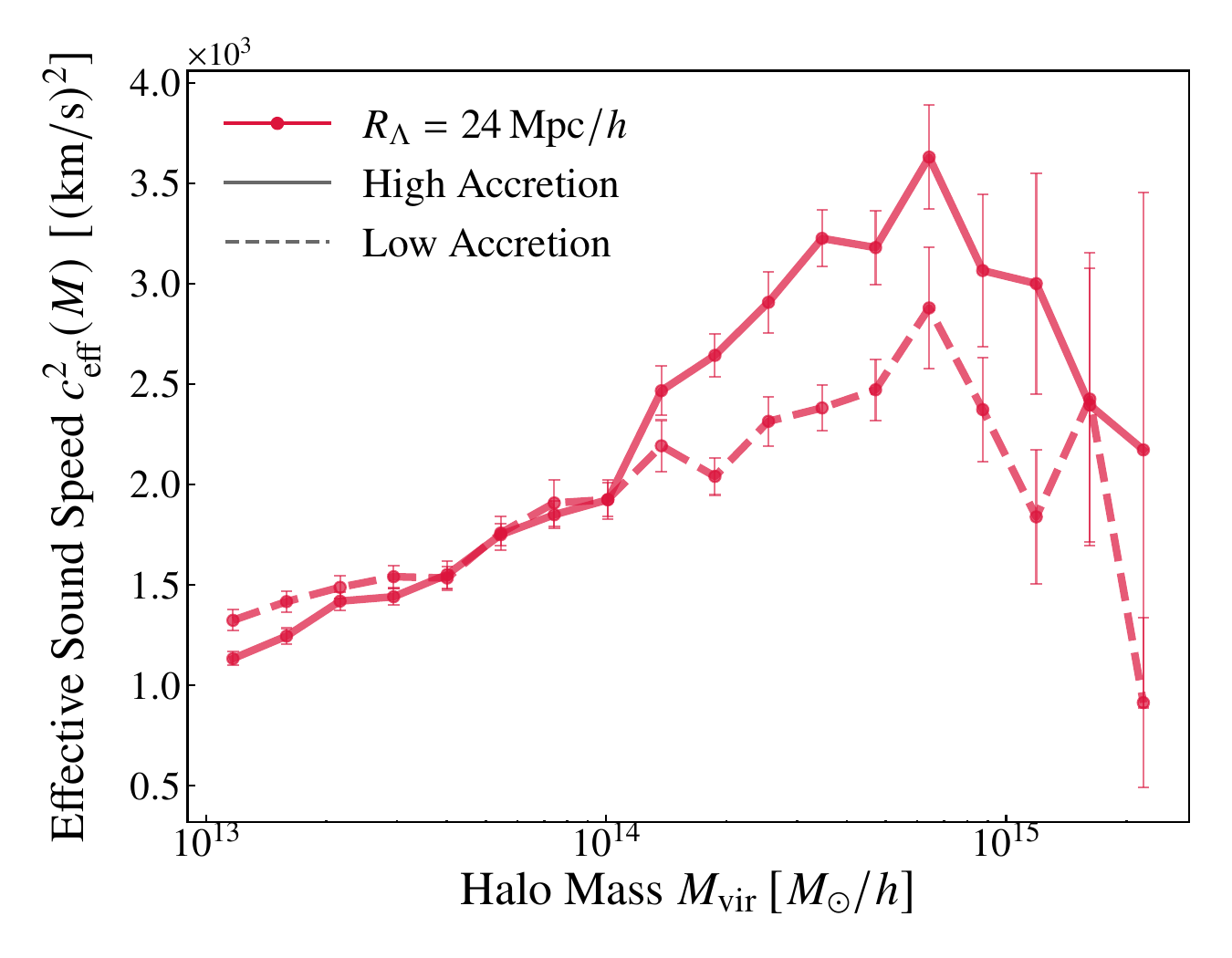}
        \label{fig:ceff2accrate1b}
    \end{subfigure}

\vspace{-0.3cm}

    \begin{subfigure}{0.49\textwidth}
        \centering
        \includegraphics[width=\linewidth]{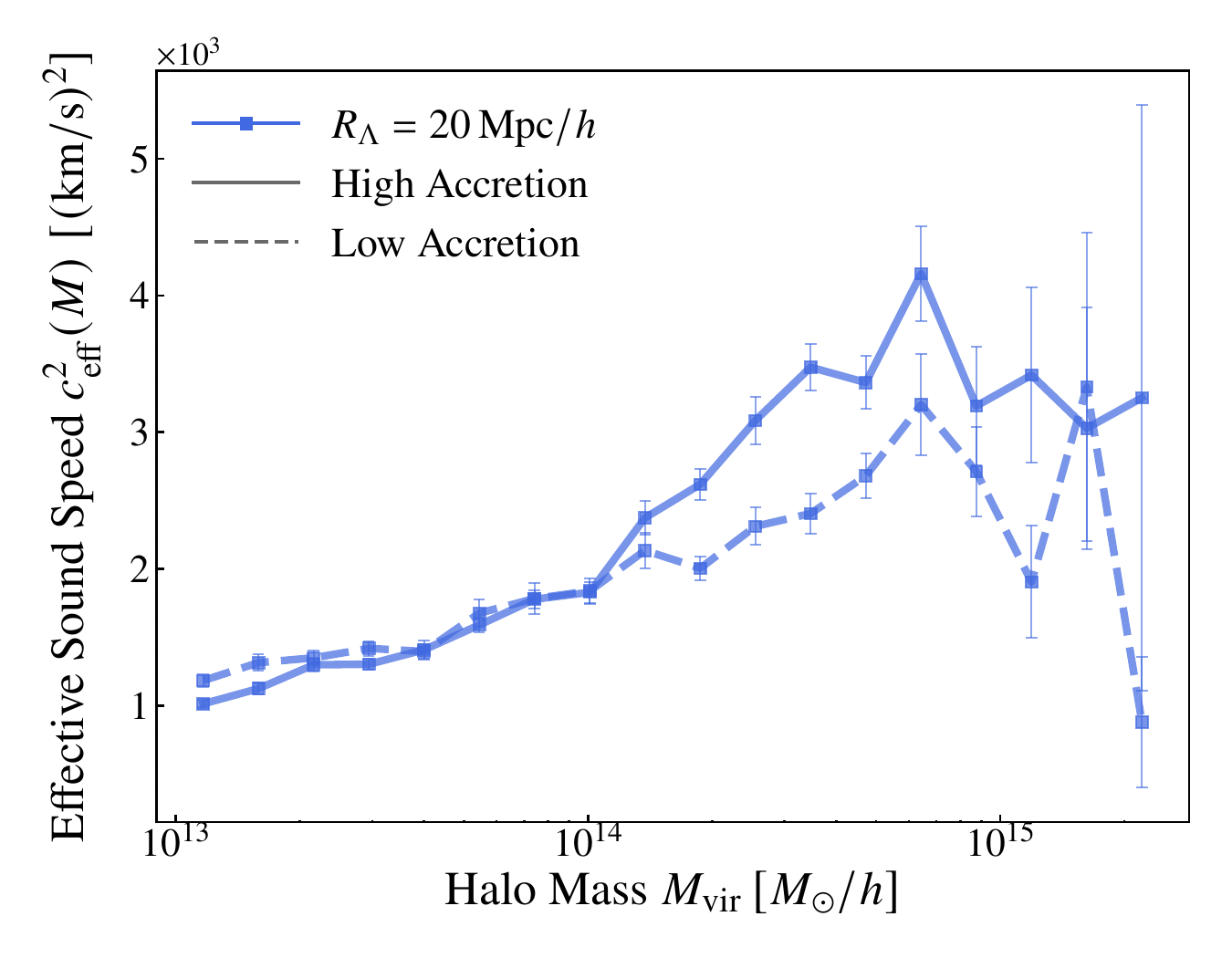}
        \label{fig:ceff2accrate1c}
    \end{subfigure}
    \hfill
    \begin{subfigure}{0.49\textwidth}
        \centering
        \includegraphics[width=\linewidth]{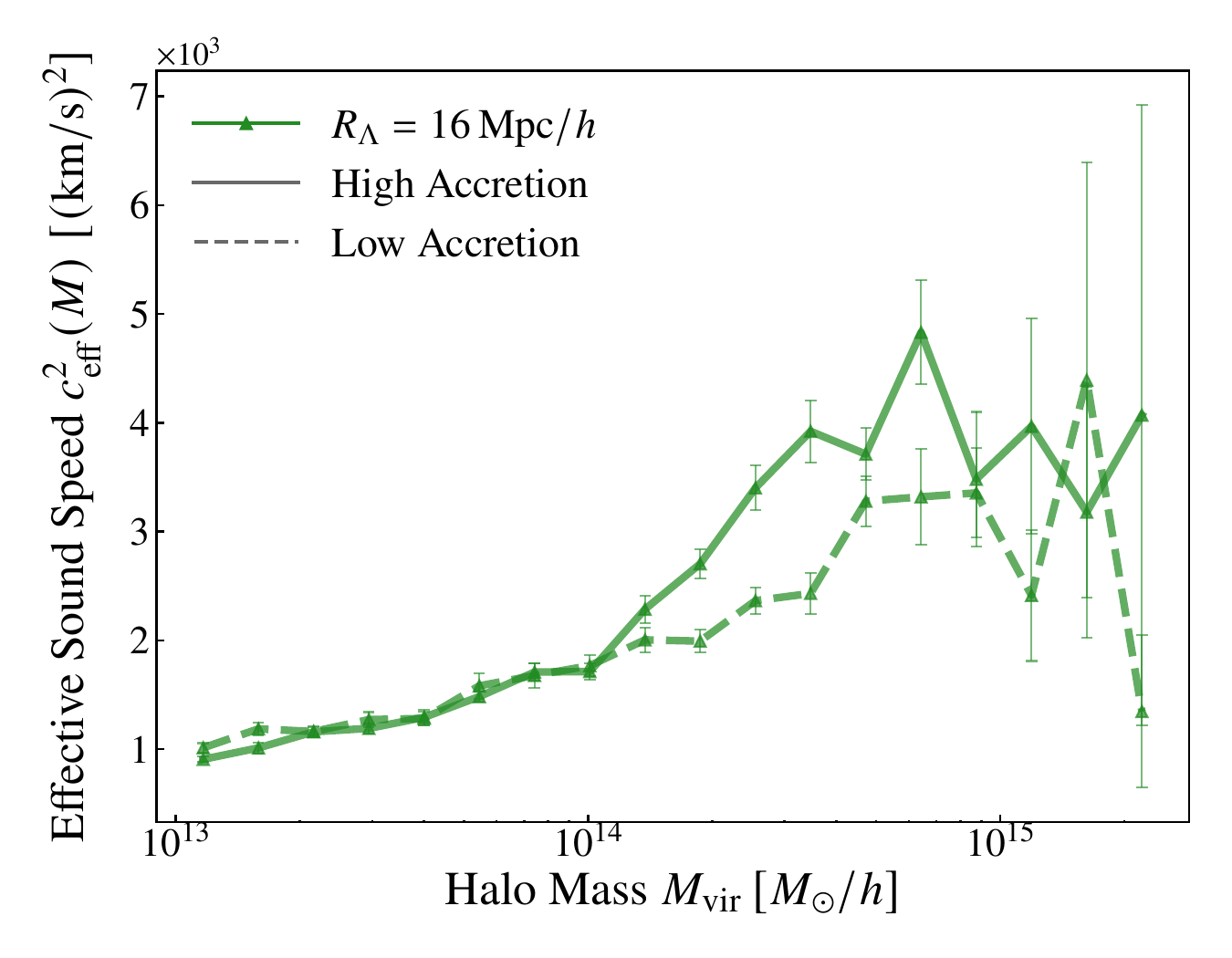}
        \label{fig:ceff2accrate1d}
    \end{subfigure}

    \caption{Contributions to \(c_{\mathrm{eff}}^2\) split by \textsc{MDPL2}'s Accretion Rate, for \(R_{\Lambda} = 16,\,20,\,24,\,28~\mathrm{Mpc}/h\). We note a split between the two populations, particularly at higher halo masses. The split is most apparent for \(R_{\Lambda}=24,\,28~\mathrm{Mpc}/h\), while the results for \(R_{\Lambda}=16,\,20~\mathrm{Mpc}/h\) are somewhat noisier.}
    \label{fig:ceff2accrate}
\end{figure}

As seen in Figure~\ref{fig:pressureaccrate}, there is clear separation between the two populations in the pressure contribution. The signal in the effective sound speed plot, shown in Figure~\ref{fig:ceff2accrate}, is weaker, but the split is visible near the peak of the contribution at cluster-scale halo masses. Halos with larger inflow rates make systematically larger contributions to both \(\bar{P}\) and \(c_{\mathrm{eff}}^2\). This trend persists in the \textsc{MDPL2} simulation, where an independently defined instantaneous halo mass accretion rate produces a clear separation in the pressure contribution, with a corresponding, albeit noisier, separation in the effective sound speed around the cluster-scale halo masses. The persistence of this qualitative behaviour across different simulations and accretion rate definitions supports the case that accretion-driven, out-of-equilibrium dynamics of dark matter halos are a key process underlying the physical origin of these counterterms.

\subsubsection{Shear Viscosity}\label{sec:counter2}

In this section we focus on computing the counterterms related to the traceless part of the effective stress tensor. Specifically, we will analyse the physical origins of the shear viscosity counterterm, \(c_{sv}^2\). From the long-wavelength expansion of the effective stress tensor, the traceless part \(\hat\tau_{ij}\) can be written as 
\begin{equation}
\left\langle [\hat\tau^{ij}]_{\Lambda} \right\rangle_{\delta_l} =- \frac{3}{4} \frac{\bar{\rho}\hspace{0.1cm} c_{sv}^2}{Ha} \left( \partial^j v_l^i + \partial^i v_l^j - \frac{2}{3} \delta^{ij} \partial_k v_l^k \right) = - \frac{3}{4} \frac{\bar{\rho} \hspace{0.1cm} c_{sv}^2}{Ha} \sigma^{ij}\,,
\end{equation}
where \(\sigma^{ij}\) is the long-wavelength velocity shear tensor. In the following, we have denoted \(\left\langle [\hat\tau^{ij}]_{\Lambda} \right\rangle_{\delta_l}\) as \(\hat\tau^{ij}\) for simplicity. 

\begin{figure}[!t]
    \centering
    \includegraphics[width=0.7\textwidth]{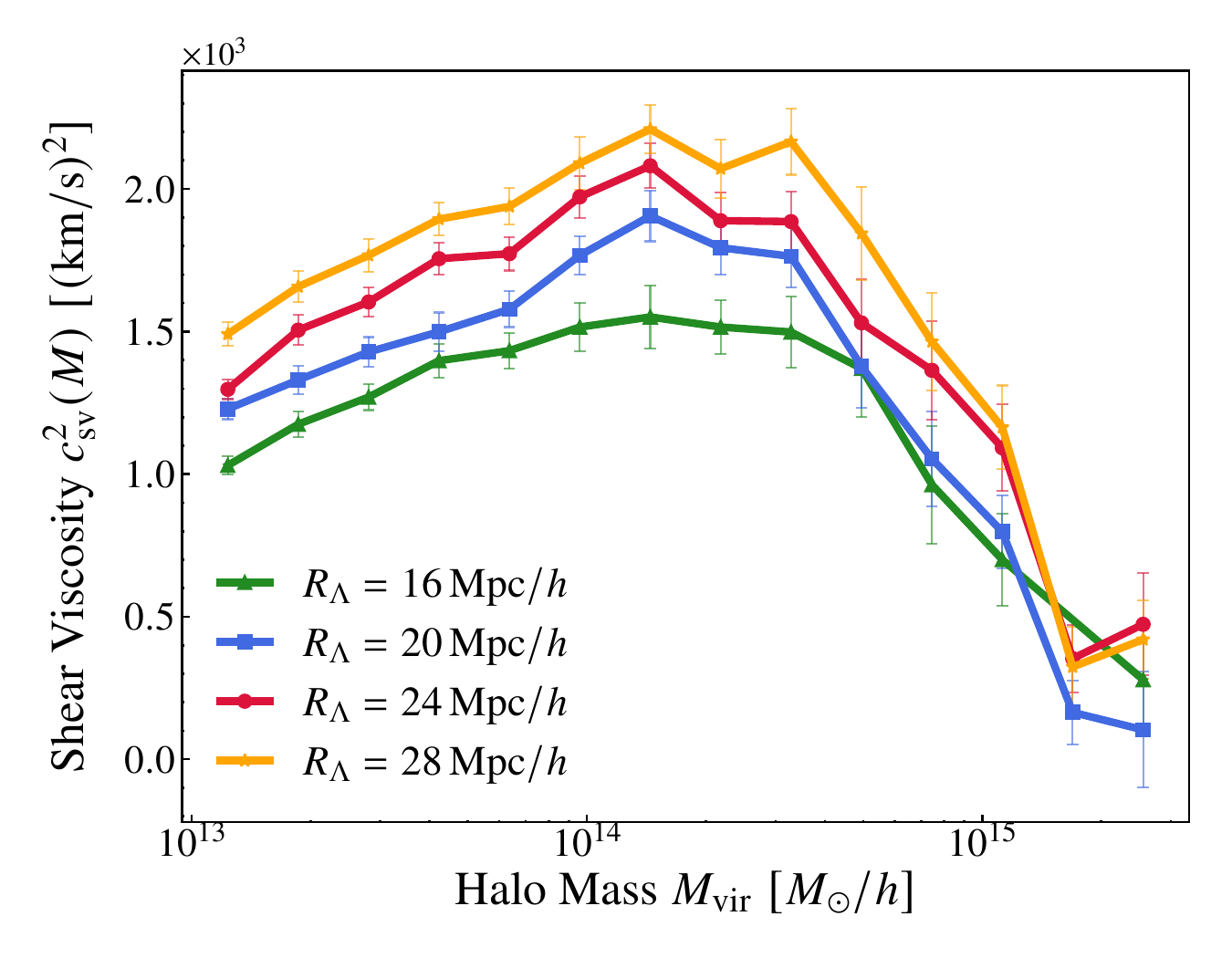}
    \caption{Contributions to \(c_{sv}^2\) by Halo Mass, for various \(R_{\Lambda}\). The peak in the contributions at cluster-scale halo masses is not as clear as in Figure~\ref{fig:halomass1}, but is still visible.}
    \label{fig:shear1}
\end{figure}

To begin the measurement, we construct a velocity field from the particle velocities, on a \(1024^3\) grid. This velocity field is then smoothed by a real-space Top Hat filter, to obtain the long-wavelength velocity field. To construct the long-wavelength velocity shear tensor, we Fourier Transform the long-wavelength velocity field, and take the necessary derivatives in Fourier space, before performing an inverse Fourier transform to obtain \(\sigma^{ij}\) in real space. Following the previous subsection, we smooth the effective stress tensor by distributing 100,000 random spheres of radius \(R_{\Lambda}\) throughout the simulation box. Within each sphere, the total traceless effective stress tensor due to the halos, \(\hat\tau_{ij}\) is computed as: 
\begin{equation}
\hat\tau_{ij} =
\frac{\displaystyle
\sum_{\text{halos} \hspace{0.1cm}\in\hspace{0.1cm}\text{sphere}}
(\hat\tau^{\mathrm{halo}}_{ij} \cdot V^{\mathrm{halo}})
}{V_{\mathrm{sphere}}} \hspace{0.5 cm}\mathrm{, where}\hspace{0.2cm} \hat\tau^{\mathrm{halo}}_{ij} = \tau^{\mathrm{halo}}_{ij} - \frac{1}{3}Tr(\tau_{ij}^{\mathrm{halo}}) \hspace{0.1cm}\delta_{ij}\,.
\end{equation}
Thus, in each sphere, we have a measurement for \((\hat\tau^{ij},\sigma^{ij})\) and the shear viscosity \(c_{sv}^2\) can be extracted as: 
\begin{equation}
    c^2_{sv} = -\frac{4 H a}{3\bar{\rho}} \frac{\langle\hat\tau^{ij} {\sigma^{ij}}\rangle}{\langle{\sigma^{ij}} {\sigma^{ij}}\rangle}\,.
\end{equation}

Following the previous subsection, we split the halo population into mass bins and perform the above calculation for each mass bin separately. The off-diagonal elements of the effective stress tensor for dark matter halos are, in general, smaller in magnitude than the diagonal elements. In addition, the velocity shear tensor involves derivatives of the velocity field, which up-weights small-scale fluctuations by an additional factor of \(k\) in Fourier space. This results in a lower signal-to-noise ratio for the shear viscosity measurement as compared to the other counterterms.

\begin{figure}[!t]
    \centering
    \begin{subfigure}{0.49\textwidth}
        \centering
        \includegraphics[width=\linewidth]{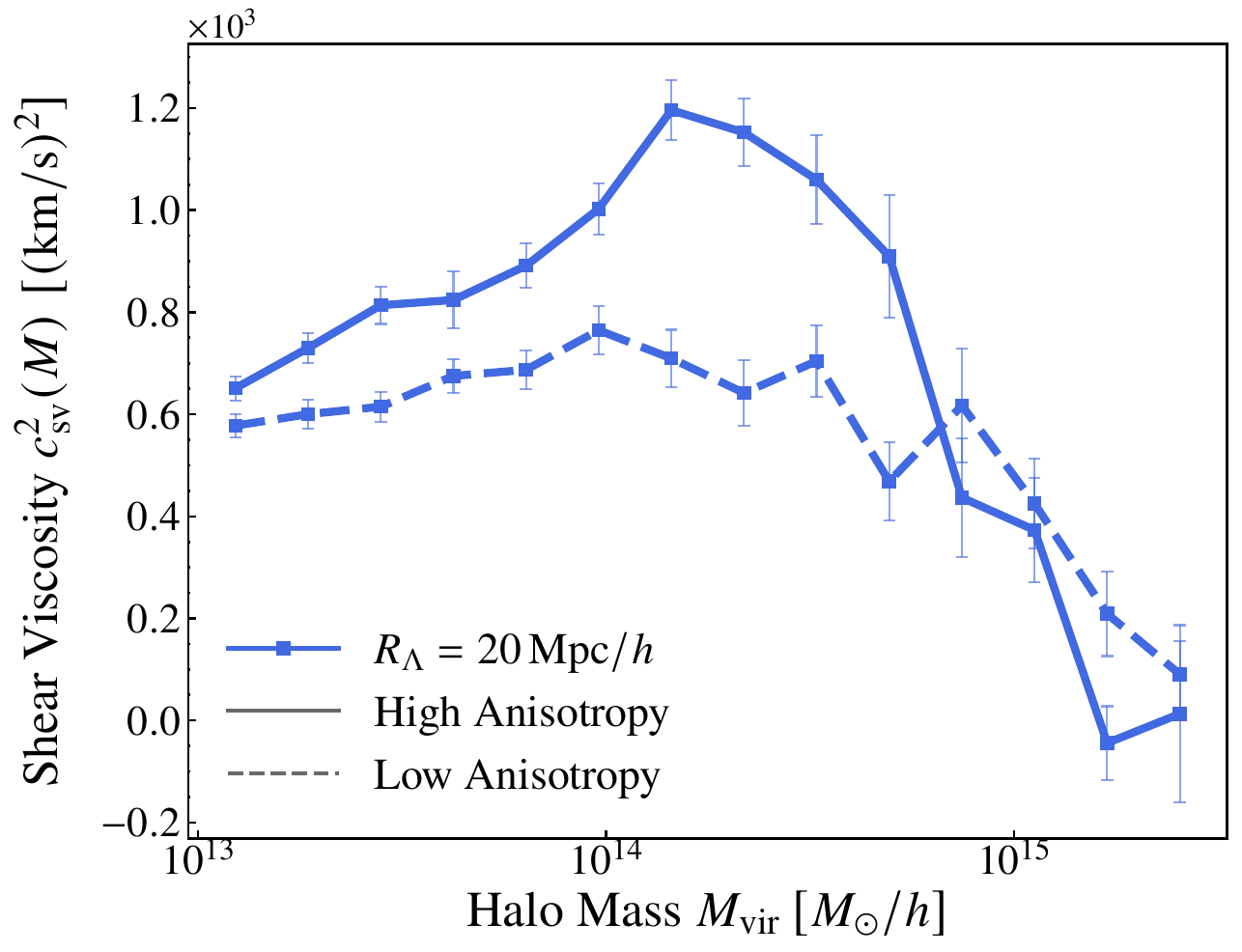}
    \end{subfigure}
    \hfill
    \begin{subfigure}{0.49\textwidth}
        \centering
        \includegraphics[width=\linewidth]{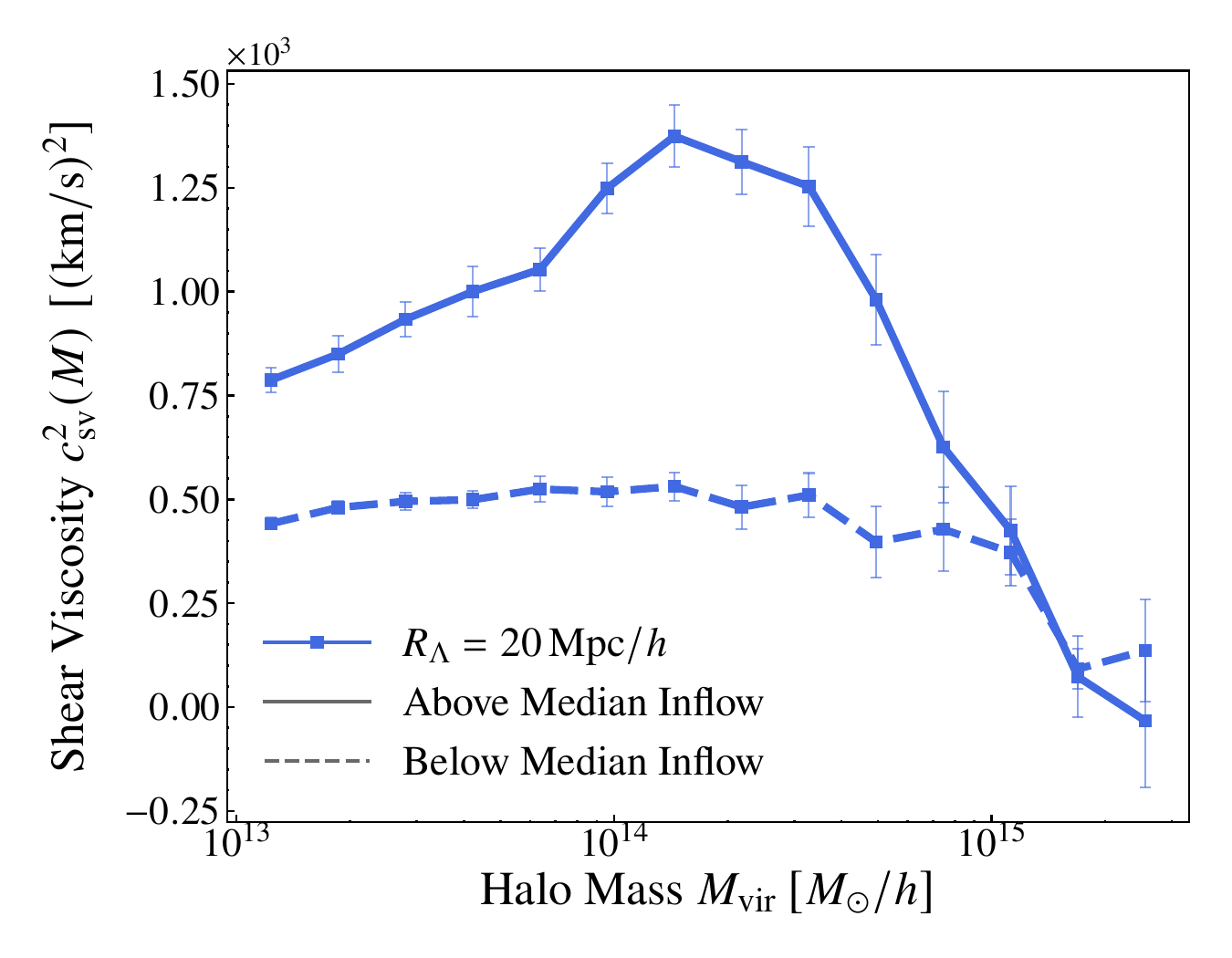}
    \end{subfigure}
    \caption{Contributions to \(c_{sv}^2\) split by the magnitude of the Anisotropic Flux Tensor \(|A^{TF}|\)  (\textit{left}) and total inflow rate (\textit{right}) for \(R_\Lambda=20\,\mathrm{Mpc}/h\). We observe a clear split between the two populations by total inflow rate, hinting at a similar underlying physical process as \(\bar{P}\) and \(c_{\mathrm{eff}}^2\). The excess contribution by the above median population of \(|A^{TF}|\) allows us to demonstrate that anisotropic accretion underlies the generation of \(c_{sv}^2\).}
    \label{fig:shear2} 
\end{figure}

\begin{figure}[!t]
    \centering
    \begin{subfigure}{0.49\textwidth}
        \centering
        \includegraphics[width=\linewidth]{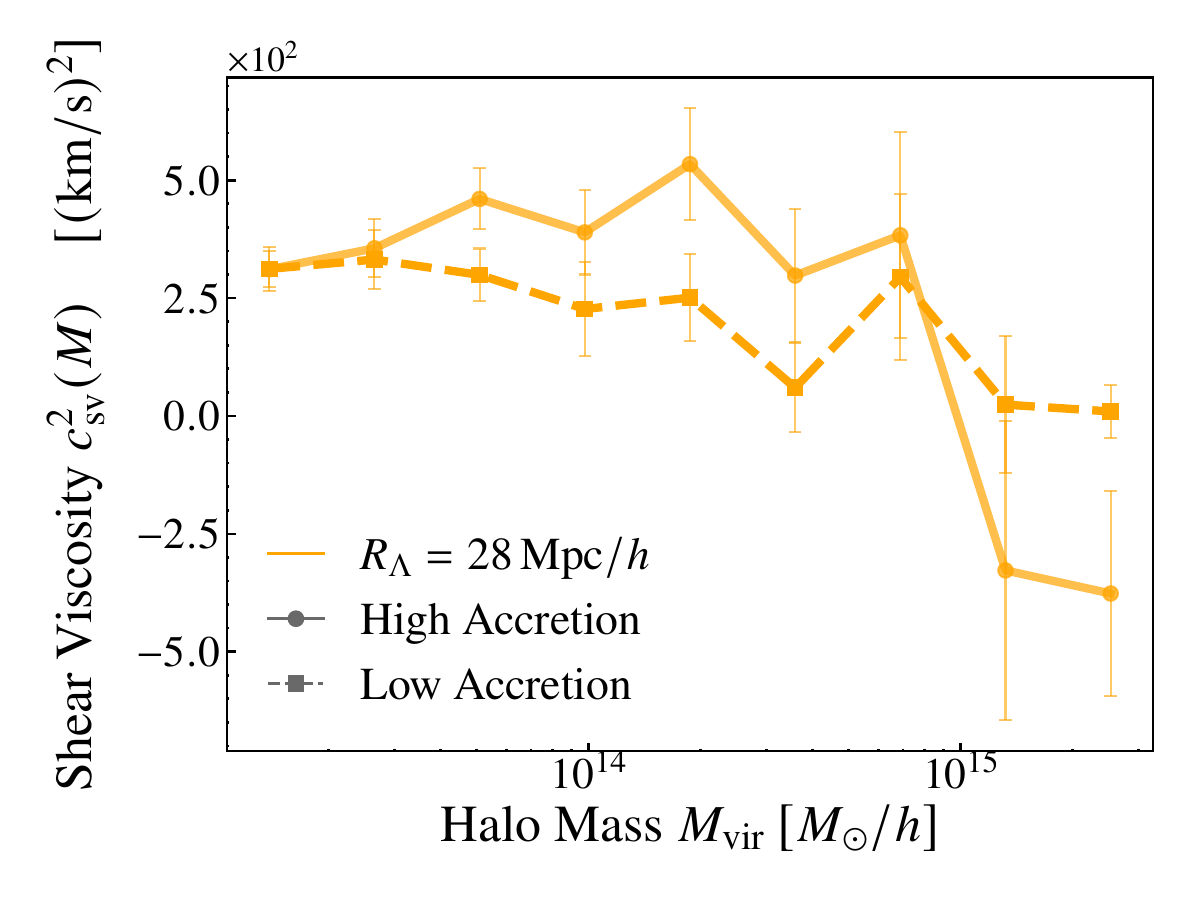}
        \label{fig:shear4a}
    \end{subfigure}
    \hfill
    \begin{subfigure}{0.49\textwidth}
        \centering
        \includegraphics[width=\linewidth]{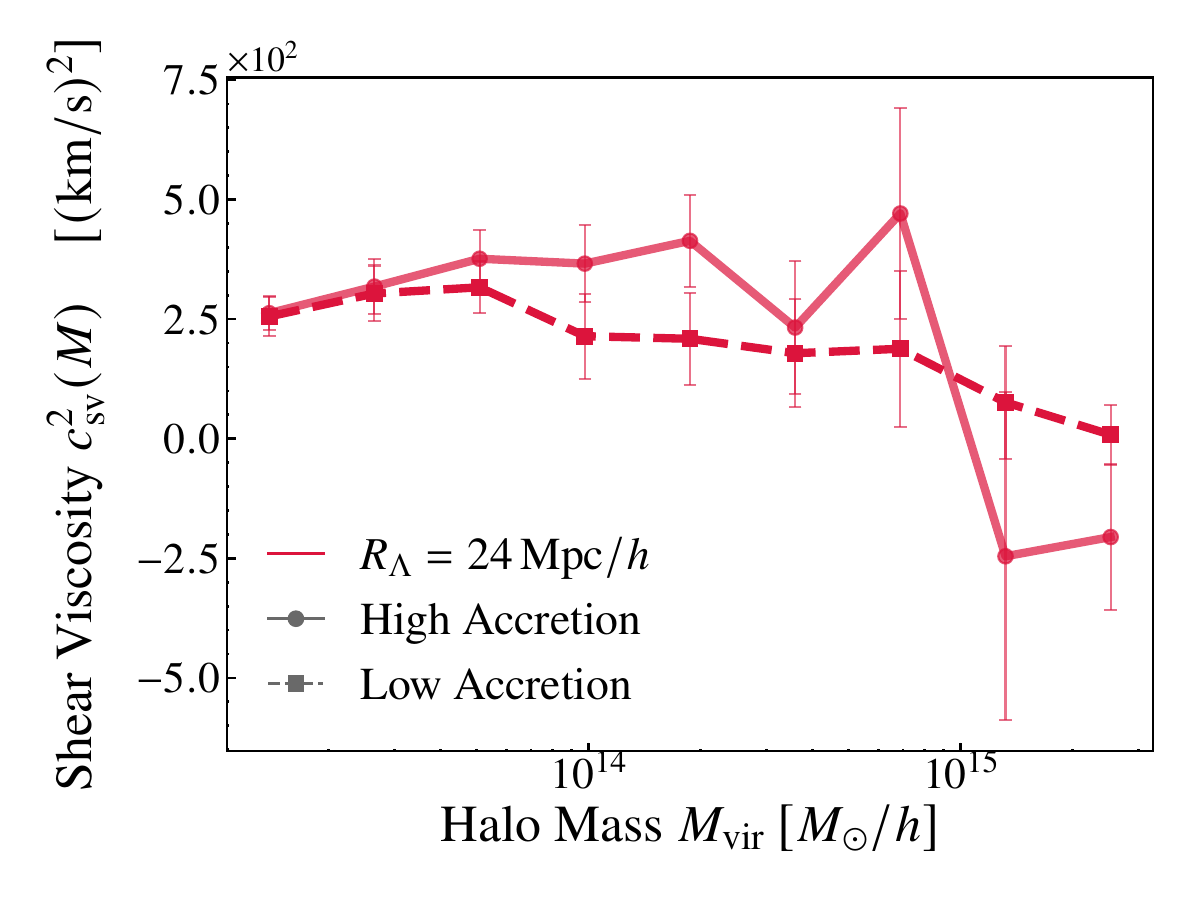}
        \label{fig:shear4b}
    \end{subfigure}

    \vspace{-0.3cm}

    \begin{subfigure}{0.49\textwidth}
        \centering
        \includegraphics[width=\linewidth]{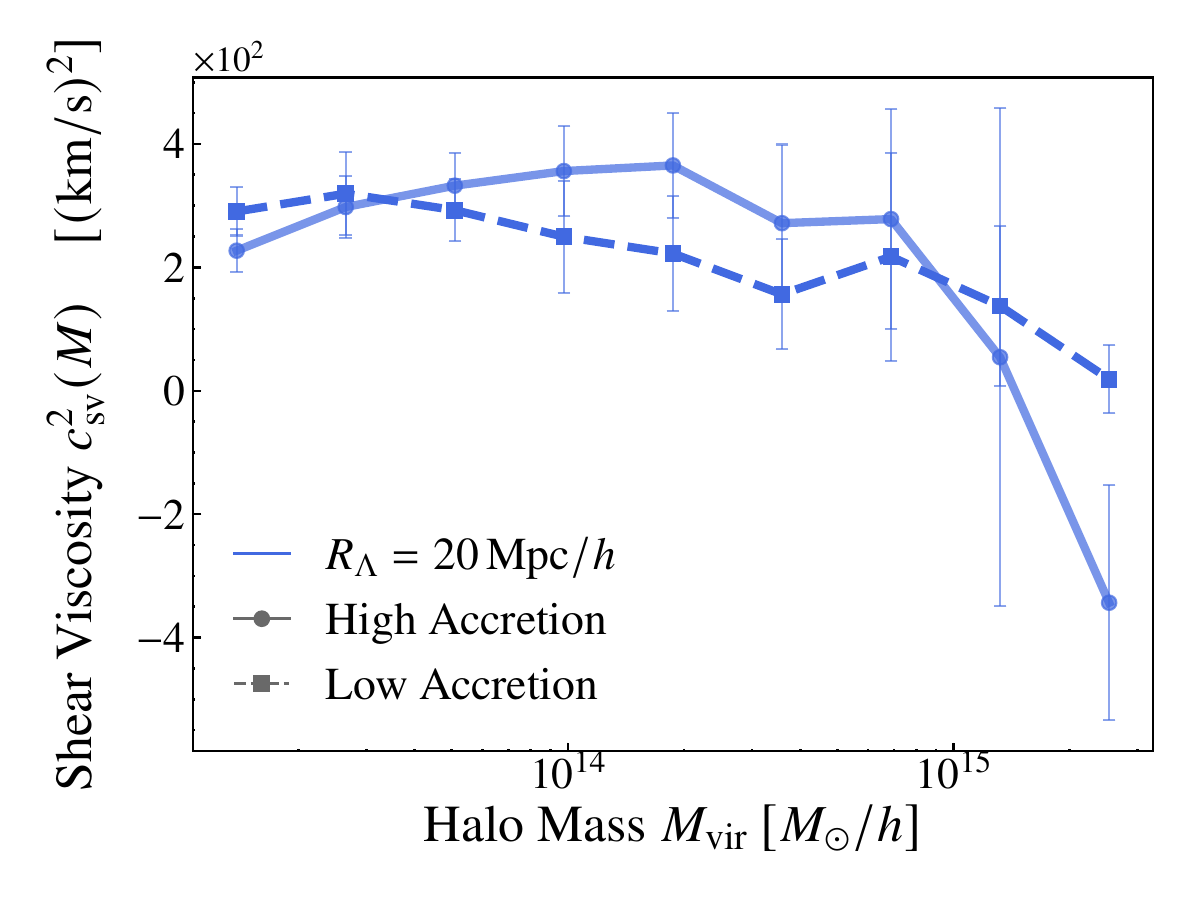}
        \label{fig:shear4c}
    \end{subfigure}
    \hfill
    \begin{subfigure}{0.49\textwidth}
        \centering
        \includegraphics[width=\linewidth]{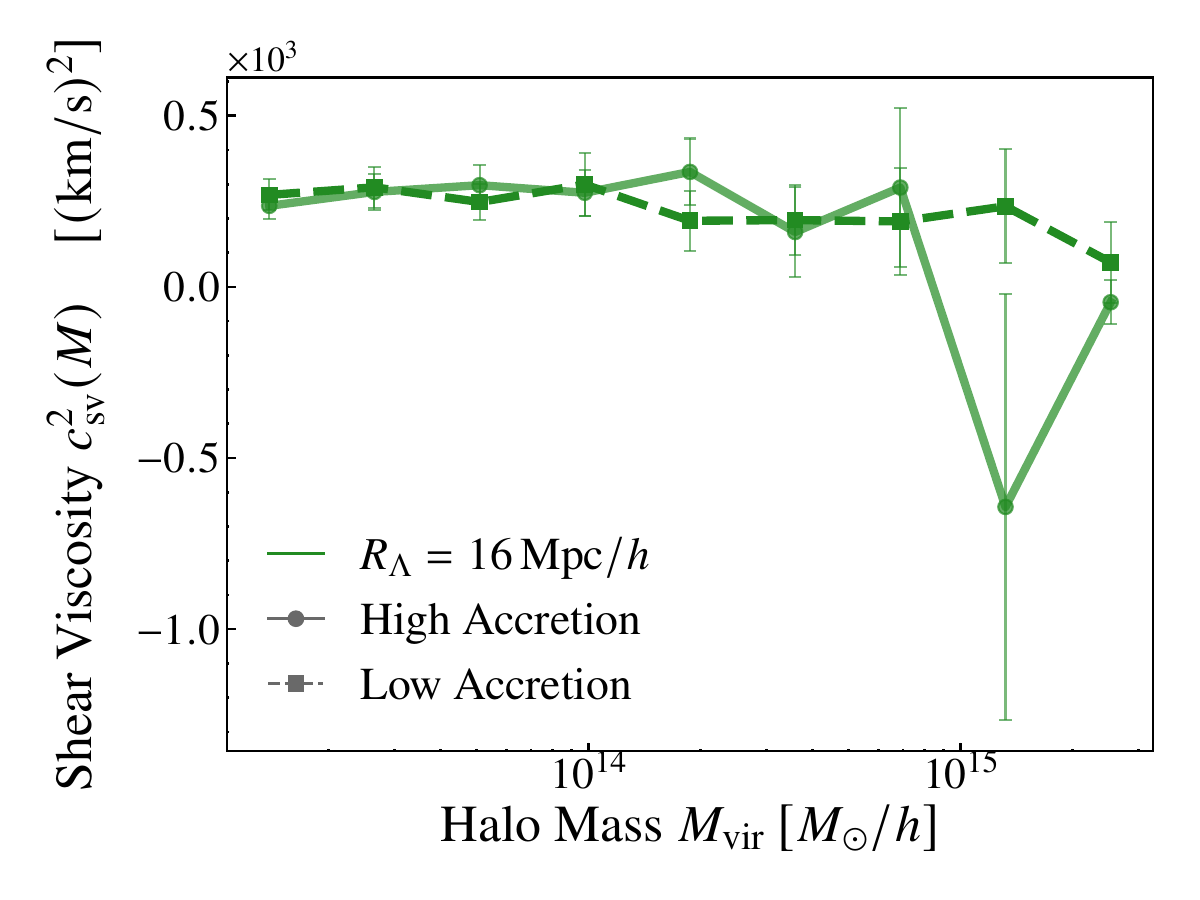}
        \label{fig:shear4d}
    \end{subfigure}

    \caption{Contributions to \(c_{sv}^2\) split by \textsc{MDPL2}'s Accretion Rate, for \(R_{\Lambda} = 16,\,20,\,24,\,28~\mathrm{Mpc}/h\). We note that the signal is present, but is noisier than the corresponding measurements for \(\bar{P}\) and \(c_{\mathrm{eff}}^2\). The split between high and low accretion rates is significantly weaker at \(R_{\Lambda}=16,\,20~\mathrm{Mpc}/h\), where the separation is not statistically significant.}
    \label{fig:cvis2accrate}
\end{figure}

Figure~\ref{fig:shear1} shows the contributions to $c_{sv}^2$ as a function of halo mass. Similar to $\bar{P}$ and $c^2_{\rm eff}$ (see Fig.~\ref{fig:halomass1}), the contribution peaks at cluster-scale halo masses and is suppressed towards lower masses. However, we also note that both features are less pronounced here. The results of the analysis performed on a smaller simulation box, and hence for lower halo masses, can be found in Appendix~\ref{app:smallbox1}. 

We note that, unlike the results of the previous subsection, the shear viscosity does not show any clear signal when split by virial departure. Since virial departure is characterised by the scalar diagnostic \(T/|U|\), the absence of a significant dependence on this quantity suggests that the shear viscosity is sensitive instead to the tensorial, anisotropic structure of the stress, rather than to the overall scalar departure from virial equilibrium.
This expectation is supported by Figure~\ref{fig:shear2}, where the above and below median populations show clear separation when split by the magnitude of the Anisotropic Flux Tensor, and portrays the even clearer split by total inflow rate. The Anisotropic Flux Tensor measures the magnitude of the anisotropy in the radial infall around a halo, and therefore provides a tensorial diagnostic that is not captured by the scalar virial diagnostic \(T/|U|\). The details of the related computation can be found in Appendix~\ref{app:aniso}. From these results, we expect that \textit{anisotropic} accretion by dark matter halos is an essential physical process underlying the generation of the shear viscosity counterterm. Yet again, this is supported by the intuition gained in Appendix~\ref{app:collide}, where we found that off-axis collision scenario generates the related off-diagonal element of the stress tensor. 

\begin{figure}[!t]
    \centering
    \includegraphics[width=1\textwidth]{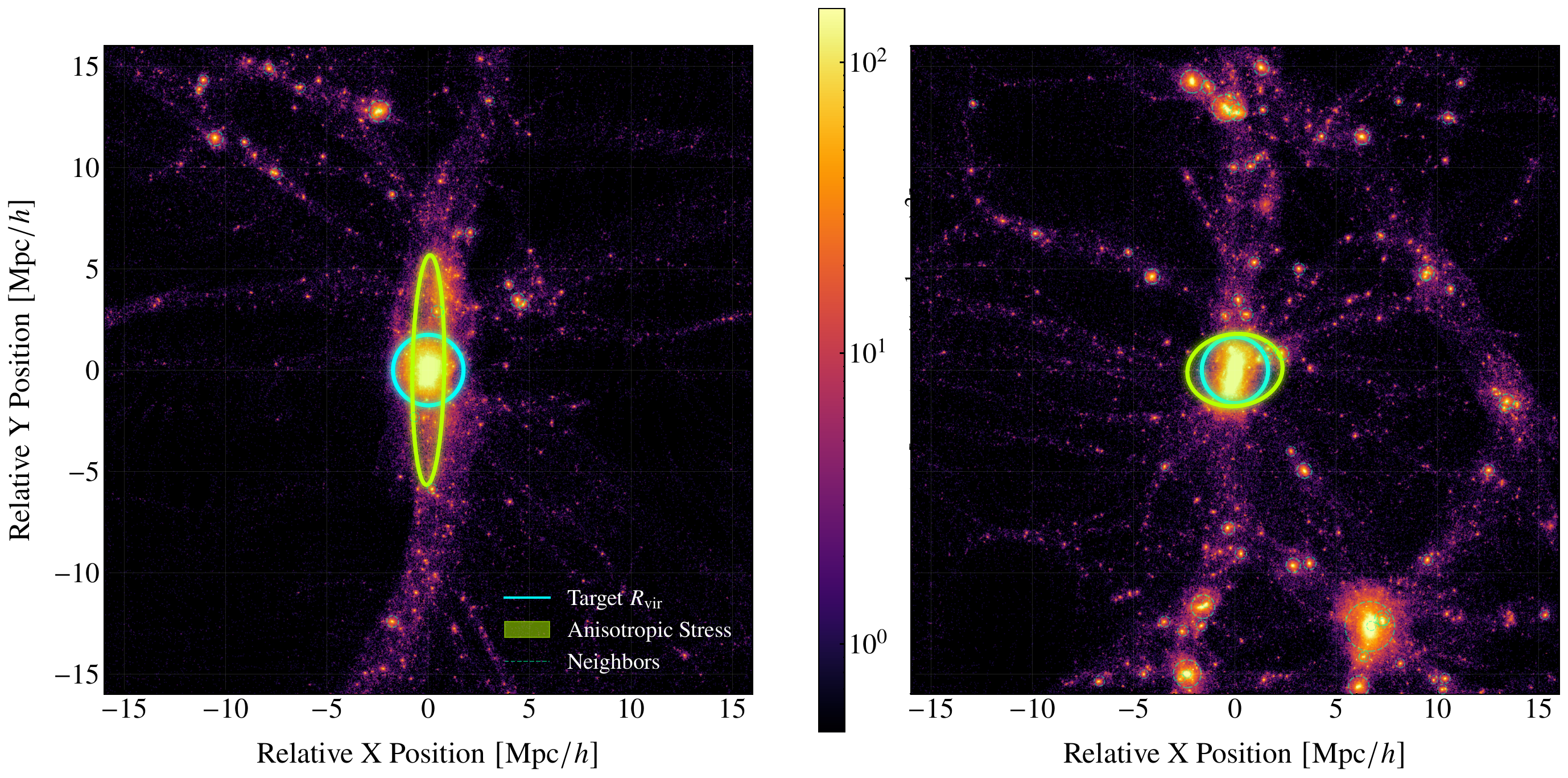}
    \caption{Orientation of the traceless anisotropic stress tensor (green ellipse), relative to a halo's local accretion geometry, for two halos. (\textit{left}) A halo embedded in a single filament, with the anisotropic stress ellipse elongated along it. (\textit{right}) A halo at a multi-filament node, where net inflow (and hence the anisotropic stress ellipse) is closer to isotropic.}
    \label{fig:haloalign}
\end{figure}

Figure~\ref{fig:cvis2accrate} depicts the split by instantaneous accretion rate from the MDPL2 analysis. We observe that the above-median population has larger contributions than the below-median one for intermediate halo masses, with the difference becoming more pronounced at larger smoothing scales. The populations are indistinguishable at the lower mass bins, and the highest mass bins exhibit large uncertainties, due to their low halo counts. The corresponding splits for smaller smoothing scales are comparatively noisier, consistent with the lower signal-to-noise ratio of the shear viscosity measurement. Nevertheless, the persistence of the trend under an independently defined accretion rate provides further evidence that accretion is an important physical source of the effective stress tensor.

In contrast to the background pressure and effective sound speed, the shear viscosity is sensitive not simply to the magnitude of accretion, but to its anisotropy. Matter accreting preferentially along filaments or other large-scale structures therefore provides a natural mechanism for generating non-zero traceless stress. This idea is illustrated directly in Figure~\ref{fig:haloalign} for two representative halos. The axes of the green ellipses around both halos point in the eigendirections of their traceless stress tensors. The axes are scaled by the eigenvalues and the volume of the ellipse is held to be the same as the volume of their respective halo. The halo in the left panel accretes primarily along a single filament, and the anisotropic stress ellipse elongates along that axis. On the other hand, the halo in the right panel sits at a node where several filaments converge. Despite the more active environment, inflow from multiple directions leads to the ellipse being more circular. This reinforces that it is the geometry, not merely the magnitude, of accretion that sources the shear viscosity counterterm.

The results presented in this section suggest the following physical picture. Dark matter halos are continuously driven away from equilibrium by the accretion of matter from their surroundings. The associated momentum flux contributes to the effective stress tensor when the halo is coarse-grained on scales larger than the halo itself. When the accretion is approximately isotropic, this contribution appears in the trace of the stress tensor, giving rise to the background pressure and effective sound speed. When the accretion is anisotropic, the momentum flux has a preferred direction and consequently generates a traceless component, contributing to the shear viscosity. In this picture, the three counterterms considered here are thus different manifestations of the same underlying short-scale physics, with their relative importance determined by the geometry of halo accretion. 

However, it is worth emphasising that the shear viscosity measurement is substantially noisier than the measurements of the other counterterms. Consequently, the separation between the different populations is less statistically clean than for the background pressure and effective sound speed. Nevertheless, the consistency between the splits by anisotropic flux, total inflow rate, and instantaneous mass accretion rate provides evidence for a common physical origin.

\section{Full Box Analysis}\label{sec:fullbox}

In this section, we compute the effective stress tensor and the counterterms for the full simulation box. The calculation will not require the explicit construction of short-scale fields and the smoothing procedure will be analogous to the procedure used for smoothing the halo-based effective stress tensor. As a cross-validation of this computation we test it against the Layzer-Irvine cosmic energy equation, in Appendix~\ref{app:cee}. Finally, we compare and analyse the halo-based and full box estimates for the counterterms.

\subsection{Full Box Stress Tensor}

In the following, we compute the full box effective stress tensor for the same N-body simulations as used in the halo-based analysis. We begin with the calculation of the kinetic term. First, we construct the density field \(\rho\), momentum field \(\rho v_i\) and kinetic field \(\rho v_i v_j\) on \(1024^3\) grids. Each is constructed by a mass-weighted CIC deposit of the corresponding per-particle quantity (1, \(v^i_p\), \(v^i_p v^j_p\), respectively). We then smooth these fields using the real space Top Hat filter to obtain the long-wavelength fields, \(\rho^l\), \((\rho v_i)^l\) and \((\rho v_i v_j)^l\), respectively. The long-wavelength velocity field \(v^l_i\) is then constructed as the ratio of the long-wavelength momentum and density fields, \(v^l_i = (\rho v_i)^l /\rho^l\). Thus, the kinetic term is computed as 
\begin{equation}
     K_{ij} = \left\{(\rho v_i v_j)^l - \rho^l v^l_i v^l_j \right\}\,.
\end{equation}
To construct the potential term, we first construct the overdensity field, \(\delta\), and then use it to solve for the gravitational potential through Poisson's equation. Poisson's equation is solved and the derivatives of the gravitational potential, \(\partial_i \phi\), are computed in Fourier space, analogously to the halo-based calculation. The long-wavelength fields, \((\partial_i \phi)^l\) and \((\partial_i \phi \partial_j \phi)^l\), are obtained by smoothing with the real space Top Hat filter. The potential term is then computed as
\begin{equation}
P_{ij}
=
\frac{1}{8\pi G}
\left\{2
\left((\partial_i\phi\,\partial_j\phi)^l
-
(\partial_i\phi^l\,\partial_j\phi^l)\right)
-
\delta_{ij}
\left(
(\partial_k\phi\,\partial_k\phi)^l
-
\partial_k\phi^l\,\partial_k\phi^l
\right)
\right\}\,.
\end{equation}

Finally, the full box stress tensor is \(\tau_{ij} = K_{ij} + P_{ij}\).

\subsection{Extracting the Full Box Counterterms}

The computation begins with the construction of the large-scale overdensity and velocity fields, \(\delta^l\) and \(v^l\). The large-scale velocity shear tensor, \(\sigma_{ij}\), is then constructed by taking the appropriate derivatives of the velocity field. The details of these constructions are the same as those described in Section~\ref{sec:extract}.

We smooth the full box effective stress tensor using a real space Top Hat filter and then sample it at 100,000 points, randomly distributed throughout the simulation box. A comparison with the standard Fourier space smoothing technique can be found in Appendix~\ref{app:fullbox}. We note that this counterterm extraction method is analogous and equivalent to the methods used in \cite{carrasco2012effective,baumann2012cosmological,Karandikar_2024}. Thus, at each of these points we have a measurement of \(\tau_{ij}\), the long-wavelength pressure \(P^l = \frac{1}{3} \mathrm{Tr}(\tau_{ij})\), the long-wavelength overdensity \(\delta^l\) and the large-scale velocity shear tensor \(\sigma_{ij}\). The counterterms are then extracted analogously to the halo-based analysis: 
\begin{equation}
    {\langle P^l \rangle = \bar{P},} \hspace{0.5 cm} {c_{\mathrm{eff}}^2 = \frac{\langle P^l \delta^l\rangle}{\bar{\rho} \langle \delta^l \delta^l\rangle}} \hspace{0.5 cm}\mathrm{and} \hspace{0.5 cm} c^2_{sv} = -\frac{4 H a}{3\bar{\rho}} \frac{\langle\hat\tau^{ij}\sigma^{ij}\rangle}{\langle \sigma^{ij}\sigma^{ij}\rangle}\,.
\end{equation}

\begin{table}[!t]
\centering
\begin{tabular}{|c|c|c|c|}
\hline
\(R_{\Lambda}\) (in \(\mathrm{Mpc}/h\))  & Halo-Based Estimate & Full Box Estimate & Percentage Contribution \\
\hline
16 & $2.72\times10^{4}$ & $2.85\times10^{4}$ & $95.39\%$ \\
20 & $2.72\times10^{4}$ & $2.84\times10^{4}$ & $95.90\%$ \\
24 & $2.72\times10^{4}$ & $2.85\times10^{4}$ & $95.31\%$ \\
28 & $2.72\times10^{4}$ & $2.87\times10^{4}$ & $94.69\%$ \\
\hline
\end{tabular}
\caption{\label{tab:pressure}
Comparison of the halo-based and full box estimates of \(\bar{P}/\bar{\rho}\), in units of \((\mathrm{km/s})^2\), for different EFT cutoff scales \(R_{\Lambda}\) (in \(\mathrm{Mpc}/h\)). The halo-based estimate is obtained from halos with \(M_\mathrm{vir} > 10^{12} M_{\odot}/h\), and the final column gives it as a percentage of the full box value. This contribution is stable at \(\sim 95\%\) across all smoothing scales.}
\end{table}

\begin{table}[!t]
\centering
\begin{tabular}{|c|c|c|c|}
\hline
\(R_{\Lambda}\) (in \(\mathrm{Mpc}/h\)) & Halo-Based Estimate & Full Box Estimate & Percentage Contribution \\
\hline
16 & $9.35\times10^{4}$ & $1.07\times10^{5}$ & $87.80\%$ \\
20 & $9.03\times10^{4}$ & $1.03\times10^{5}$ & $87.30\%$ \\
24 & $8.81\times10^{4}$ & $1.02\times10^{5}$ & $86.27\%$ \\
28 & $8.65\times10^{4}$ & $1.01\times10^{5}$ & $85.20\%$ \\
\hline
\end{tabular}
\caption{\label{tab:ceff2}
Comparison of the halo-based and full box estimates of the effective sound speed \(c_{\mathrm{eff}}^2\), in units of \((\mathrm{km/s})^2\), for different EFT cutoff scales \(R_{\Lambda}\) (in \(\mathrm{Mpc}/h\)). The halo-based estimate is obtained from halos with \(M_\mathrm{vir} > 10^{12} M_{\odot}/h\), and the final column gives it as a percentage of the full box value. This contribution is \(\sim 85\%\), decreasing only mildly with increasing \(R_{\Lambda}\).}
\end{table}

\begin{table}[!t]
\centering
\begin{tabular}{|c|c|c|c|}
\hline
\(R_{\Lambda}\) (in \(\mathrm{Mpc}/h\)) & Halo-Based Estimate & Full Box Estimate & Percentage Contribution \\
\hline
16 & $1.52\times10^{4}$ & $1.62\times10^{4}$ & $93.81\%$ \\
20 & $1.76\times10^{4}$ & $2.21\times10^{4}$ & $79.86\%$ \\
24 & $2.04\times10^{4}$ & $2.78\times10^{4}$ & $73.34\%$ \\
28 & $2.25\times10^{4}$ & $3.46\times10^{4}$ & $65.10\%$ \\
\hline
\end{tabular}
\caption{\label{tab:cvis2}
Comparison of the halo-based and full box estimates of the shear viscosity \(c_{sv}^2\), in units of \((\mathrm{km/s})^2\), for different EFT cutoff scales \(R_{\Lambda}\) (in \(\mathrm{Mpc}/h\)). The halo-based estimate is obtained from halos with \(M_\mathrm{vir} > 10^{12} M_{\odot}/h\), and the final column gives it as a percentage of the full box value. In contrast to \(\bar{P}\) and \(c_{\mathrm{eff}}^2\), this contribution falls from \(\sim 94\%\) to \(\sim 65\%\) as \(R_{\Lambda}\) increases, the strongest smoothing-scale dependence of the three counterterms.}
\end{table}

\subsection{Comparison with Halo-Based Estimates}\label{sec:compare}

In this subsection, we compare and discuss the halo-based and full box estimates of the counterterms. As seen in Tables~\ref{tab:pressure},~\ref{tab:ceff2} and~\ref{tab:cvis2}, the halos account for \(\sim 95 \%\) of the equation of state \(\bar{P}/\bar{\rho}\), \(\sim 85 \%\) of the effective sound speed and \(\sim 70 \%\) of the shear viscosity, depending on the smoothing scale. We note that these numbers are quoted from the results of the analysis on halos with \(M_\mathrm{vir} > 10^{12} M_{\odot}/h\) from the \(250\,\mathrm{Mpc}/h\) box (see Appendix~\ref{app:smallbox1}). We emphasise that, since \(R_h = 3 \times R_{\mathrm{vir}}\) already extends into the near-halo environment, the halo-based estimate is not blind to filamentary structure: the anisotropic stress sourced by accretion along filaments close to a halo (see Figure~\ref{fig:haloalign}) is already included in these percentages. The unaccounted residual is therefore more likely dominated by the diffuse filament backbone well outside any halo's \(R_h\), together with the sub-threshold halo population discussed in Appendix~\ref{app:smallbox1}.

Both the background pressure and effective sound speed percentage contributions are quite stable under a change in smoothing scale, whereas the shear viscosity contribution shows some fluctuations. We also note that, in general, the shear viscosity is a factor of a few smaller than the effective sound speed, which we attribute to the fact that the off-diagonal elements of the stress tensor are generally smaller than the diagonal elements, as found in Appendix~\ref{app:plummer}. If instead we extend the analysis to halos with \(M_\mathrm{vir} > 10^{11} M_{\odot}/h\) (see Appendix~\ref{app:smallbox1}), the halo-based estimate for \(\bar{P}\) exceeds the full box estimate. Dark matter halos are found in overdense regions, which contribute positively to the background pressure. Analogously, underdense regions are expected to provide a negative contribution. The contributions from lower mass halos and other components of the cosmic web, including voids and filaments, will be investigated in future work. 

From these results and the indications of Section~\ref{sec:extract}, we assert that halo accretion is an important physical process underlying the coupling of the large-scale and short-scale modes and hence the counterterms themselves.

\begin{table}[t]
\centering
\footnotesize
\renewcommand{\arraystretch}{1.4}
\setlength{\tabcolsep}{4pt}
\begin{tabular}{@{}
  >{\raggedright\arraybackslash}p{0.16\textwidth}
  >{\raggedright\arraybackslash}p{0.16\textwidth}
  >{\raggedright\arraybackslash}p{0.24\textwidth}
  >{\raggedright\arraybackslash}p{0.35\textwidth}
@{}}
\toprule
\textbf{Large-scale counterterm} & \textbf{Part of $\tau^{ij}$} & \textbf{Short-scale halo process}
  & \textbf{Relevant Figures} \\
\midrule
$\bar{P}$\newline (background pressure)
  & Trace, $\tfrac{1}{3}\mathrm{Tr}\,\tau^{ij}$ (mean value)
  & Momentum flux of accreting matter, halos driven out of virial equilibrium
  & Above-median accretion rate dominates (Figs.~\ref{fig:press1},~\ref{fig:pressureaccrate});
    on-axis Plummer collision sources $\tau^{zz}$ (Fig.~\ref{fig:onaxis})
 \\
\addlinespace
$c^2_{\rm eff}$ \newline (effective sound speed)
  & Trace, response to $\delta^l$
  & Halo accretion-driven stress, correlated with $\delta^l$ as halos sit in overdense regions
  & Splits by virial departure and accretion rate (Figs.~\ref{fig:ceff2virsplit} and ~\ref{fig:ceff2accrate}); rises and turns over with
    structure growth (Fig.~\ref{fig:evolve})
 \\
\addlinespace
$c^2_{sv}$\newline (shear viscosity)
  & Traceless part $\hat{\tau}^{ij}$, response to $\sigma^{ij}$
  & Anisotropic accretion; direction-dependent momentum flux, e.g.\ along a filament feeding the halo
  & Splits by $|A^{TF}|$ and accretion rate (Figs.~\ref{fig:shear2},~\ref{fig:cvis2accrate});
    stress ellipsoid aligned with filament (Fig.~\ref{fig:haloalign});
    off-axis Plummer collision sources $\tau^{yz}$ (Fig.~\ref{fig:offaxis})
 \\
\bottomrule
\end{tabular}
\caption{Dictionary between the large-scale EFT counterterms and the
short-scale halo dynamics that source them. The trace of the halo stress tensor
tracks the \emph{magnitude} of accretion, while the traceless part also tracks its
\emph{accretion geometry}.}
\label{tab:dictionary}
\end{table}

\begin{figure*}[!t]
\centering

\begin{subfigure}{0.48\textwidth}
    \centering
    \includegraphics[width=\linewidth]{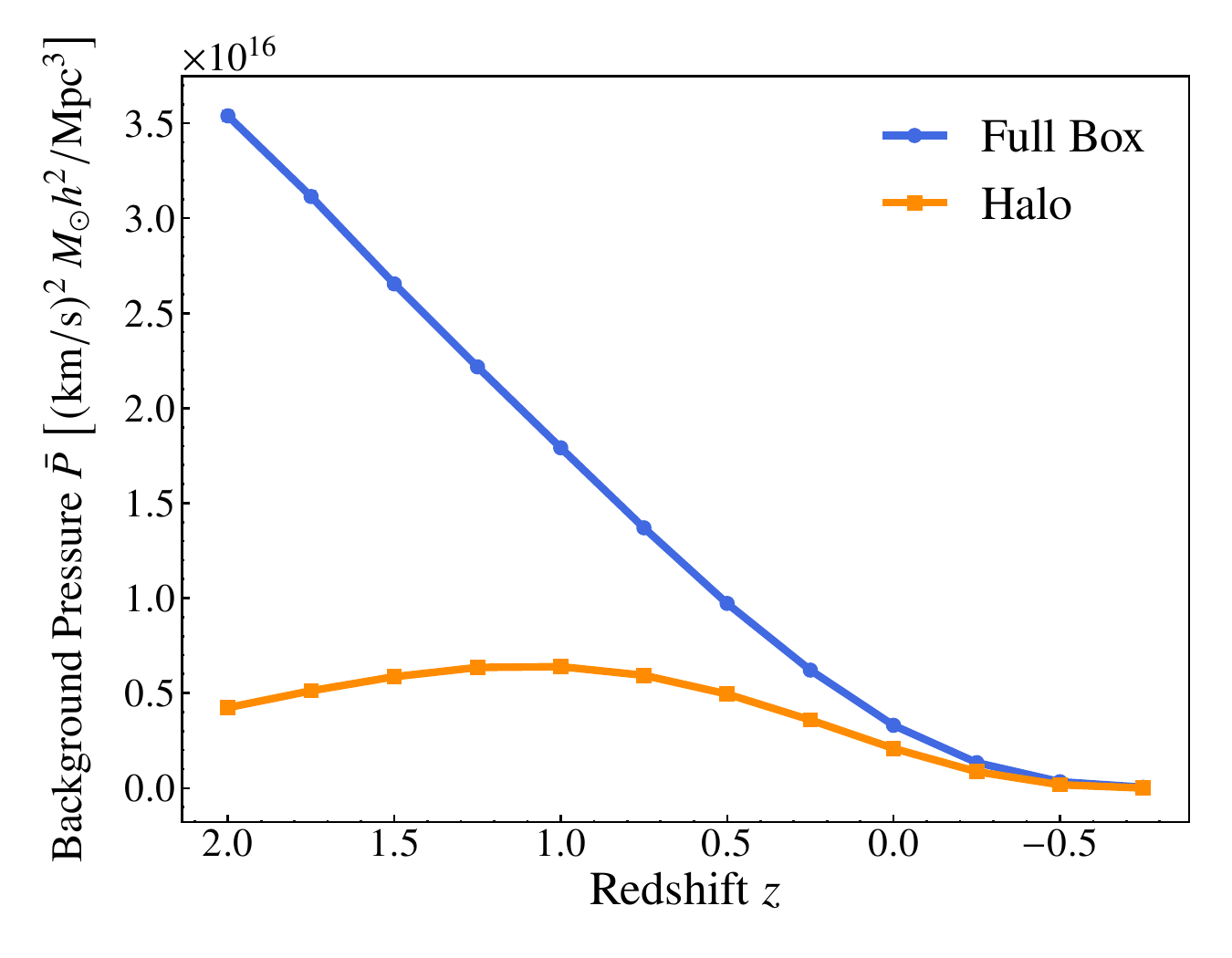}
\end{subfigure}
\hfill
\begin{subfigure}{0.48\textwidth}
    \centering
    \includegraphics[width=\linewidth]{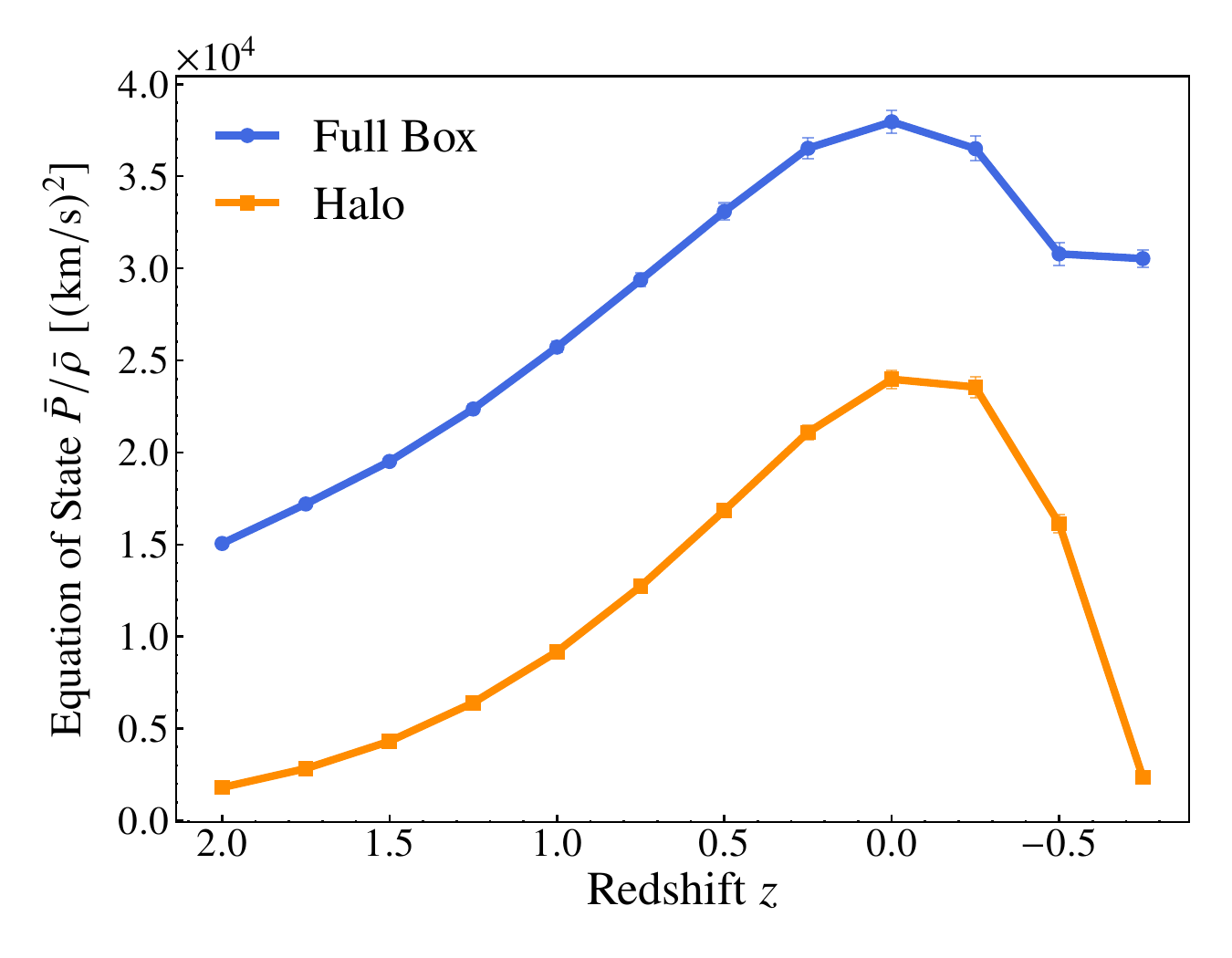}
\end{subfigure}

\vspace{-0.7em}

\begin{subfigure}{0.48\textwidth}
    \centering
    \includegraphics[width=\linewidth]{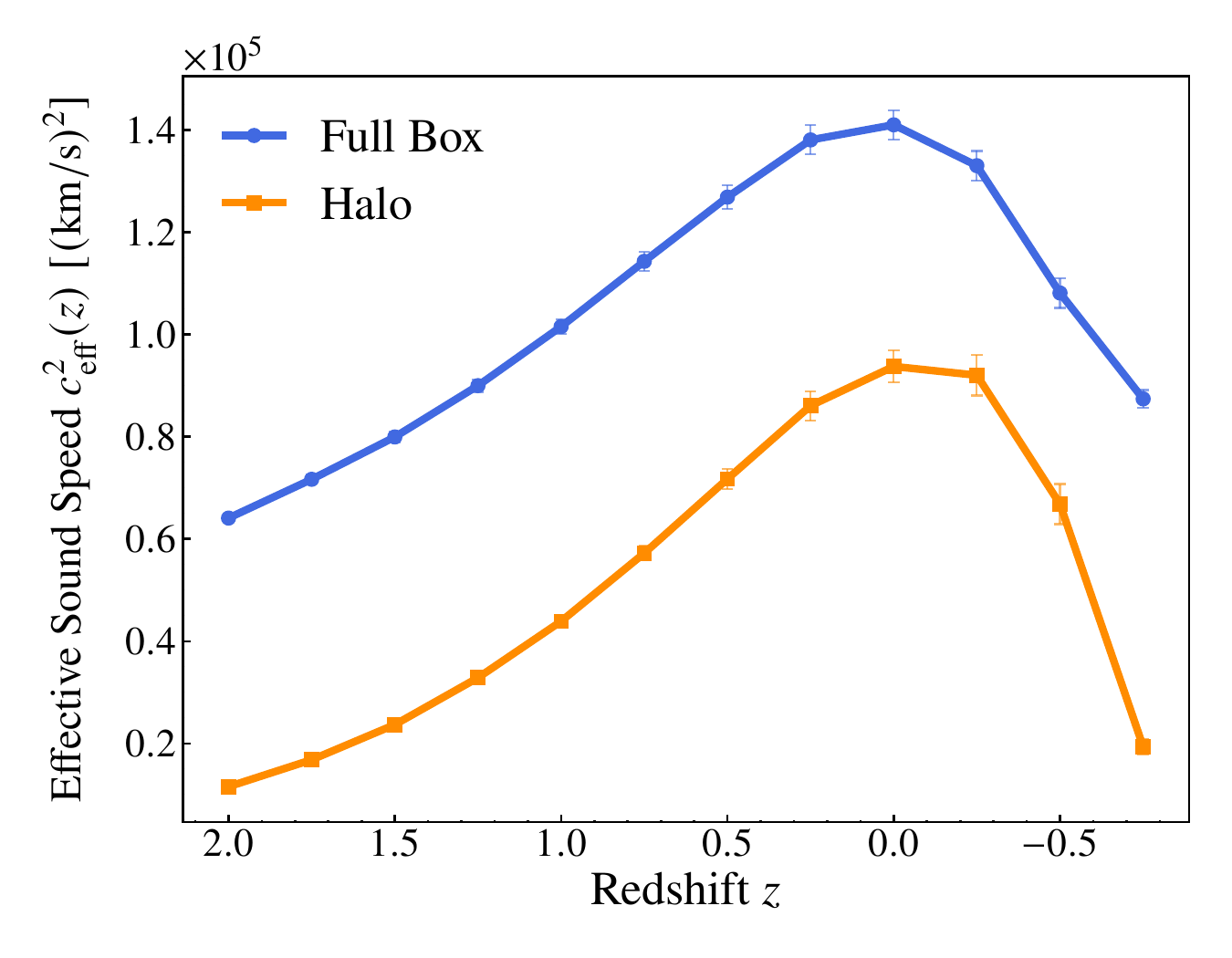}
\end{subfigure}
\hfill
\begin{subfigure}{0.48\textwidth}
    \centering
    \includegraphics[width=\linewidth]{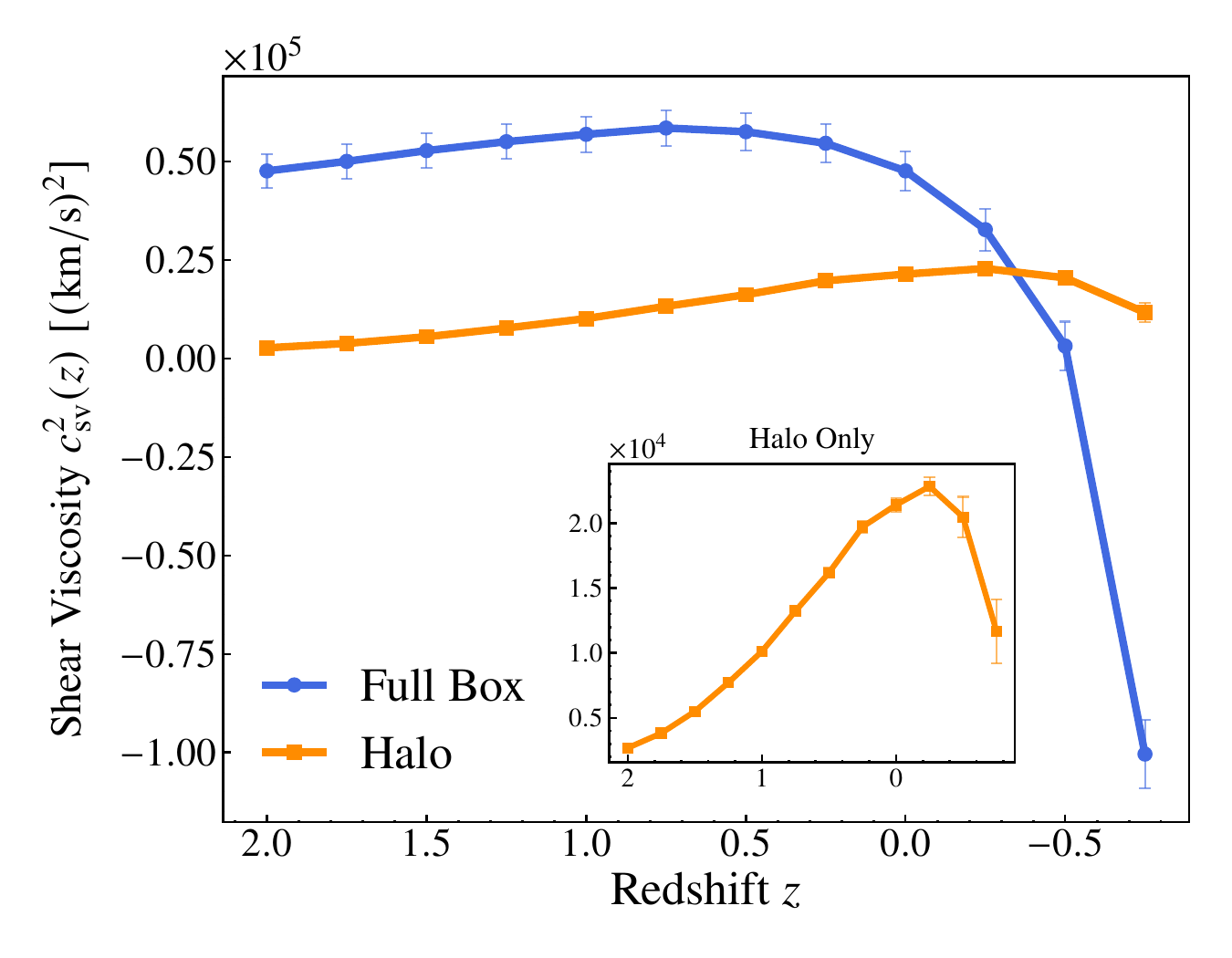}
\end{subfigure}

\caption{Redshift evolution of the halo-based and full box counterterms: the full box \(\bar{P}\) (top left), the equation of state \(\bar{P}/\bar{\rho}\), which scales out the dilution of \(\bar\rho\) (top right), the effective sound speed \(c_{\mathrm{eff}}^2\) (bottom left) and the shear viscosity \(c_{sv}^2\) (bottom right). The inset in the bottom right panel shows the halo-based \(c_{sv}^2\) shown alone, since the full box curve swings negative at the final snapshot and obscures the halo-based trend on the shared axis. This analysis is performed for \(R_{\Lambda} = 28\,\mathrm{Mpc}/h\) in the \(1000\,\mathrm{Mpc}/h\) box for all halos with \(M_\mathrm{vir} > 10^{13} M_{\odot}/h\).}
\label{fig:evolve}

\end{figure*}

\subsection{Evolution of the Counterterms}\label{sec:evolve}

In this subsection, we analyse the redshift evolution of the counterterms. During matter domination, structure formation proceeds through the hierarchical growth and accretion of dark matter halos \cite{lacey1993merger,wechsler2002concentrations,mcbride2009mass,fakhouri2010merger,correa2015accretion}. If halo accretion is an essential physical process in the generation of the counterterms, we expect the counterterms to increase during this epoch. Near the end of matter domination, the universe becomes increasingly dominated by dark energy, leading to a suppression of the growth of structure. Halo accretion is driven by the continuous infall of surrounding matter, and over time the rate of accretion is overcome by the accelerated expansion of the background. This provides a natural regime to test our expectation, as we expect the counterterms to decrease as the universe evolves deeper into dark energy domination. 

In Figure~\ref{fig:evolve}, we depict the results of our analysis on twelve snapshots between \(z = 2\) and \(z = -0.75\), equally spaced in redshift. The top left panel portrays the full box background pressure \(\bar{P}\), which decreases monotonically as the universe expands, reflecting the dilution of the background matter density. We scale out this effect in the top right panel by plotting the equation of state, \(\bar{P}/\bar{\rho}\), instead. With this dilution scaled out, both \(\bar{P}/\bar{\rho}\) and the effective sound speed \(c_{\mathrm{eff}}^2\) (bottom left panel) rise throughout matter domination for both the full box and halo-based estimates, turning over close to \(z = 0\) before declining as dark energy begins to dominate. This turnover coincides with the epoch at which structure stops growing, and hence the rate of halo accretion begins to be suppressed, consistent with our expectation.

The shear viscosity \(c_{sv}^2\) (bottom right panel) shows qualitatively similar behaviour, but is considerably noisier, consistent with the lower signal-to-noise ratio discussed in Section~\ref{sec:counter2}. The full box estimate rises during matter domination but then swings sharply negative at the final snapshot, falling below the halo-based estimate. As \(c_{sv}^2\) is extracted as a regression slope, it is not required to remain positive, and a sign change of this kind is possible, particularly once the velocity shear field becomes small and noise-dominated deep in dark-energy domination. As the large dip in the full box curve makes the halo-based trend difficult to discern on the same axis, we show the halo-based \(c_{sv}^2\) separately as an inset in the bottom right panel of Figure~\ref{fig:evolve}. There, the same qualitative trend seen in \(\bar{P}/\bar{\rho}\) and \(c_{\mathrm{eff}}^2\) is clearly visible. Taken together, the redshift evolution of all three counterterms is consistent with our expectation. They grow while dark matter halos are actively accreting during matter domination, and decline once the accelerated expansion suppresses further accretion. This reinforces the picture developed in Section~\ref{sec:extract} that these counterterms are
sourced by the out-of-equilibrium dynamics of accreting halos.

\section{Conclusion}

In this work, we laid the foundations for a ``Kinetic Theory of the EFTofLSS'', connecting the physical origins of EFTofLSS counterterms to the short-scale dynamics of dark matter halos. We have developed a framework to compute the effective stress tensor of halos in cosmological N-body simulations, allowing us to quantify their contribution to the different counterterms and determine which halo properties control their magnitude.

We found that halos with \(M_{\rm vir}>10^{12}M_\odot/h\), the population considered in this work, account for \(\sim 95 \%\) of the background pressure, \(\sim 85\%\) of the effective sound speed, and \(\sim 70\%\) of the shear viscosity counterterm, depending on smoothing scale, as summarised in Tables~\ref{tab:pressure}, \ref{tab:ceff2} and~\ref{tab:cvis2}; the close resemblance between the halo-sourced and full box pressure fields in Figure~\ref{fig:visual} gives a visual impression of this dominance. These contributions are dominated by cluster-scale halos (Figures~\ref{fig:halomass1} and~\ref{fig:shear1}), as the combination of halo abundance and the magnitude of their effective stress tensor peaks at these masses.

We demonstrated that halo accretion underlies the generation of the trace of the halo-based effective stress tensor, as a result contributing to the background pressure and effective sound speed, while anisotropic accretion sources the traceless part and hence the shear viscosity. Accretion results in mass and momentum fluxes which drive halos away from virial equilibrium. The extent to which these out-of-equilibrium conditions persist depends on the comparison between the accretion and dynamical timescales. Halos may accrete material faster than they can gravitationally relax, sustaining departures from virial equilibrium and contributing to the halo-based effective stress tensor. This provides a physical picture in which the different counterterms arise as distinct manifestations of the same underlying short-scale process, with their relative contributions determined by the geometry of halo accretion. Furthermore, as Figures~\ref{fig:press1} and~\ref{fig:ceff2virsplit} show, halos with large inflow rates contribute more to the background pressure and effective sound speed, in support of the interpretation that accretion-driven, out-of-equilibrium dynamics are an important physical source of these counterterms. The redshift evolution in Figure~\ref{fig:evolve} provides further support for this picture: both the halo-based and full box estimates of \(\bar{P}/\bar{\rho}\) and \(c_{\mathrm{eff}}^2\) grow during matter domination, while halos are actively accreting, and turn over close to \(z = 0\), as dark energy begins to suppress further accretion.

Although the shear viscosity measurement is substantially noisier than those of the diagonal counterterms, we find evidence that its contribution is sensitive not only to the magnitude of accretion but also to its geometry. Figure~\ref{fig:shear2} shows that halos experiencing more anisotropic inflow produce larger contributions to the traceless part of the effective stress tensor, and Figure~\ref{fig:haloalign} illustrates this geometry for two halos: the anisotropic stress is elongated along the filament feeding a halo, and closer to isotropic for a halo at a multi-filament node. Table~\ref{tab:dictionary} compiles the physical picture we have developed, in the form of a dictionary between the large-scale counterterms and the short-scale halo processes that generate them.

Our results suggest a short-scale, Kinetic Theory interpretation of the macroscopic parameters of the EFTofLSS. Rather than viewing the counterterms as solely phenomenological parameters to be fitted and marginalised over, they can be connected to identifiable short-scale physical processes. These results can be organised compactly: each counterterm can be interpreted as an integral over the halo mass function, weighted by an ``efficiency factor'' that encodes the contribution from a halo of a given mass. This opens the possibility of replacing completely agnostic priors with physics-informed priors based on the properties and dynamics of the short-scale structures contributing to the effective stress tensor (but see footnote~\ref{fn:caveats}), complementary to priors calibrated on HOD mock catalogues \cite{zhang2025hod,desi2025hod,ivanov2025analytic}.

There are several natural directions for future work. Our analysis of halo properties focuses on halos with \(M_{\rm vir}>10^{13}M_\odot/h\), with the full box comparison extending to \(M_{\rm vir}>10^{12}M_\odot/h\); extending the calculation to lower halo masses would allow us to determine more precisely the contribution of the full halo population. This extension alone, however, will not close the gap between the halo-based and full box estimates: as shown in Section~\ref{sec:compare}, extending the halo-based estimate of \(\bar{P}\) down to \(M_{\rm vir}>10^{11}M_\odot/h\) makes it exceed the full box value. Halos trace overdense regions, so the full box average must include a compensating negative contribution from underdense regions. A complete accounting of the counterterms will therefore require the other components of the cosmic web, voids and filaments \cite{bond1996filaments,hahn2007properties,libeskind2018tracing,galarraga2020populations,galarraga2022relative}, as well as the low-mass halo population. In Figure~\ref{fig:haloalign} we found that the anisotropic stress of halos is oriented along the filamentary structures feeding them, suggesting that these structures may provide a non-negligible contribution to the effective stress tensor. Our analysis included contributions from infall regions around the halos, but we expect considerable contributions from the more diffuse filamentary structures, which are not captured by our halo-based estimate.

It would also be interesting to investigate the dependence of the counterterms on additional properties of the halo environment and to develop a more quantitative mapping between halo accretion histories and the time dependence of the EFTofLSS parameters. At fixed mass, a halo's contribution depends strongly on its inflow rate and environment (Figures~\ref{fig:visual}, \ref{fig:press1} and~\ref{fig:ceff2virsplit}), reminiscent of halo assembly bias, where clustering at fixed mass depends on formation time and concentration \cite{gao2005age,wechsler2006dependence,dalal2008halo}. As halo accretion rates can already be inferred observationally through probes such as the splashback radius \cite{diemer2014dependence,adhikari2014splashback,more2015splashback,diemer2017sparta, chang2018splashback,shin2019measurement, more2016detection,baxter2017halo}, an especially promising direction for future work is to inform these physics-based priors using observationally measured accretion rates, rather than ones measured from simulations.

Thus, the framework developed here provides a route toward connecting the short-scale dynamics of nonlinear structure formation to the macroscopic description of the Universe used by the EFTofLSS. We hope that such a connection can ultimately lead to a more physically motivated understanding of the counterterms and to improved theoretical priors for large-scale cosmological observables.

\acknowledgments

OKH thanks Rahul S Hejjaji and Arhum Ansari for helpful discussions. AB thanks Susmita Adhikari, Sachin Jain and Ravi Sheth for helpful discussions. The authors acknowledge the PARAMBrahma Facility under the National Supercomputing Mission, Government of India, at the Indian Institute of Science Education and Research, Pune, for providing the computing resources for this work. The CosmoSim database used in this paper is a service by the Leibniz-Institute for Astrophysics Potsdam (AIP). The MultiDark database was developed in cooperation with the Spanish MultiDark Consolider Project CSD2009-00064. The authors gratefully acknowledge the Gauss Centre for Supercomputing e.V. and the Partnership for Advanced Supercomputing in Europe for funding the MultiDark simulation project by providing computing time on the GCS Supercomputer SuperMUC at Leibniz Supercomputing Centre. Research at Perimeter Institute is supported in part by the Government of Canada through the Department of Innovation, Science and Economic Development Canada and by the Province of Ontario through the Ministry of Economic Development, Job Creation and Trade. We also acknowledge the use of the Pylians \cite{villaescusa2018pylians} library for numerical analysis, and thank its developers for making it publicly available. This work also made extensive use of NumPy \cite{Harris_2020}, SciPy \cite{Virtanen_2020}, and Matplotlib \cite{hunter2007matplotlib} for data processing and visualisation.

\appendix

\section{Plummer Spheres: A Toy Model}\label{app:plummer}

In this appendix, we apply our methodology to compute the EFTofLSS effective stress tensor on various N-body simulation set-ups involving Plummer spheres, whose simplicity and favourable analytical properties make them a natural testing ground for our calculation. Isolated Plummer spheres satisfy virial equilibrium \cite{Plummer1911}, which we will use to demonstrate the robustness of our stress tensor calculation against the EFTofLSS non-renormalization theorem \cite{baumann2012cosmological}. We also simulate and analyse scenarios involving colliding Plummer spheres, one in which they collide along the same axis, and another where the collision is off-axis. We compute the following expression for the effective stress tensor, in terms of the short-scale fields: 

\begin{equation}
\tau_{ij} = \overbrace{\rho_m v_i^s v_j^s}^{\text{Kinetic Term}} + \underbrace{\frac{2 \partial_i\phi^s \partial_j\phi^s - \delta_{ij} \partial_k\phi^s \partial_k\phi^s}{8 \pi G}}_{\text{Potential Term}}\,.
\end{equation}

For the purposes of this analysis, we primarily consider the volume-averaged form of the stress tensor
\begin{equation}
\left\langle \tau^{ij} \right\rangle_V
=
\frac{1}{V}
\int d^3x\,\tau^{ij}(\mathbf{x})\,,
\end{equation}
where $V$ denotes the volume of the region over which the averaging is performed. Henceforth, \(\tau^{ij}\) will refer to the volume-averaged effective stress tensor, unless explicitly specified otherwise. In the following subsections, we describe the details of the stress tensor calculation and the results in different scenarios. As the Plummer sphere scenarios under consideration are all isolated, no large-scale modes exist in these toy models, and hence all fields under consideration will be short-scale. As a result, the following analysis will not involve any mention of the EFT cutoff scale.

\subsection{The Effective Stress Tensor}

We begin with a description of the computation of the kinetic term, which is quite straightforward. The computation involves the following sum over all particles: 

\begin{equation}
K_{ij} = \frac{1}{V_{\mathrm{box}}}\sum_p m_p v_i v_j\,.
\end{equation}
Here we have divided by the volume of the simulation box \(V_{\mathrm{box}}\).

We will be interested in the ratios between the kinetic and potential terms and the comparison between the magnitudes of the various stress tensor components. Hence the overall scaling by the simulation box volume, as opposed to the volume of the Plummer sphere itself, will not be relevant here. The potential term is slightly more involved to compute, as it requires derivatives of the gravitational potential. We first construct the density field on a grid from the particle positions and use it to source Poisson's equation, which we solve in Fourier space.

\begin{figure*}[!t]
\centering
\begin{subfigure}{0.48\textwidth}
    \centering
    \includegraphics[width=\linewidth]{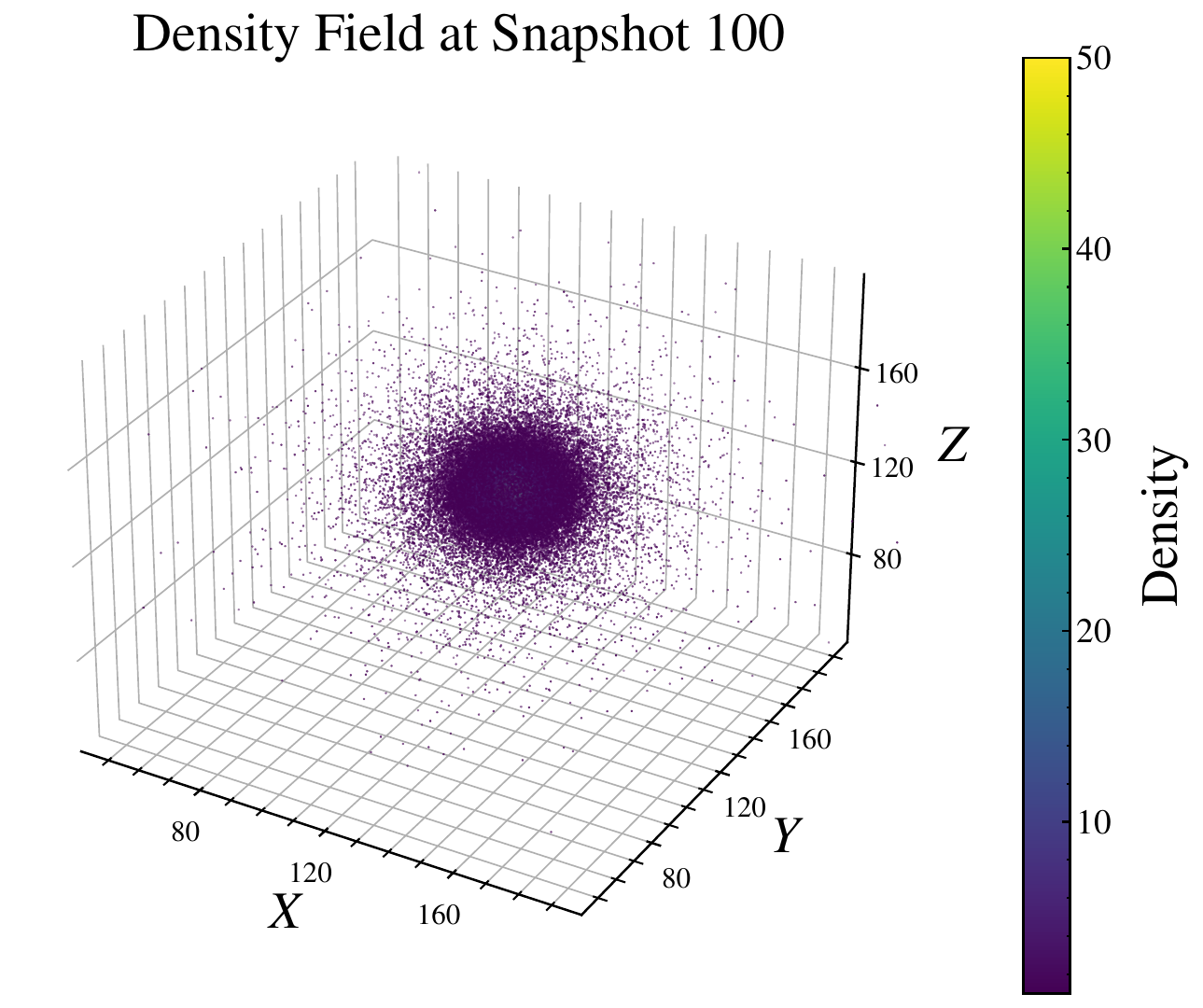}
    \label{fig:singleplummer1}
\end{subfigure}
\hfill
\begin{subfigure}{0.48\textwidth}
    \centering
    \includegraphics[width=\linewidth]{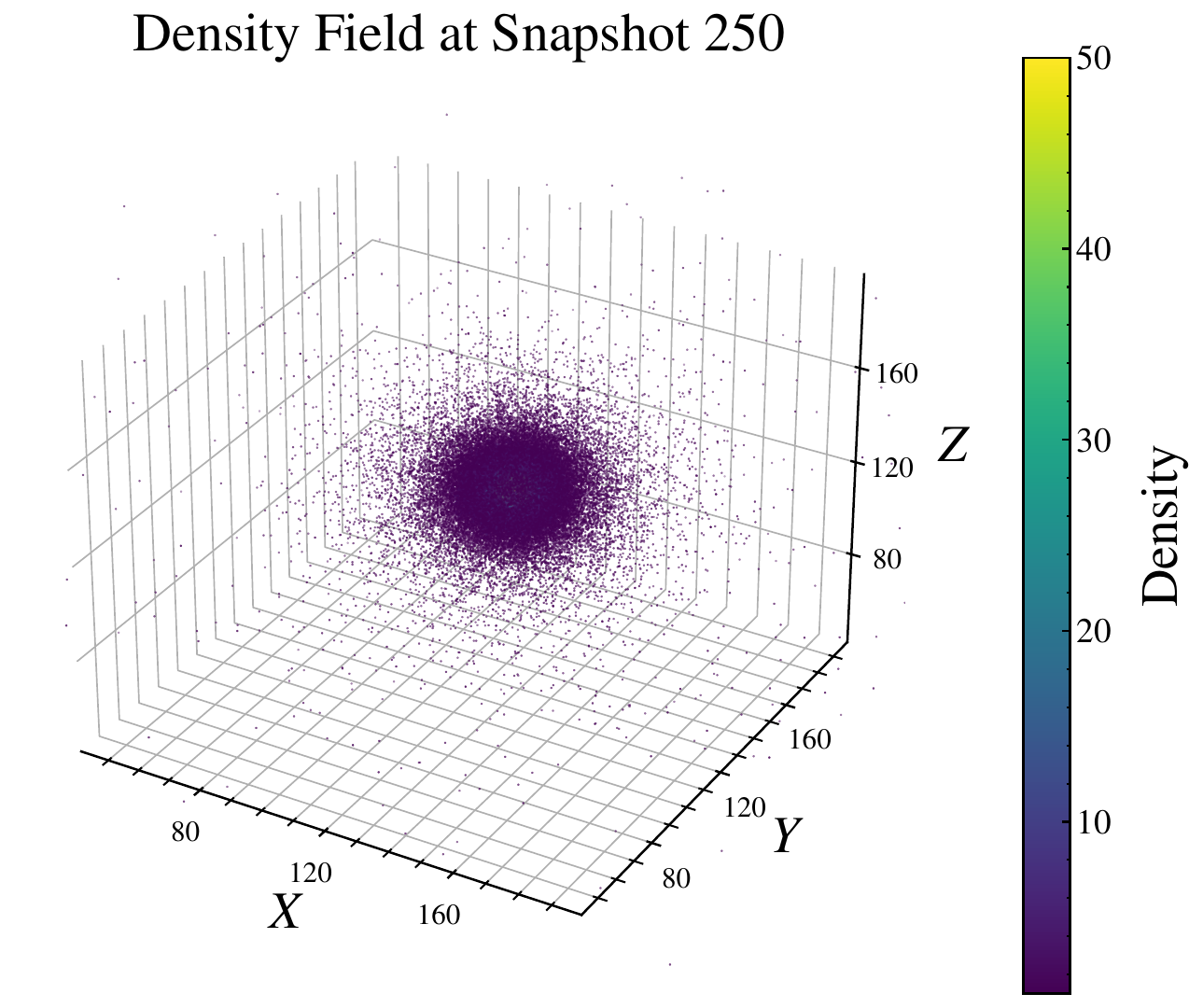}
    \label{fig:singleplummer2}
\end{subfigure}\

\vspace{0.5em}

\begin{subfigure}{0.48\textwidth}
    \centering
    \includegraphics[width=\linewidth]{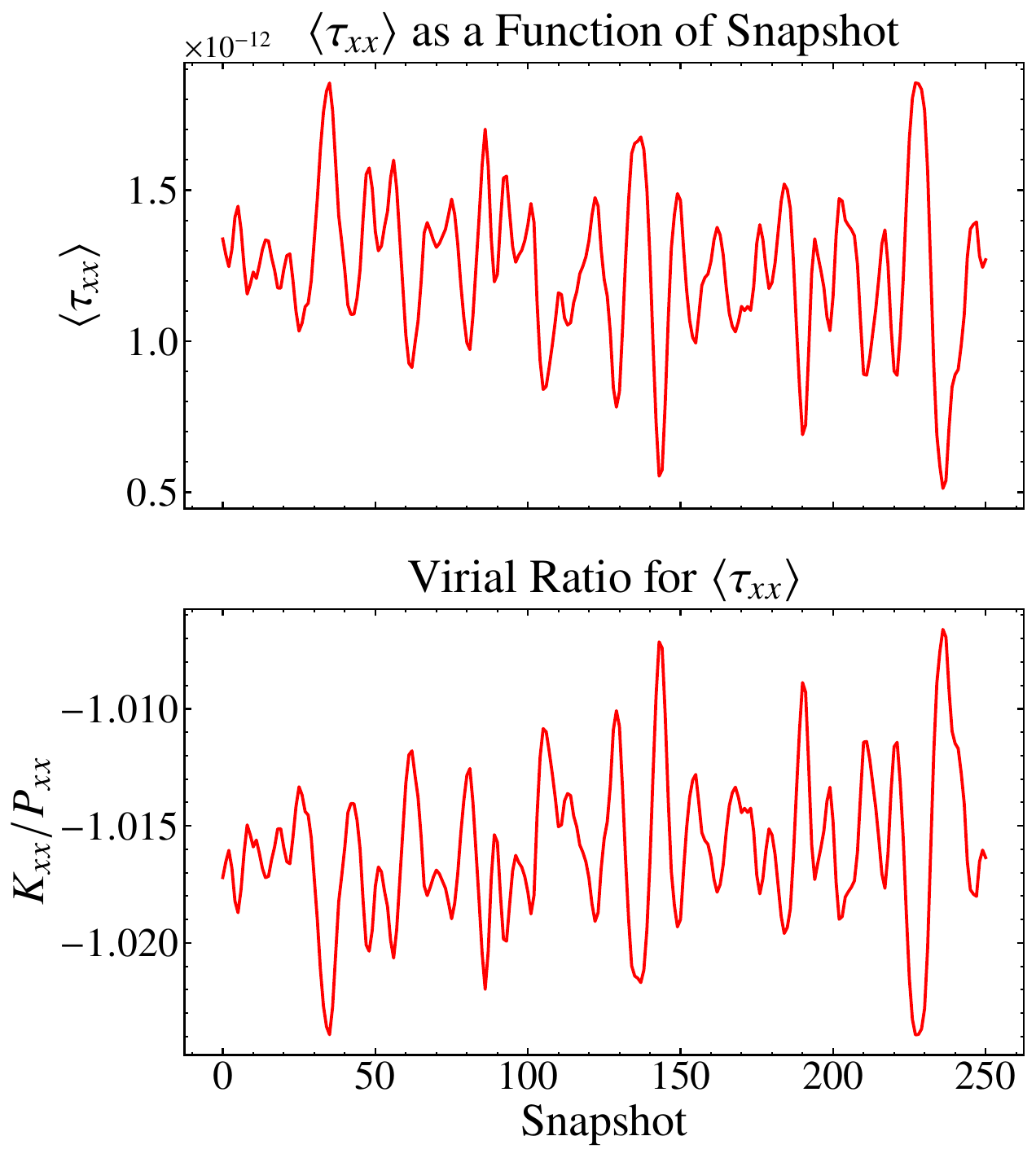}
    \caption{Evolution of \(\langle \tau_{xx} \rangle\) and the Virial Ratio}
    \label{fig:txx}
\end{subfigure}
\hfill
\begin{subfigure}{0.48\textwidth}
    \centering
    \includegraphics[width=\linewidth]{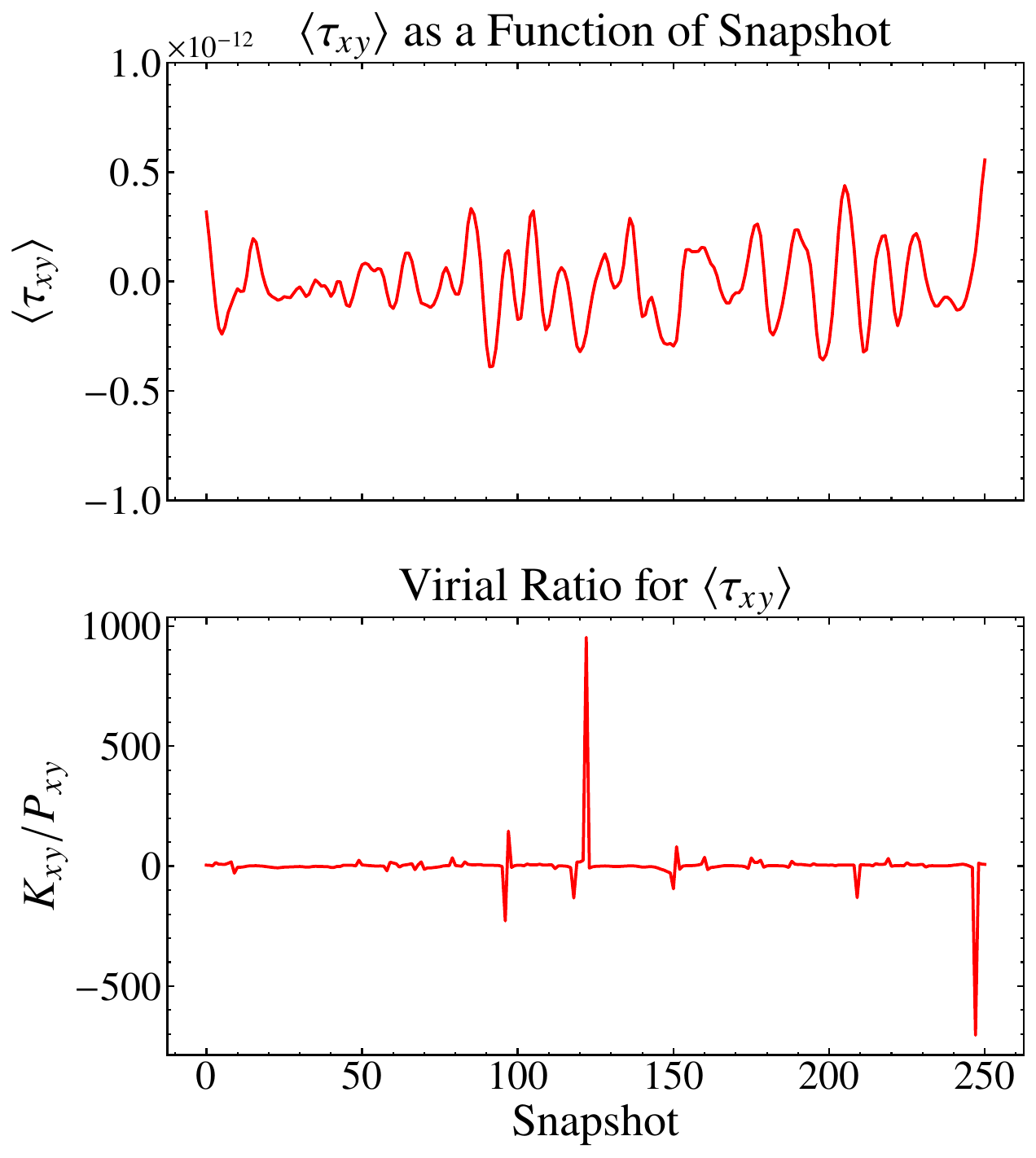}
    \caption{Evolution of \(\langle \tau_{xy} \rangle\) and the Virial Ratio}
    \label{fig:txy}
\end{subfigure}

\caption{Evolution of \(\langle \tau_{xx} \rangle\) and \(\langle \tau_{xy} \rangle\) for a single isolated Plummer sphere, used as a cross-check of the EFTofLSS non-renormalization theorem. The ratio \(K_{xx}/P_{xx}\) remains consistently close to \(-1\) throughout the evolution, as expected from cancellation between the kinetic and potential terms. The off-diagonal component \(\tau_{xy}\) is at least an order of magnitude smaller than \(\tau_{xx}\) and is consistent with numerical noise. See text for details.}
\label{fig:single}

\end{figure*}

\begin{equation}
\nabla^2 \phi = 4 \pi G (\rho - \bar\rho) \implies \phi(k) = - \frac{4 \pi G}{k^2} \,\delta\rho(k)\,, \qquad \delta\rho \equiv \rho - \bar\rho\,.
\end{equation}

\vspace{0.1 cm}

In practice, one constructs the density field \(\rho\) directly from the particle positions, and the contribution of the homogeneous background is removed simply by setting the \(k=0\) mode of \(\rho(k)\) to zero, which is equivalent to replacing \(\rho(k)\) by \(\delta\rho(k)\). The spatial derivatives of the gravitational potential are also constructed in Fourier space, by taking products with the respective wavenumber components: 

\begin{equation}
 \phi(k) = -\frac{4 \pi G}{k^2} \delta\rho(k) \implies ik^j \phi(k) = - \frac{4 \pi G \, ik^j}{k^2} \delta\rho(k)\,.
\end{equation}

\vspace{0.1 cm}

Finally, we perform an inverse Fourier Transform to isolate the derivatives of the gravitational potential in real space. The potential term is then calculated as a sum over contributions from all grid cells in the box: 

\begin{equation}
P_{ij} = \frac{1}{V_{\mathrm{box}}} \frac{(\Delta x)^3}{8\pi G}\sum_{\mathrm{grid}} (2 \hspace{0.05cm}\partial_i \phi \hspace{0.05cm}\partial_j \phi - \delta_{ij} \hspace{0.05cm}\partial_k \phi \hspace{0.05cm}\partial_k \phi )\,,
\end{equation}
where \((\Delta x)^3\) is the volume of a single grid cell.

Hence the volume-averaged effective stress tensor is simply \(\tau_{ij} = K_{ij} + P_{ij} \). This completes our discussion of the computation of the effective stress tensor for these toy models.

\subsection{Plummer Sphere Tests}

In this subsection, we apply the effective stress tensor calculation to several simple scenarios. We analyse snapshots from non-cosmological simulations of Plummer spheres, run with the \textsc{Gadget-4} N-body simulation code \cite{Springel_2021}.

\subsubsection{An Isolated Plummer Sphere}

For this scenario, we simulate a single Plummer sphere with \(\mathrm{Mass} = 2\), scale radius \(\mathrm{a} = 5\), truncation radius \(\mathrm{R} = 67\) (all values in code units) and \(200{,}000\) particles. We choose the truncation radius so that \( > 99 \%\) of the total mass is included. 

By the EFTofLSS non-renormalization theorem, the effective stress tensor of this simulation must vanish. We check our numerical calculation against this theorem by looking at the ratio of the kinetic and potential terms of the effective stress tensor. If this ratio is close to \(-1\), the two terms cancel, and the residual effective stress tensor is numerical noise. 

\begin{figure*}[!t]
    \centering

    \begin{subfigure}{0.32\textwidth}
        \centering
        \includegraphics[width=\linewidth]{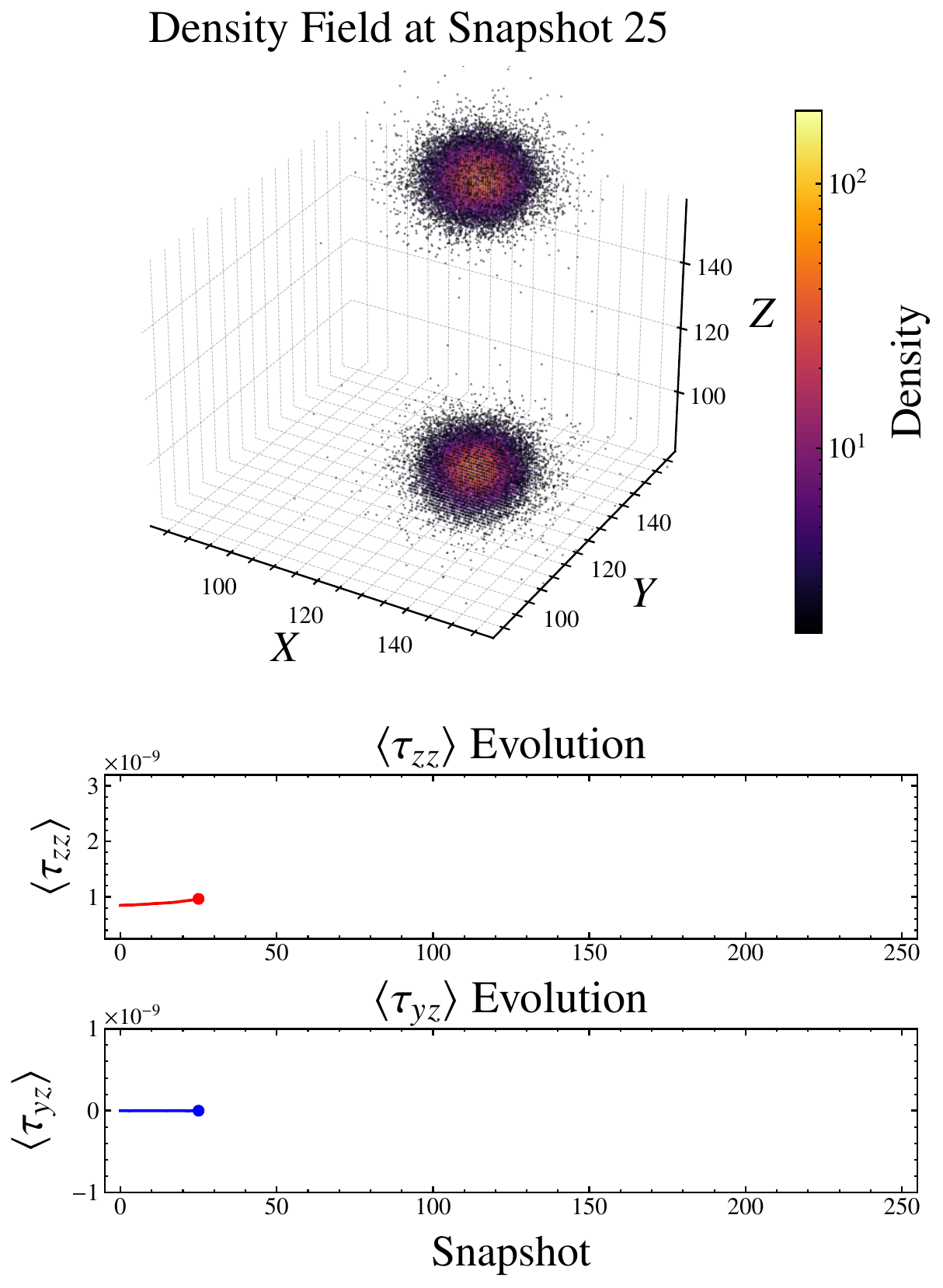}
        \caption{Approach Phase}
    \end{subfigure}
    \hfill
    \begin{subfigure}{0.32\textwidth}
        \centering
        \includegraphics[width=\linewidth]{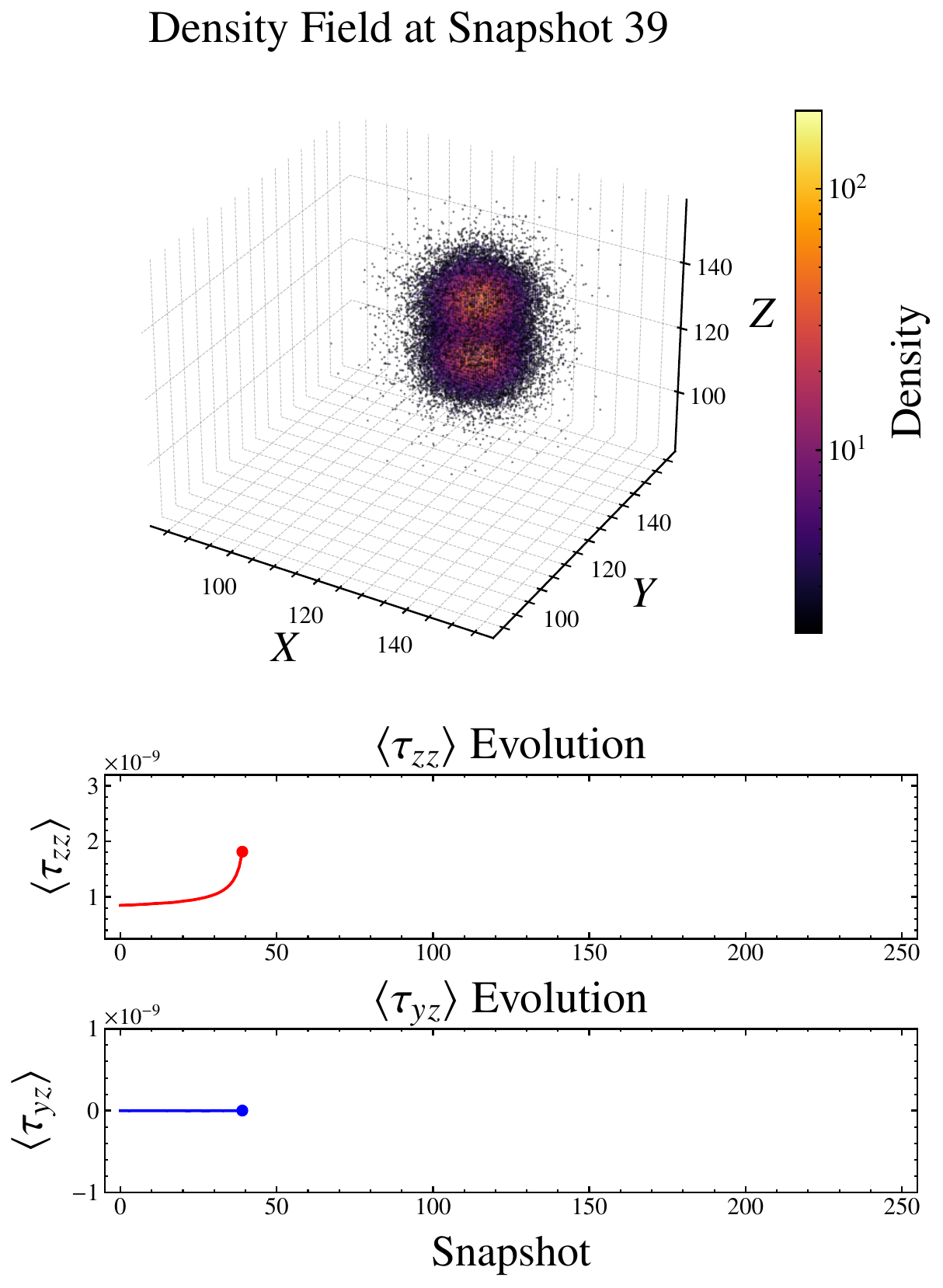}
        \caption{At Collision}
    \end{subfigure}
    \hfill
    \begin{subfigure}{0.32\textwidth}
        \centering
        \includegraphics[width=\linewidth]{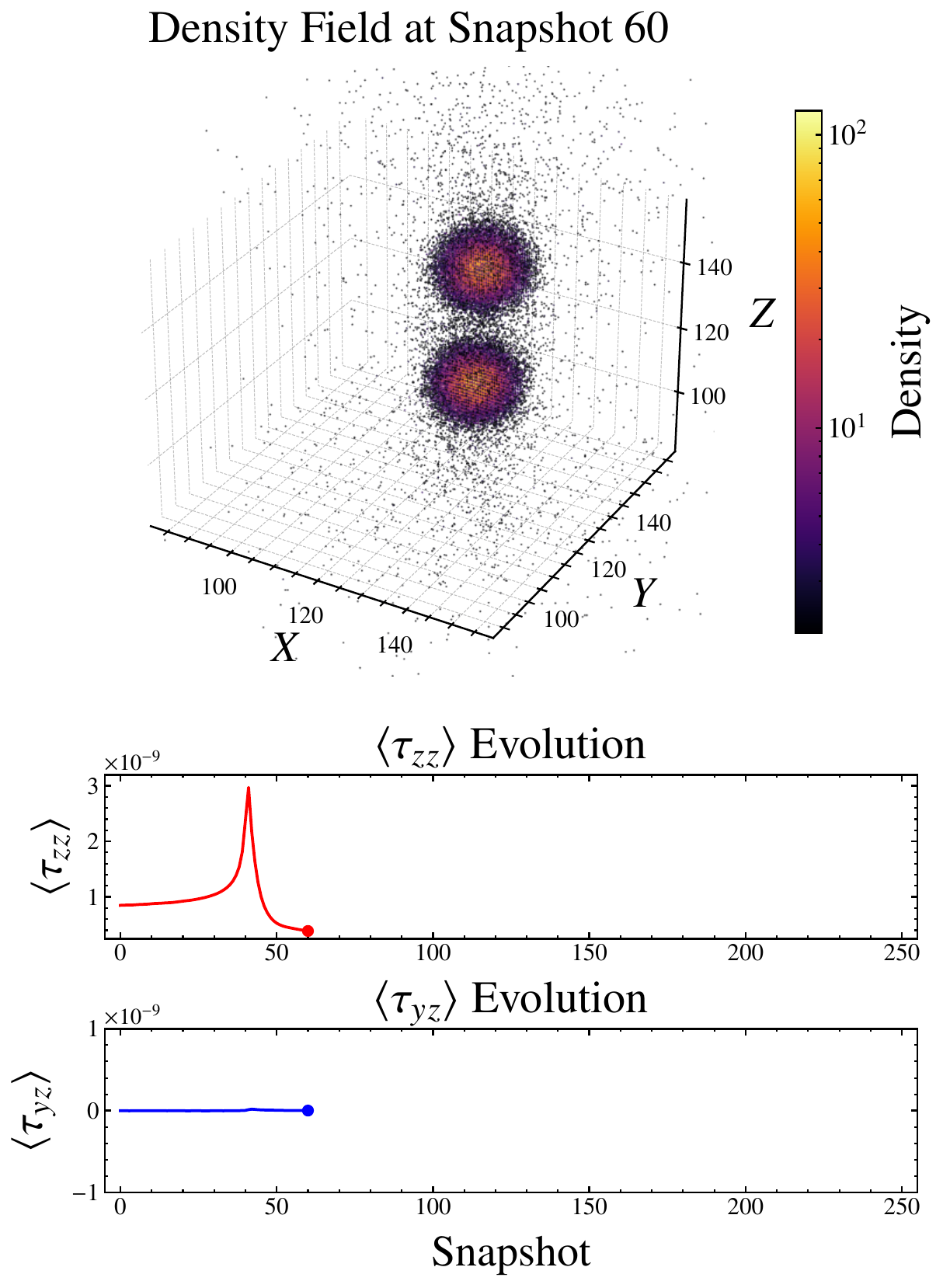}
        \caption{Turnaround Phase}
    \end{subfigure}

    \vspace{0.5em}

    \begin{subfigure}{0.32\textwidth}
        \centering
        \includegraphics[width=\linewidth]{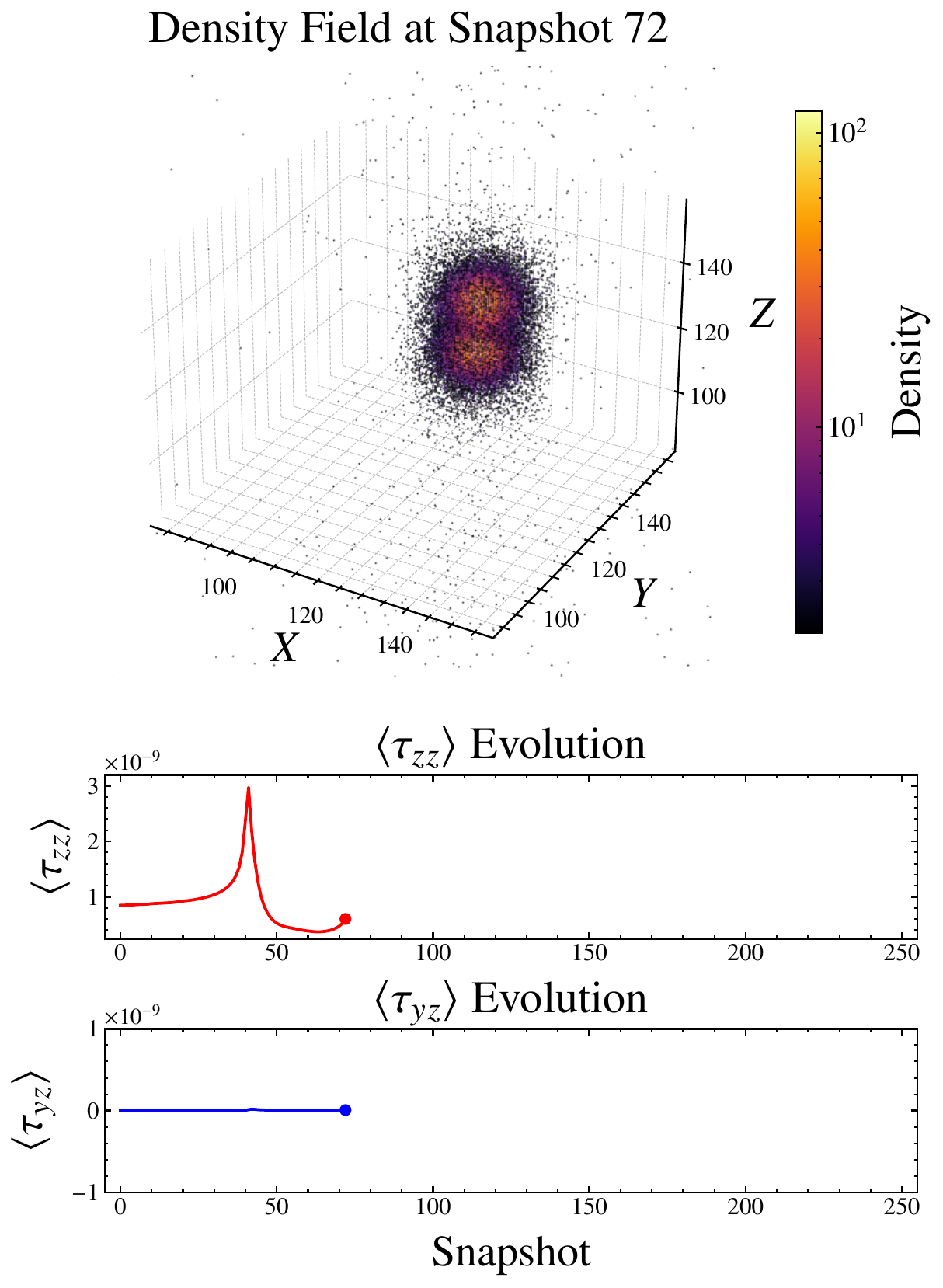}
        \caption{Recombination}
    \end{subfigure}
    \hfill
    \begin{subfigure}{0.32\textwidth}
        \centering
        \includegraphics[width=\linewidth]{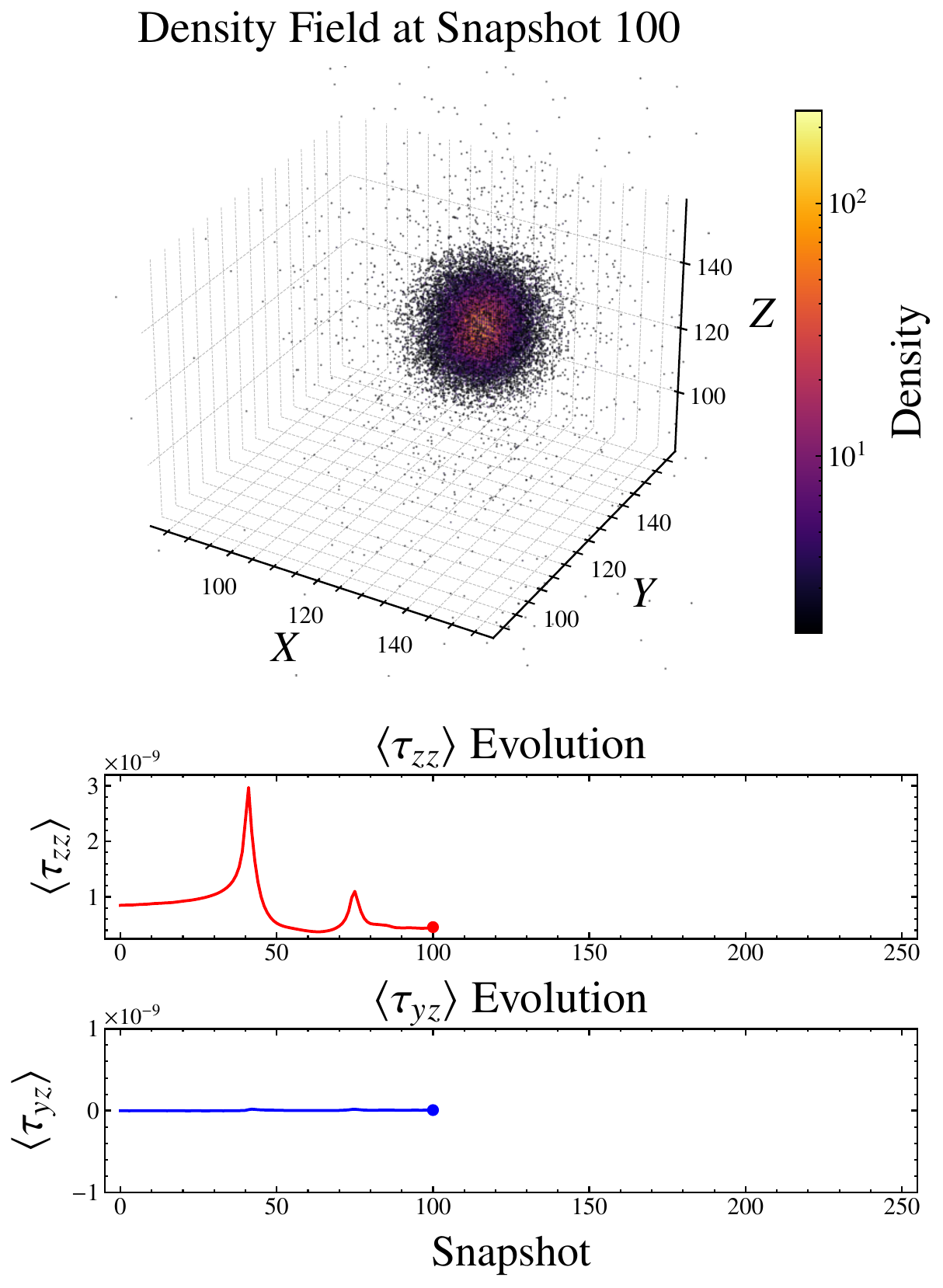}
        \caption{Ringdown Phase}
    \end{subfigure}
    \hfill
    \begin{subfigure}{0.32\textwidth}
        \centering
        \includegraphics[width=\linewidth]{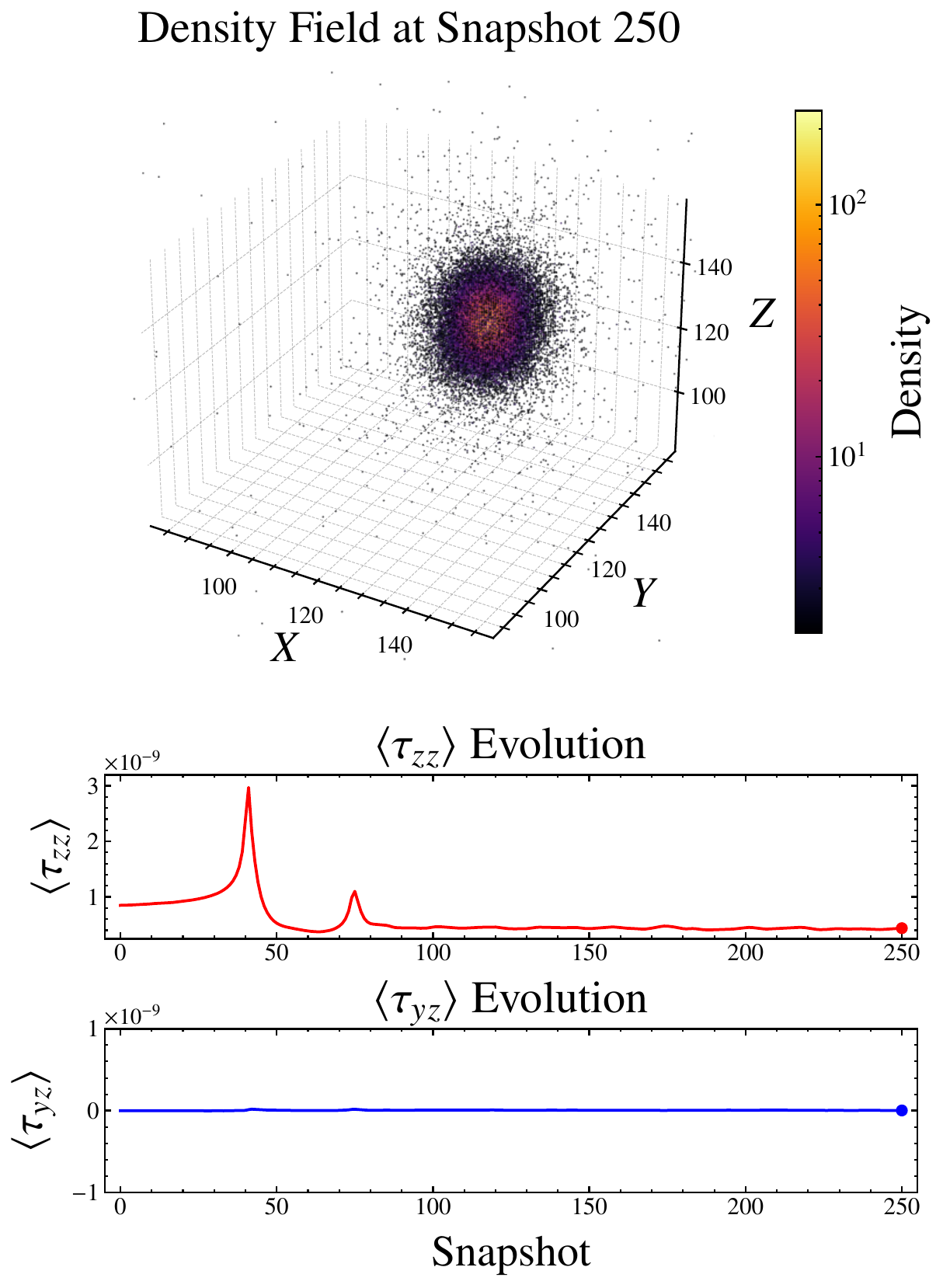}
        \caption{Approaching equilibrium}
    \end{subfigure}

    \caption{The on-axis collision of two Plummer spheres, shown at each phase of the interaction, together with the evolution of the diagonal stress tensor component \(\tau_{zz}\). The collision and later recombination appear as clear peaks in \(\tau_{zz}\), illustrating how the axially symmetric geometry of an on-axis collision sources the diagonal component of the effective stress tensor, without generating off-diagonal components.}
    \label{fig:onaxis}

\end{figure*}

\begin{figure*}[!t]
    \centering

    \begin{subfigure}{0.32\textwidth}
        \centering
        \includegraphics[width=\linewidth]{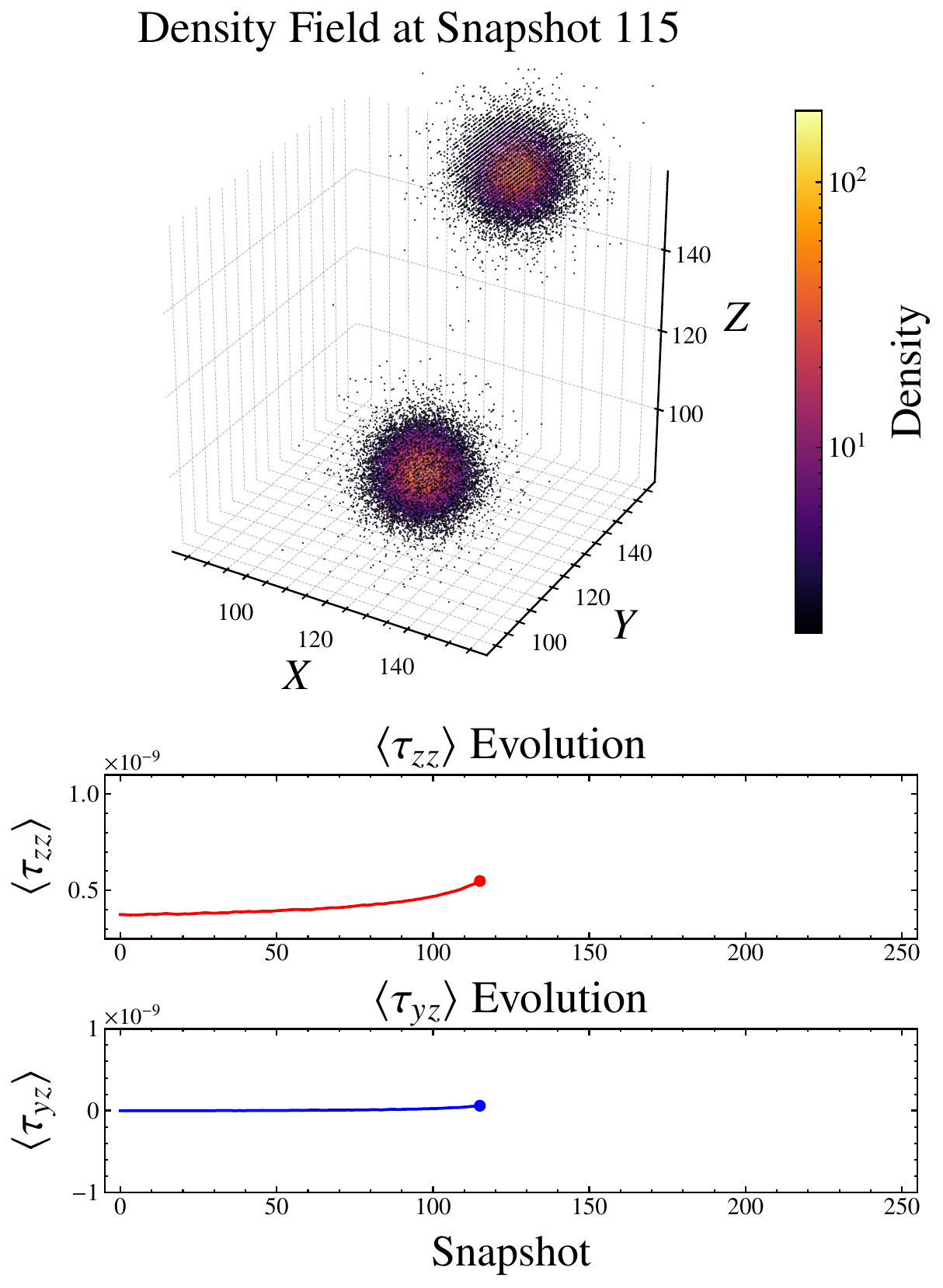}
        \caption{Approach Phase}
    \end{subfigure}
    \hfill
    \begin{subfigure}{0.32\textwidth}
        \centering
        \includegraphics[width=\linewidth]{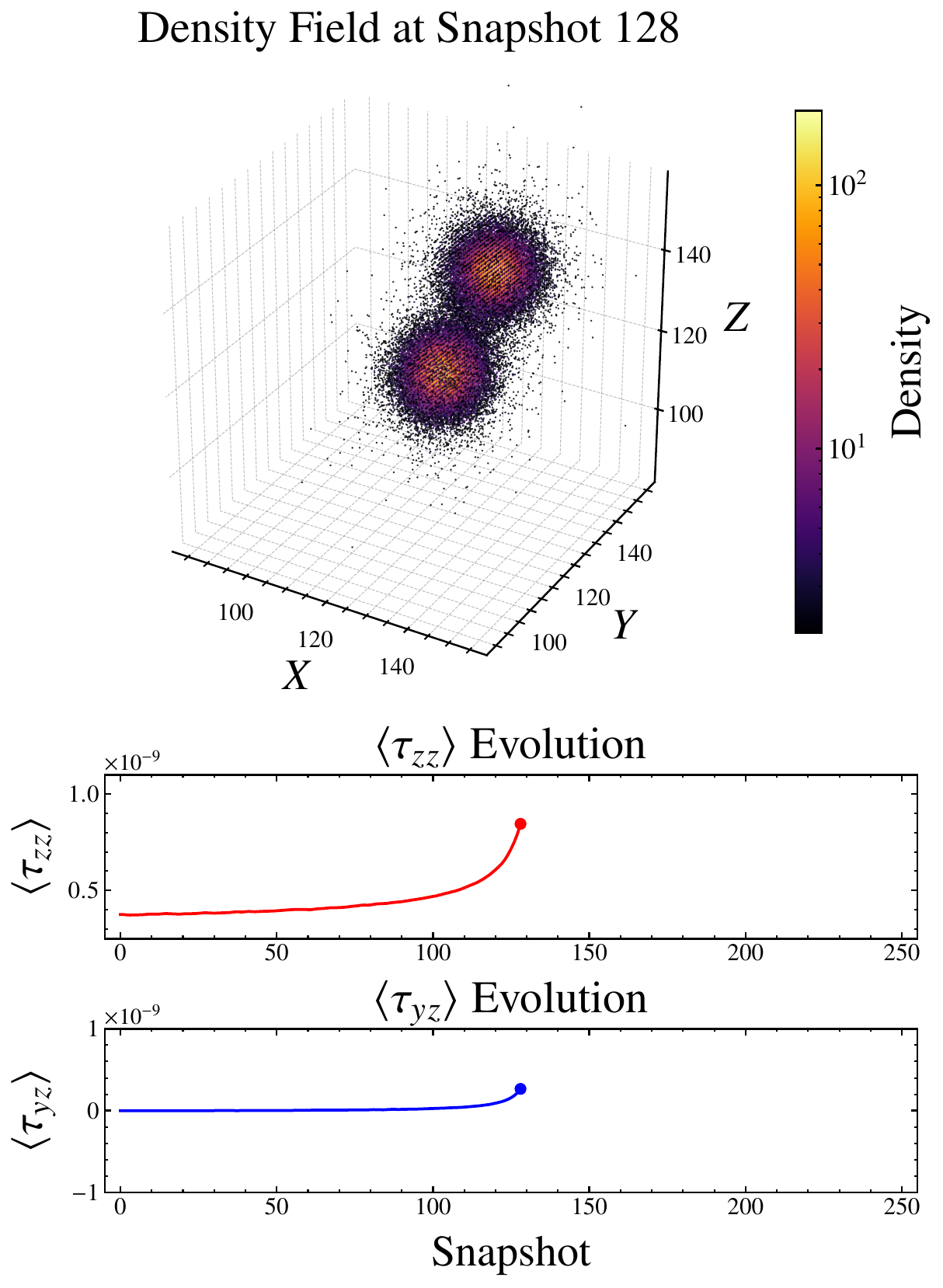}
        \caption{At Collision}
    \end{subfigure}
    \hfill
    \begin{subfigure}{0.32\textwidth}
        \centering
        \includegraphics[width=\linewidth]{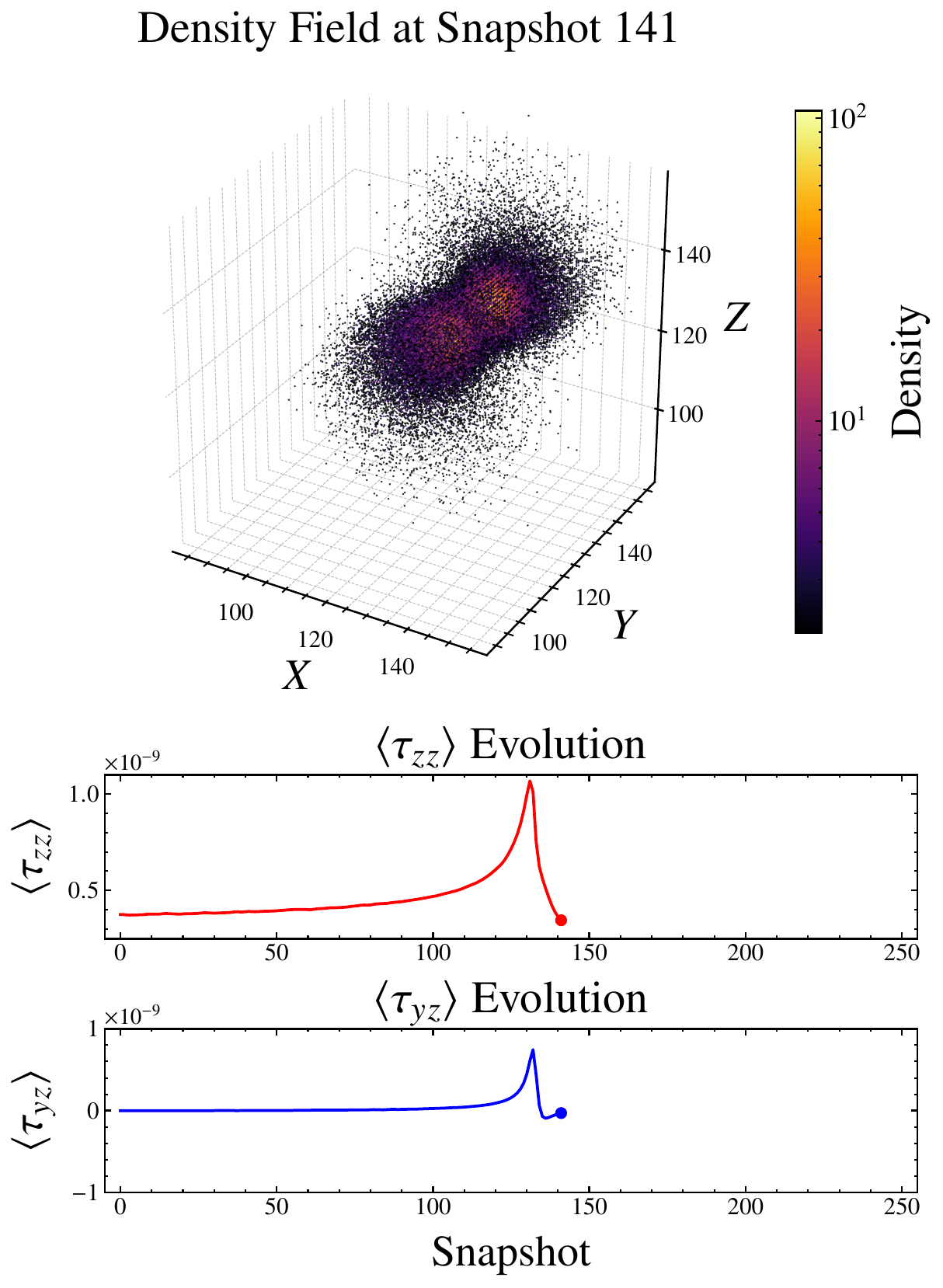}
        \caption{Turnaround Phase}
    \end{subfigure}

    \vspace{0.5em}

    \begin{subfigure}{0.32\textwidth}
        \centering
        \includegraphics[width=\linewidth]{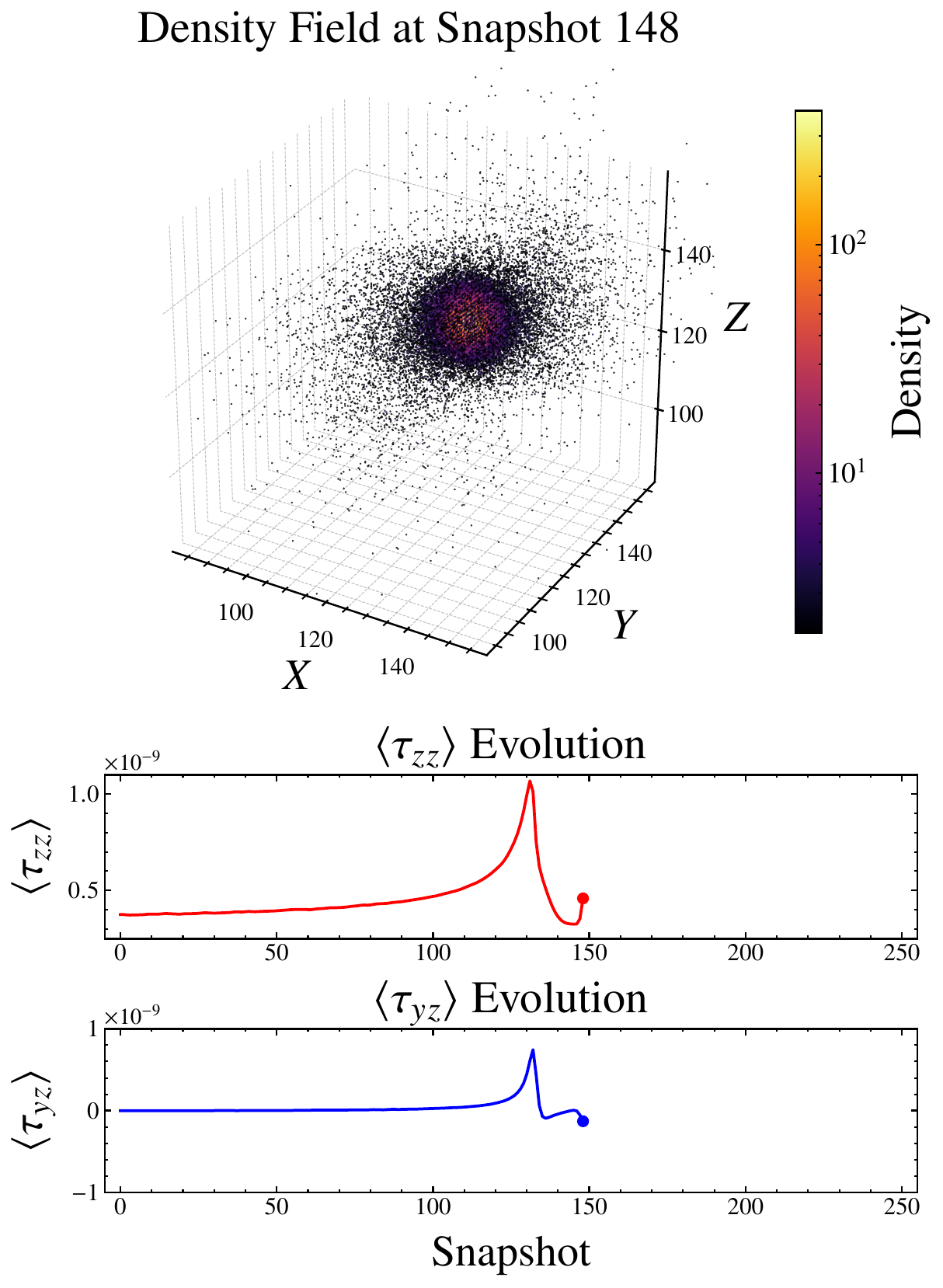}
        \caption{Recombination}
    \end{subfigure}
    \hfill
    \begin{subfigure}{0.32\textwidth}
        \centering
        \includegraphics[width=\linewidth]{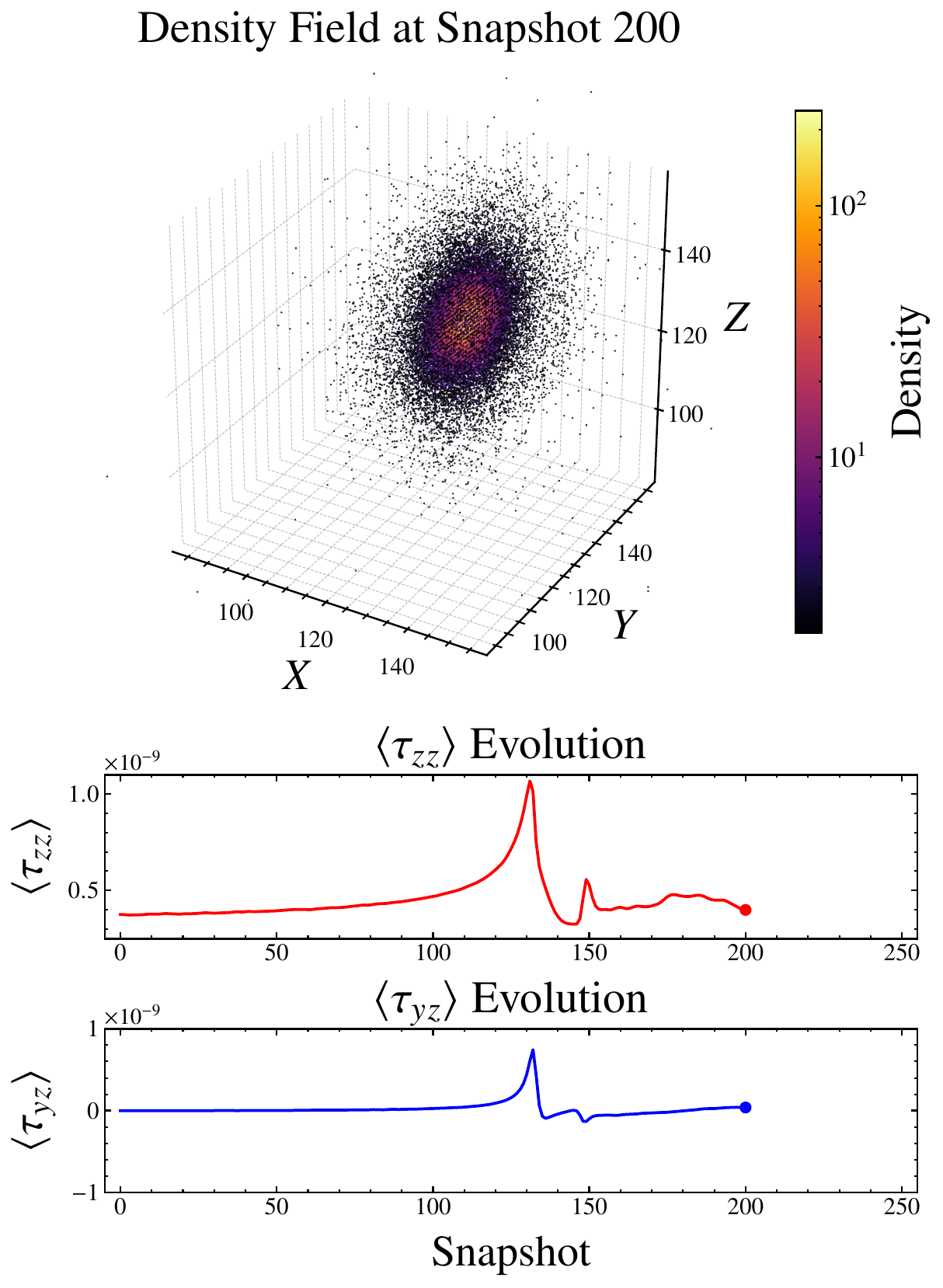}
        \caption{Ringdown Phase}
    \end{subfigure}
    \hfill
    \begin{subfigure}{0.32\textwidth}
        \centering
        \includegraphics[width=\linewidth]{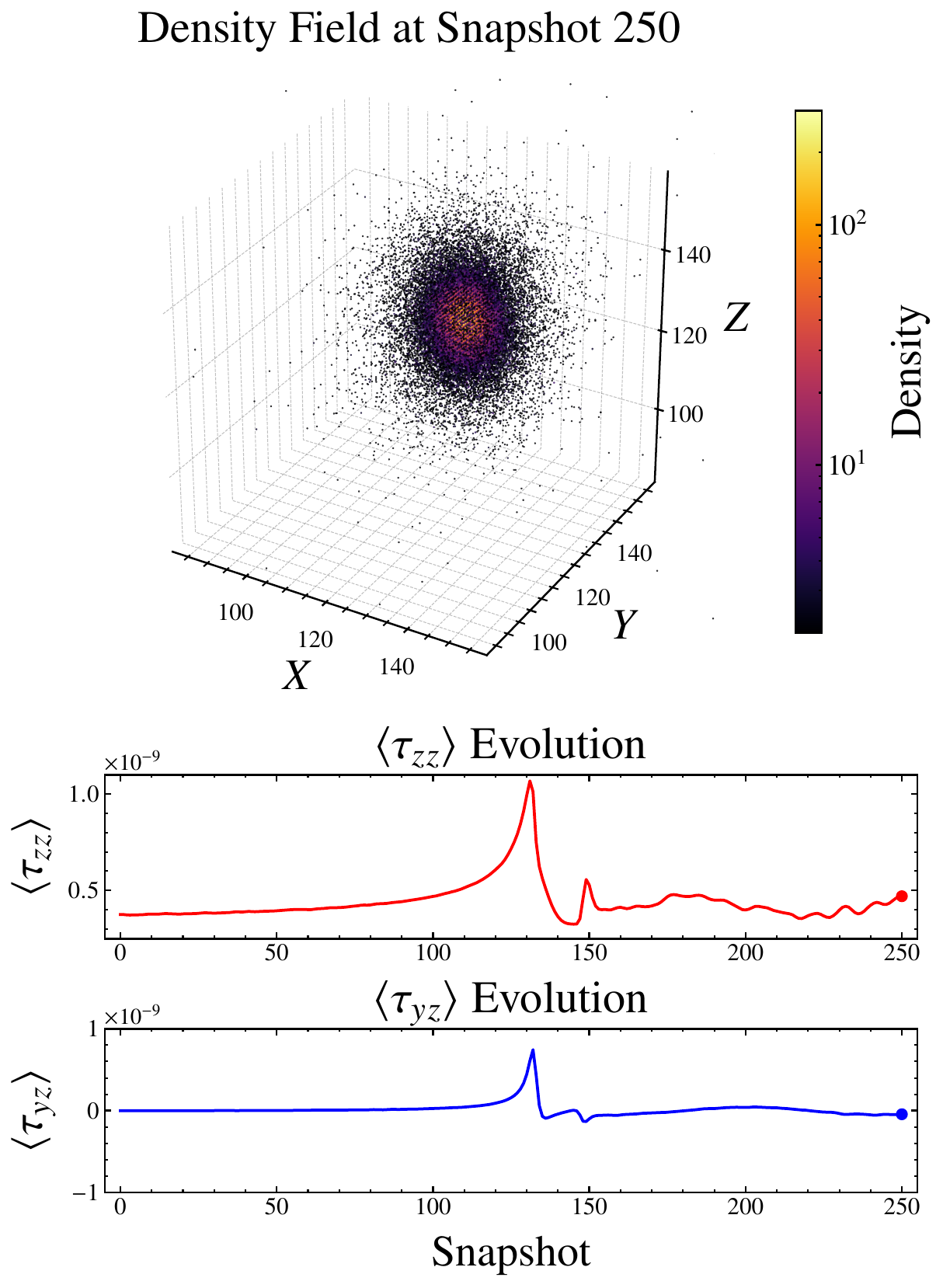}
        \caption{Approaching equilibrium}
    \end{subfigure}

    \caption{The off-axis collision of two Plummer spheres, shown at each phase of the interaction, together with the evolution of the stress tensor. Unlike the on-axis case, the anisotropic collision geometry also generates a non-zero off-diagonal component \(\tau_{yz}\), providing physical intuition for how anisotropic accretion sources the shear viscosity counterterm.}
    \label{fig:offaxis}

\end{figure*}

In Figure~\ref{fig:txx}, we show a snapshot from the simulation and the evolution of the \(xx\) component of the effective stress tensor. Although \(\tau_{xx}\) is non-zero, \(K_{xx} / P_{xx}\) hovers close to \( -1\) for the entire duration, as required. On the other hand, as seen in Figure~\ref{fig:txy}, \(\tau_{xy}\) clearly oscillates around zero, and in this case \(K_{xy} / P_{xy}\) does not stay close to \( -1\). Importantly, \(\tau_{xy}\) is at least one order of magnitude smaller than \(\tau_{xx}\) for nearly all snapshots. Thus, the apparent instability of the off-diagonal virial ratio arises from taking the ratio of two quantities that are both small. We will see that the off-diagonal elements of the stress tensor are, in general, smaller than the diagonal elements. The typical magnitudes of \(K_{xx}\) and \(P_{xx}\) are \(\sim7\times10^{-11}\), while those of \(K_{xy}\) and \(P_{xy}\) are \(\sim5\times10^{-14}\). Correspondingly, the residual \(\tau_{xx}\) is of order \(\sim10^{-12}\), i.e.\ \(\tau_{xx}/K_{xx} \sim 1\%\), consistent with the small residual expected from the imperfect virial cancellation seen in Figure~\ref{fig:txx}.

Since the kinetic and potential terms of \(\tau_{xx}\) clearly cancel, and \(\tau_{xy}\) is much smaller than \(\tau_{xx}\), we conclude that the non-zero \(\tau_{xy}\) is numerical noise. Thus, we have established the robustness of our computation by demonstrating agreement with the EFTofLSS non-renormalization theorem. Although we have only discussed the results of \(\tau_{xx}\) and \(\tau_{xy}\) here, by symmetry, the respective results for the other diagonal elements and the off-diagonal elements are nearly identical.

\subsubsection{Colliding Plummer Spheres}\label{app:collide}

In this section, we analyse two toy model set-ups. In both, the two Plummer spheres are initially separated along the z-axis, but in the second we slightly displace each of them along the y-axis, in opposite directions. The Plummer spheres are then kicked towards each other and we analyse the evolution of the effective stress tensor, during the approach, collision and ringdown phases. 

In the following, we call the first scenario the on-axis collision and the second the off-axis collision. The rationale behind including the latter is that the off-axis scenario breaks the axial symmetry of the on-axis case, which allows for the generation of off-diagonal components. For both scenarios, we simulate two Plummer spheres each with \(\mathrm{Mass} = 10\), scale radius \(\mathrm{a} = 5\), truncation radius \(\mathrm{R} = 67\) and \(100,000\) particles each. In the on-axis case, they are separated by 192 units along the z-axis and are each kicked towards each other by a velocity \(v_{\mathrm{kick}} = 0.5\).

In the off-axis case, they are separated by 212 units along the z-axis and are each displaced by 32 units along the y-axis, in opposite directions. In this case, they are kicked towards each other by a velocity \(v_{\mathrm{kick}} = 0.1\), parallel to the z-axis. Figures~\ref{fig:onaxis} and~\ref{fig:offaxis} portray the evolution of the diagonal and off-diagonal component of the stress tensor during the on-axis and off-axis collisions, respectively. In both scenarios, the spheres approach each other and collide. As we are analysing dark matter simulations, these spheres pass through each other, until their gravitational attraction forces them to turn around and recombine. Both systems then undergo a ringdown phase and settle into an equilibrium. The initial collision and later recombination are clearly seen as peaks in the diagonal stress tensor component, \(\tau_{zz}\). It's interesting to note that the off-diagonal component, \(\tau_{yz}\), shows no signal in the on-axis case but exhibits a non-zero signal in the off-axis collision. This signal captures the intuition for the physical process underlying the generation of the related counterterm and is discussed in more detail in Section~\ref{sec:counter2}. With this we conclude our analysis of the toy models. 

\section{Comparison and Validation of Results}\label{app:smallbox}

\subsection{Comparing the Approximate and Complete Calculations}\label{app:fullapprox}

\begin{figure}[!t]
    \centering
    \begin{subfigure}{0.51\textwidth}
        \centering
        \includegraphics[width=\linewidth]{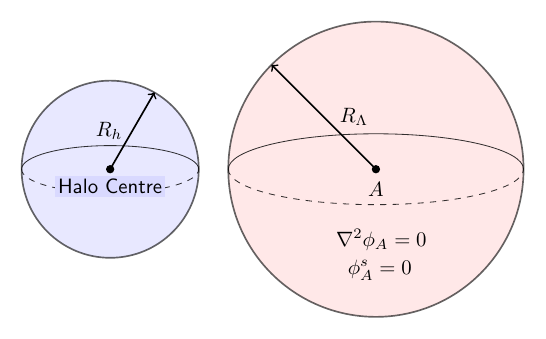}
        \caption{Non-intersecting Halo and smoothing sphere}
        \label{fig:sphere1}
    \end{subfigure}
    \hfill
    \begin{subfigure}{0.45\textwidth}
        \centering
        \includegraphics[width=\linewidth]{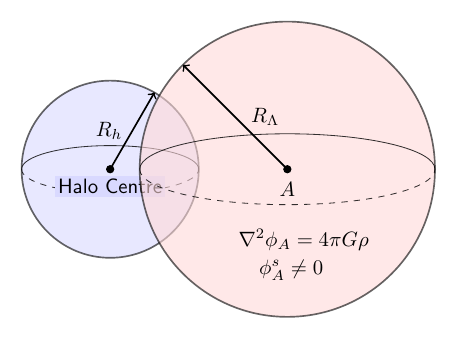}
        \caption{Intersecting Halo and smoothing sphere}
        \label{fig:sphere2}
    \end{subfigure}
    \caption{Illustration of the geometric condition determining whether a point A outside the halo contributes to the short-scale potential \(\phi^s\). (a) When the smoothing sphere of radius \(R_\Lambda\) centred at A does not intersect the halo's analysis sphere, the halo mass distribution is exterior to a region free of enclosed mass around A, so \(\phi_A\) is harmonic and \(\phi^s_A = 0\). (b) When the two spheres intersect, part of the halo mass lies within \(R_\Lambda\) of A, \(\phi_A\) is no longer harmonic, and \(\phi^s_A \neq 0\). This condition sets the minimum cube size L needed to capture all nonzero contributions.}
    \label{fig:sphere}
\end{figure}

\begin{figure}[!t]

    \centering
    \begin{subfigure}{0.49\textwidth}
        \centering
        \includegraphics[width=\linewidth]{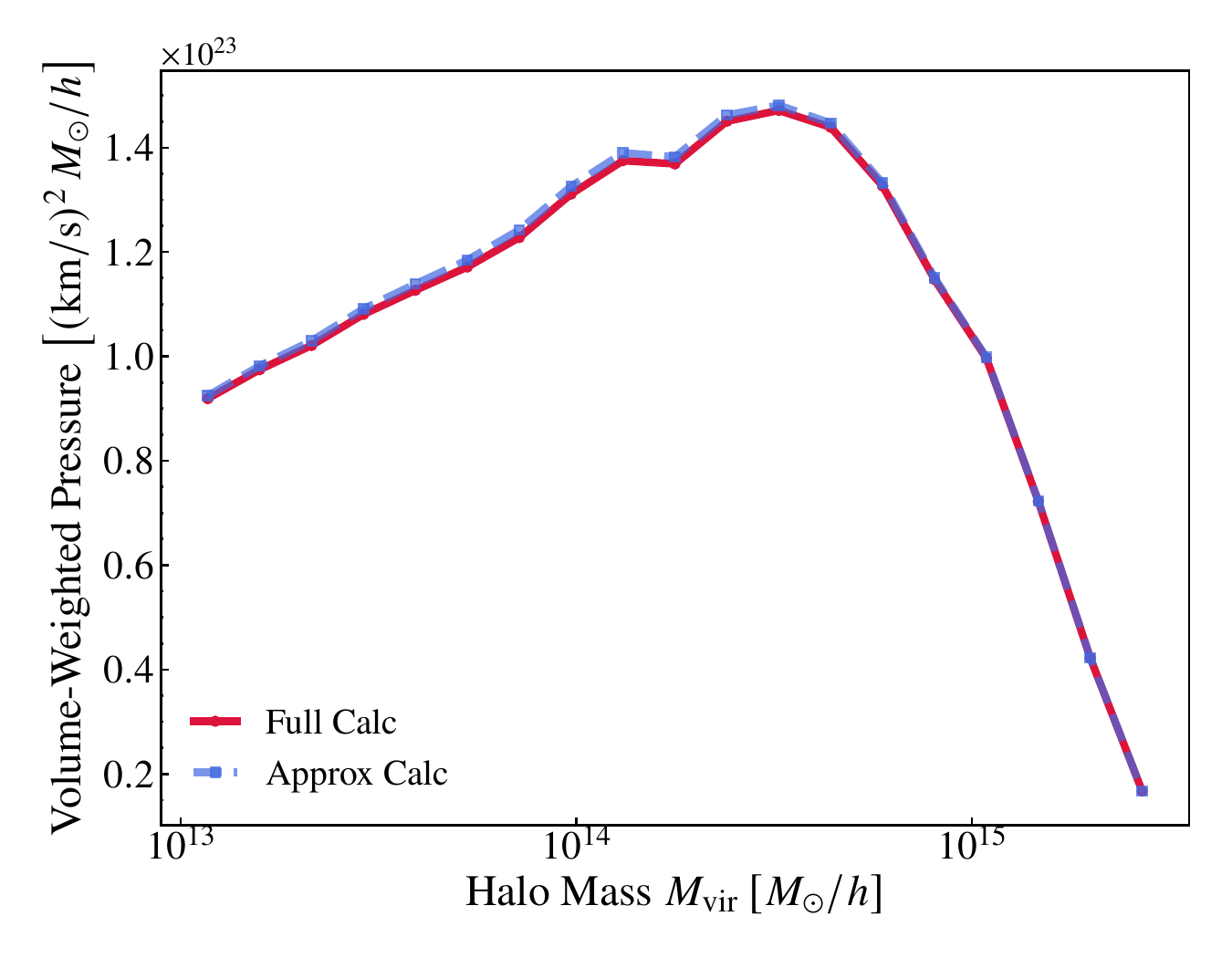}
        \label{fig:fullapprox_pressure}
    \end{subfigure}
    \hfill
    \begin{subfigure}{0.49\textwidth}
        \centering
        \includegraphics[width=\linewidth]{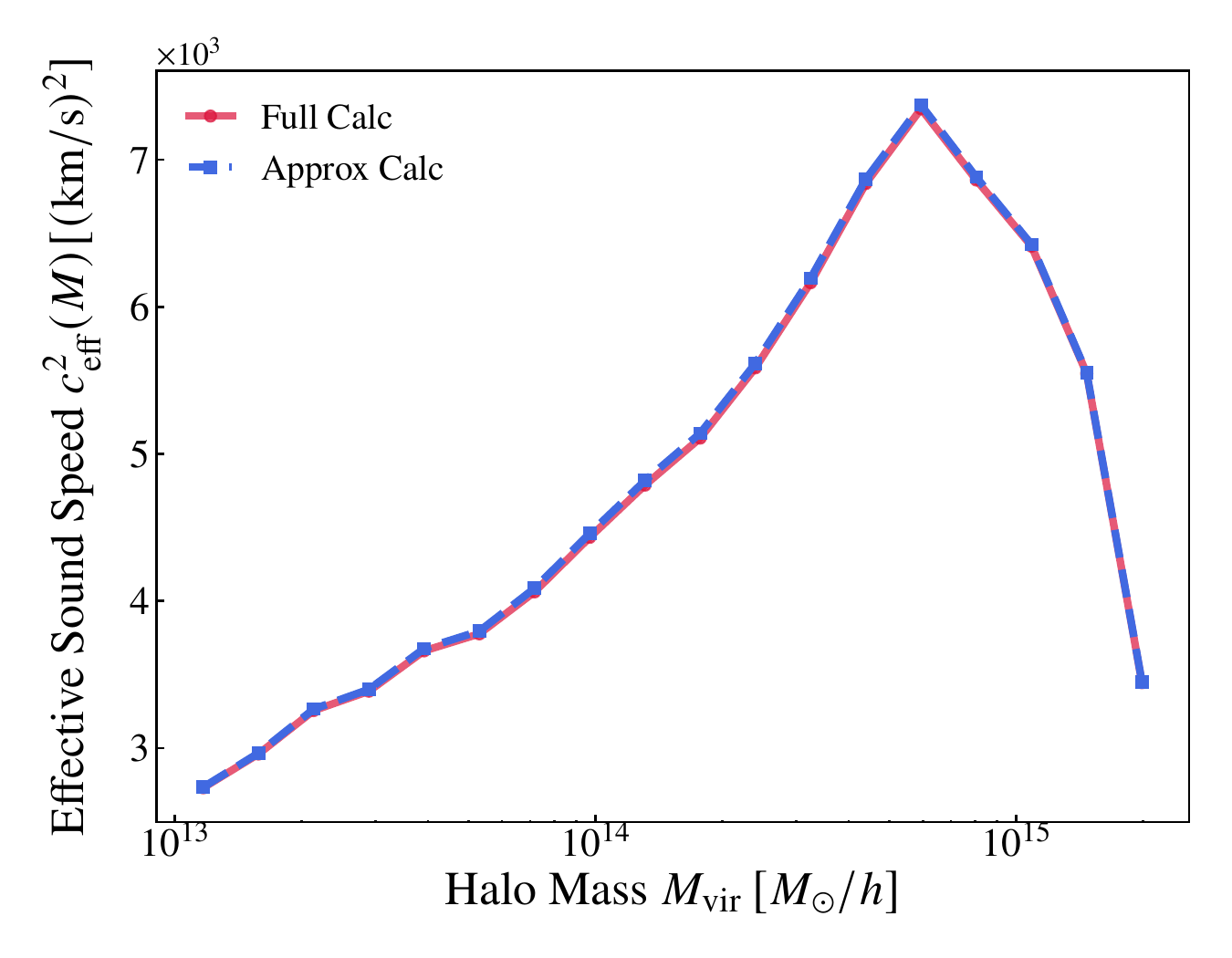}
        \label{fig:fullapprox_ceff}
    \end{subfigure}

    \begin{subfigure}{0.49\textwidth}
        \centering
        \includegraphics[width=\linewidth]{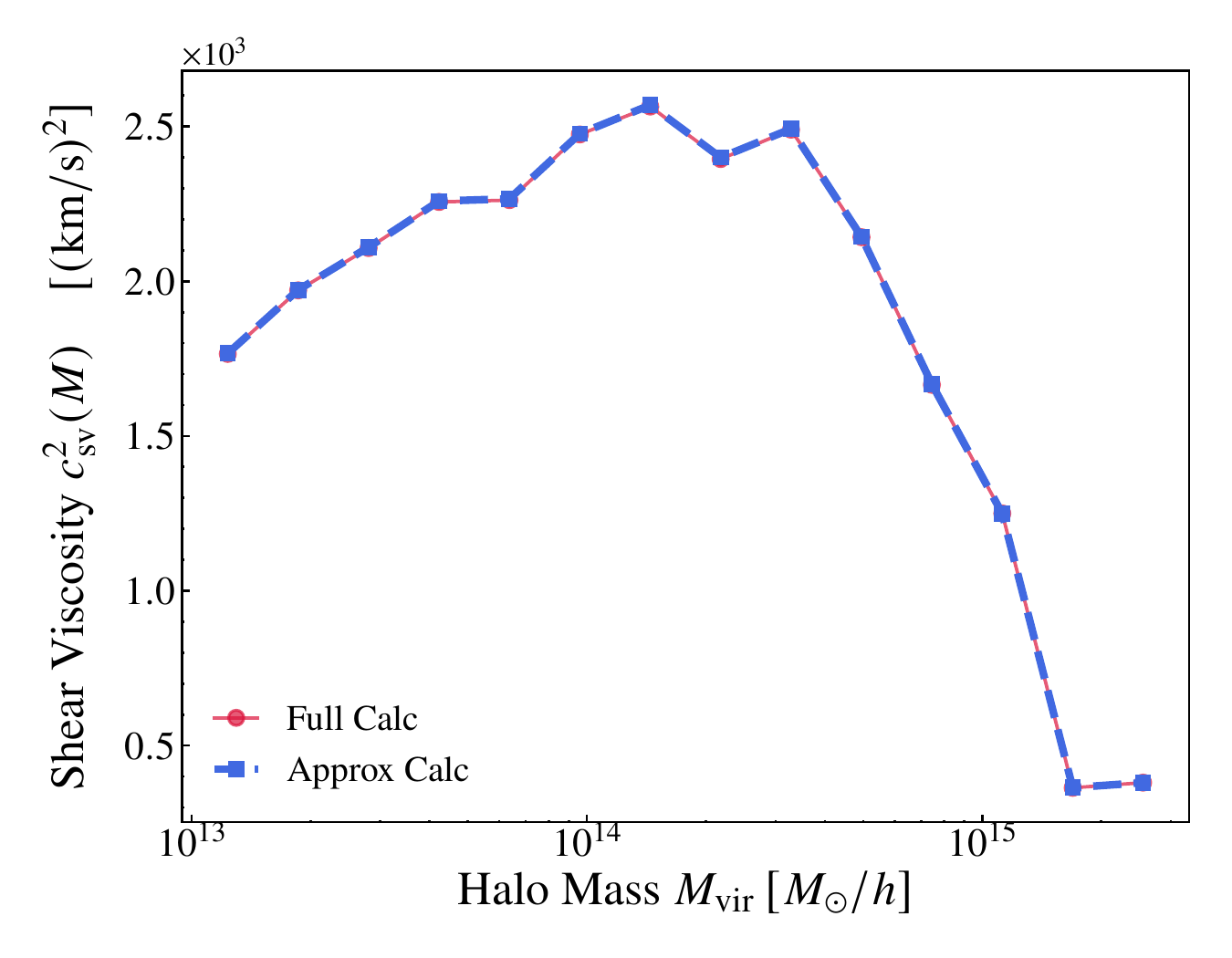}
        \label{fig:fullapprox_cvis}
    \end{subfigure}

    \caption{Comparison between the Complete and Approximate calculations for \(\bar{P}\), \(c_{\mathrm{eff}}^2\), and \(c_{sv}^2\). The excellent numerical agreement between the two calculations justifies our choice to proceed with the approximation.}

    \label{fig:fullapprox}

\end{figure}

In Section~\ref{sec:potterm}, we made the approximation L \(> 2R_{\Lambda}\) instead of L \(> 2(R_h + 2R_{\Lambda})\) and claimed that there was excellent agreement between this approximation and the complete calculation. In this brief subsection we will first argue that all contributions to \(\phi^s\) are included if we choose L \(> 2(R_h + 2R_{\Lambda})\) and then validate the approximation by comparing the results of the approximate and full calculations for \(R_{\Lambda} = 28 \hspace{0.1cm}\mathrm{Mpc}/h\). As we are interested in the gravitational potential due to the halo, naively it may seem like we would have to sum over the potential due to the halo over the entire simulation box, but as we are interested in only the short-scale potential field, this is not necessary. To establish this fact, say we are computing the short-scale potential field at some point, A, far outside the halo, such that the sphere of radius \(R_{\Lambda}\) centred at A has no intersection with the analysis sphere around the halo centre, as seen in Figure~\ref{fig:sphere1}. 

As a result, the smoothing sphere around A contains no halo particles, \(\rho_A = 0\), and the gravitational potential due to the halo within the smoothing sphere must satisfy Laplace's equation: \(\nabla^2\phi_A= 0\). In other words, \(\phi_A\) is a harmonic function. To construct \(\phi^l_A\), we must compute the average potential within the smoothing sphere, and by the mean value property of harmonic functions, the average value of a harmonic potential within a sphere is simply the value at the centre of the sphere. Thus we have: 
\begin{equation}
    \phi_A = \phi^l_A \implies \phi^s_A = 0\,.
\end{equation}
On the other hand, if A is close enough to the halo so that the smoothing sphere around A intersects with the halo, \(\phi_A\) is no longer a harmonic function and hence \(\phi^s_A \neq 0\), as depicted in Figure~\ref{fig:sphere2}. As a result, we establish that the potential term can be computed as a sum over the cells within the isolated cube, and all contributions will be included as long as L is chosen such that L \(> 2(R_h + 2R_{\Lambda})\). For the smoothing scales we will be interested in, such an L is quite large, making our computation expensive. Hence we make an approximation and choose L \(> 2R_{\Lambda}\). Although this may seem like a drastic approximation, the approximate and complete calculations show excellent numerical agreement. For a smoothing scale of \(R_{\Lambda} = 28 \hspace{0.1cm}\mathrm{Mpc}/h\), the complete calculation requires L \(> 130\hspace{0.1cm} \mathrm{Mpc}/h\). Thus, for the complete computation we choose L \(= 135\hspace{0.1cm} \mathrm{Mpc}/h\) and for the approximate calculation we choose L \(=58 \hspace{0.1cm}\mathrm{Mpc}/h\). To keep the resolution of the computation the same for both calculations, we compute on a \(256^3\) grid for the approximation and on a \(512^3\) grid for the complete case. The computational expense due to the increase in grid resolution is precisely why we prefer the approximate calculation. The results of this comparison are shown in Figure~\ref{fig:fullapprox}. Clearly, both the calculations show nearly the same results, validating our choice to use the approximate approach. 

\begin{figure}[!t]

    \centering
    \begin{subfigure}{0.49\textwidth}
        \centering
        \includegraphics[width=\linewidth]{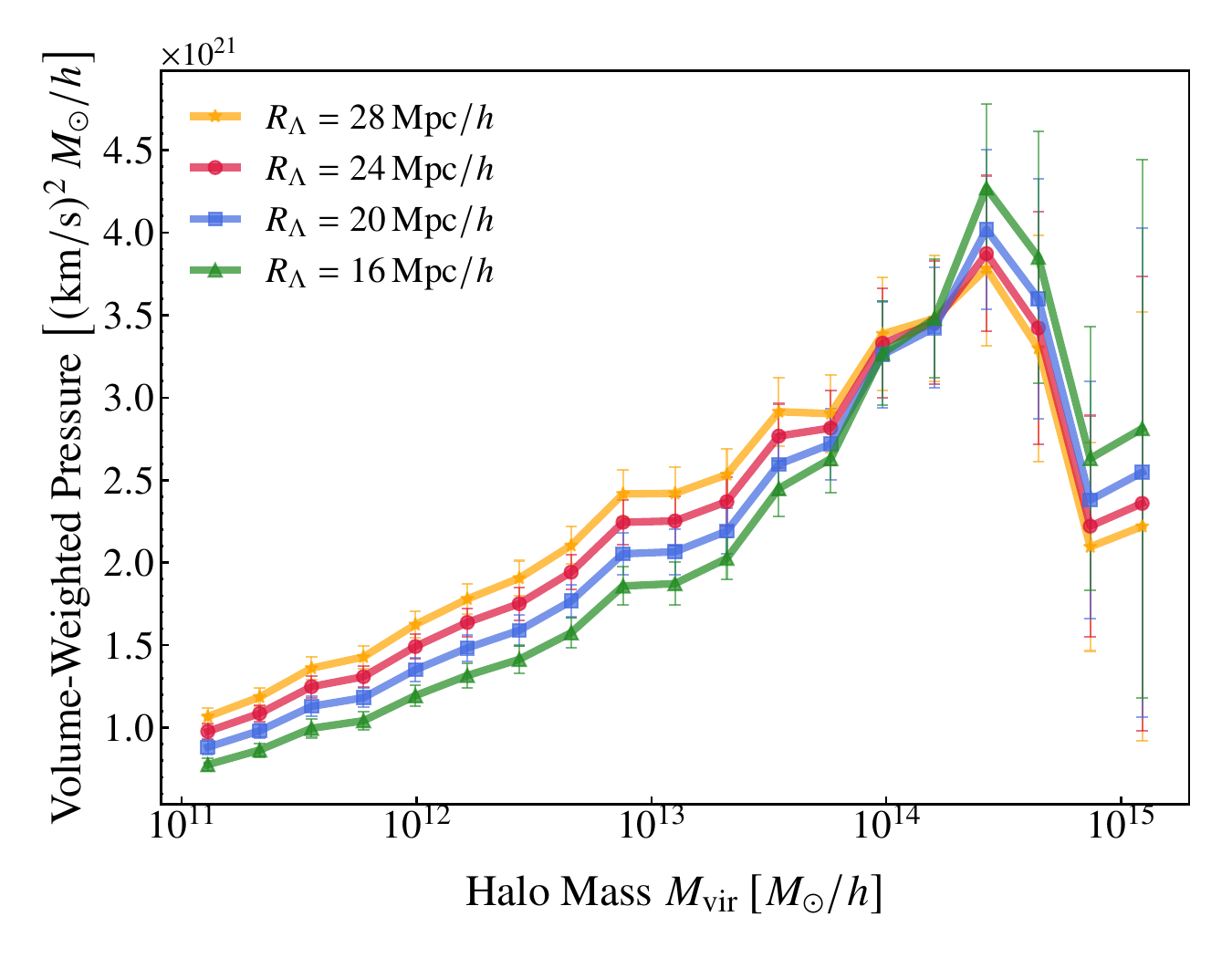}
    \end{subfigure}
    \hfill
    \begin{subfigure}{0.49\textwidth}
        \centering
        \includegraphics[width=\linewidth]{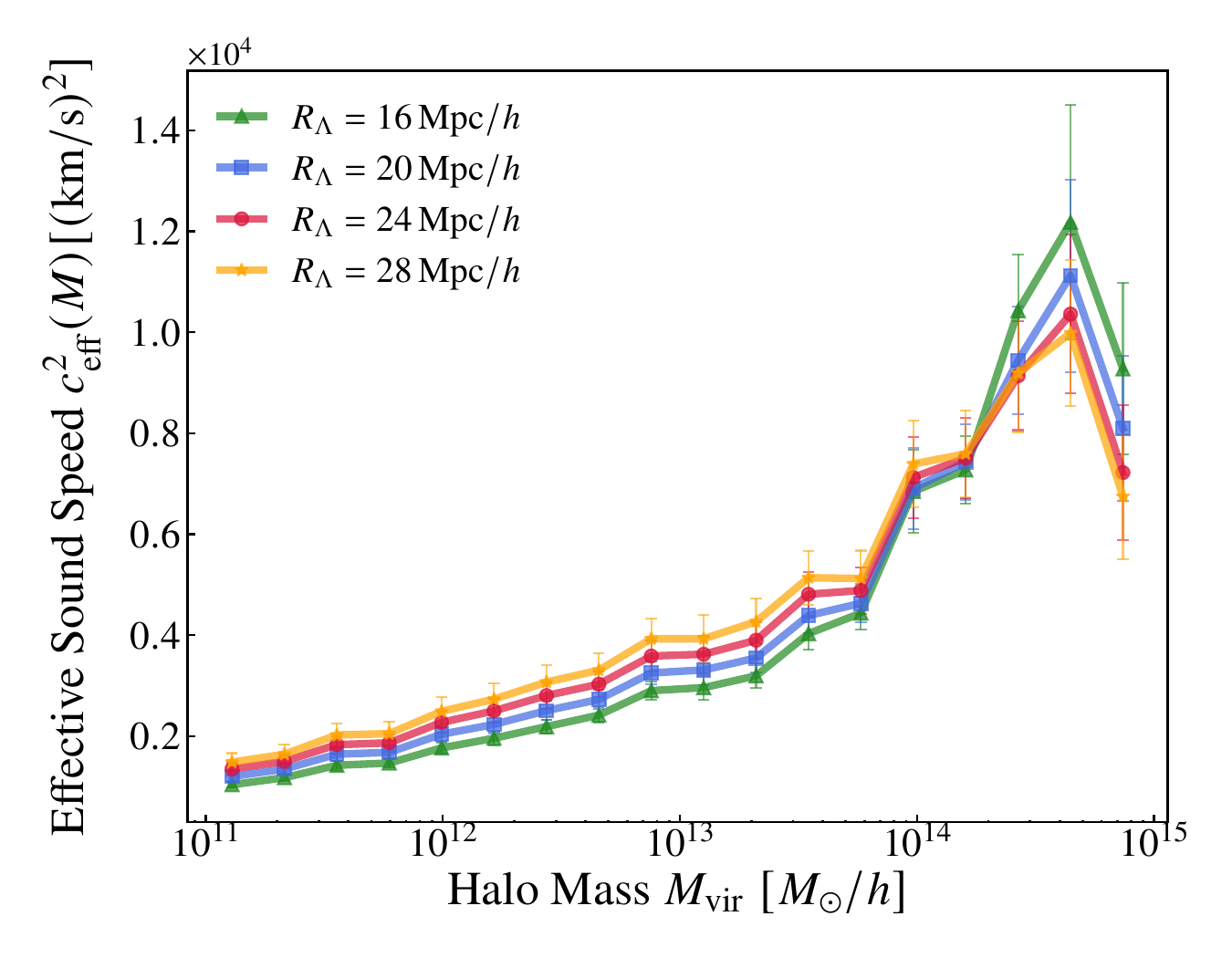}
    \end{subfigure}
    \begin{subfigure}{0.49\textwidth}
        \centering
        \includegraphics[width=\linewidth]{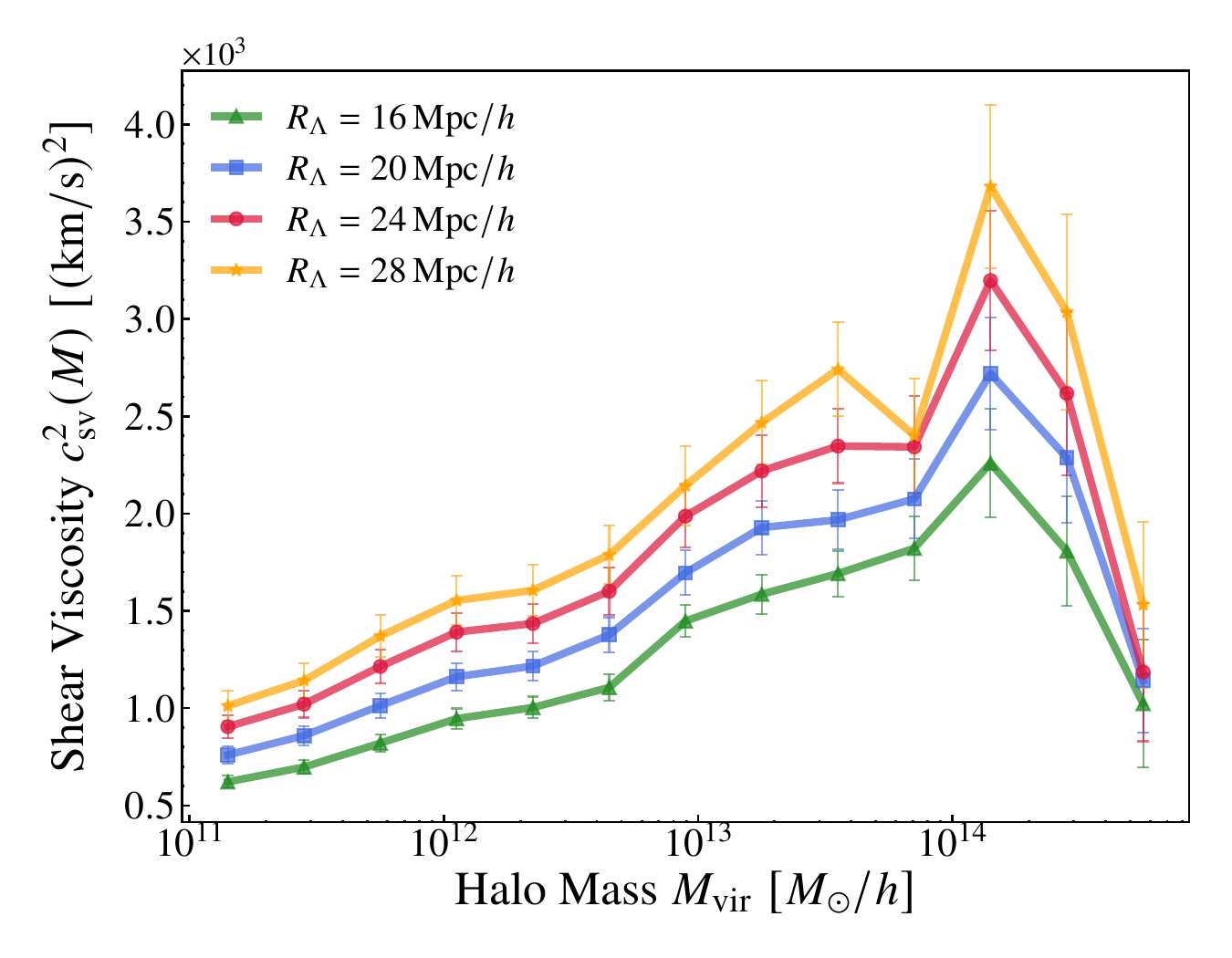}
    \end{subfigure}

    \caption{Computing \(\bar{P}\), \(c_{\mathrm{eff}}^2\) and \(c_{sv}^2\) for a smaller box allows us to extend the computation to lower halo masses. We note the suppression in the contributions and the absence of any UV divergence.}

    \label{fig:smallbox}

\end{figure}

\subsection{Small Box Simulation}\label{app:smallbox1}

In this brief section we present the results from the analysis on the \(250 \hspace{0.1cm} \mathrm{Mpc}/h\) box described in Section~\ref{sec:haloanalysis}, which allows us to extend the computation to halos with \(M_\mathrm{vir} > 10^{11} M_{\odot}/h\). The results of this analysis are shown in Figure~\ref{fig:smallbox}. Clearly there is no UV-divergence and no new peaks in the contribution for any counterterm. The contributions continue to be suppressed as we extend to lower halo masses. Thus we establish that the major contribution comes from cluster-scale halos, validating our use of the \(1000 \hspace{0.1cm} \mathrm{Mpc}/h\) box, which resolves only \(M_\mathrm{vir} > 10^{13} M_{\odot}/h\) but better samples these rare massive halos, for the analysis in Section~\ref{sec:extract}.\\~\\

\begin{figure}[!t]
    \centering
    \begin{subfigure}{0.49\textwidth}
        \centering
        \includegraphics[width=\linewidth]{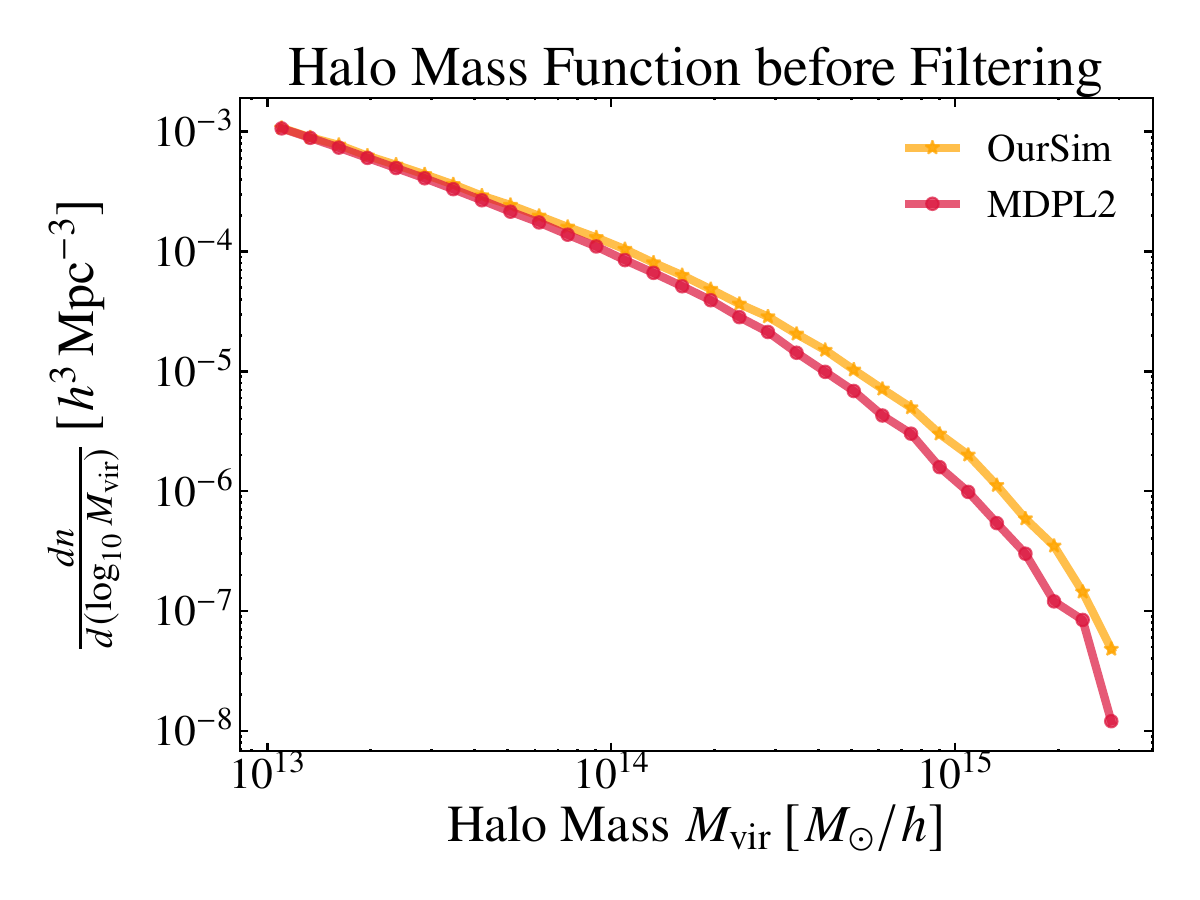}
    \end{subfigure}
    \hfill
    \begin{subfigure}{0.49\textwidth}
        \centering
        \includegraphics[width=\linewidth]{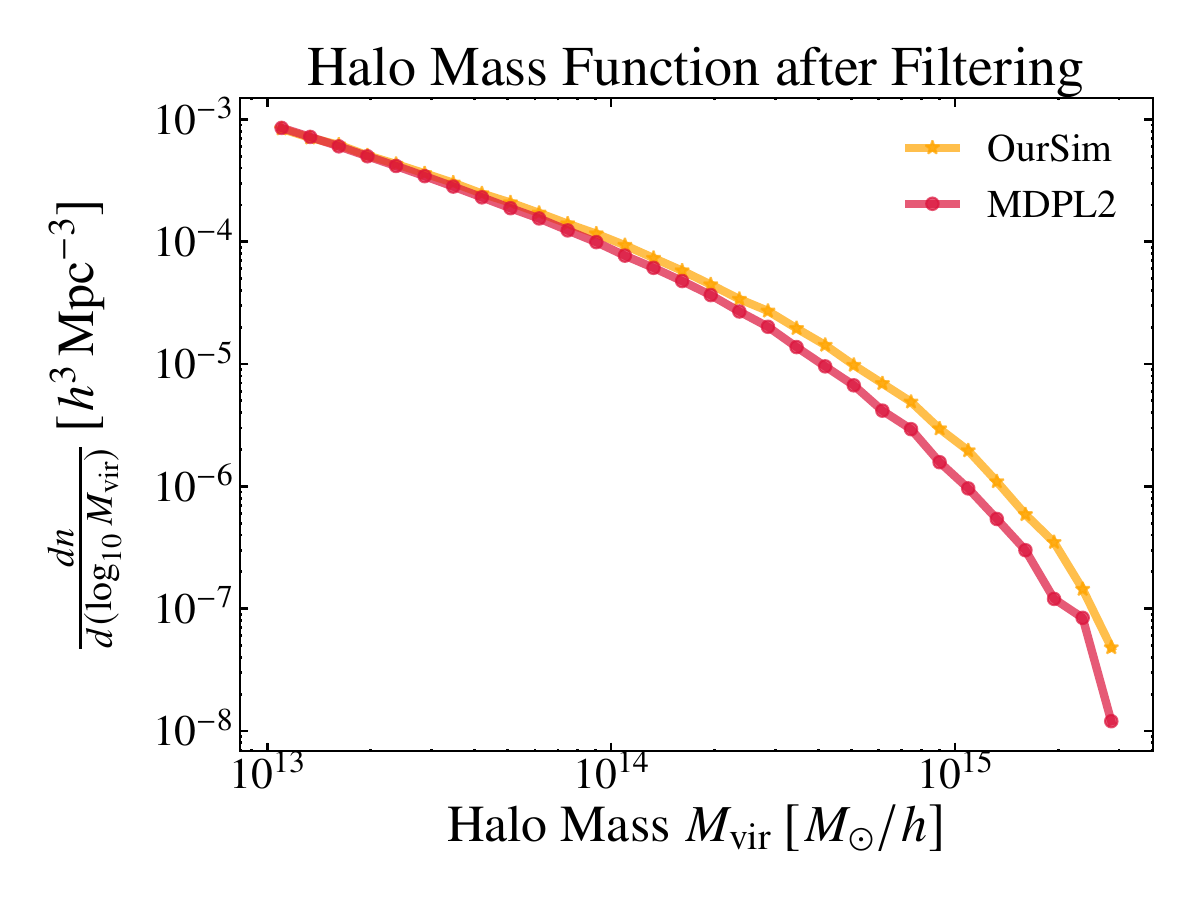}
    \end{subfigure}
    \caption{We compare the Halo Mass Functions from our simulations and \textsc{MDPL2}, before and after applying the geometric filter. Clearly our simulations contain more halos, leading to larger halo-based counterterm magnitudes.}
    \label{fig:mdpl2compare1}
\end{figure}
\vspace{-0.5 cm}
\begin{figure}[!t]
    \centering
    \begin{subfigure}{0.49\textwidth}
        \centering
        \includegraphics[width=\linewidth]{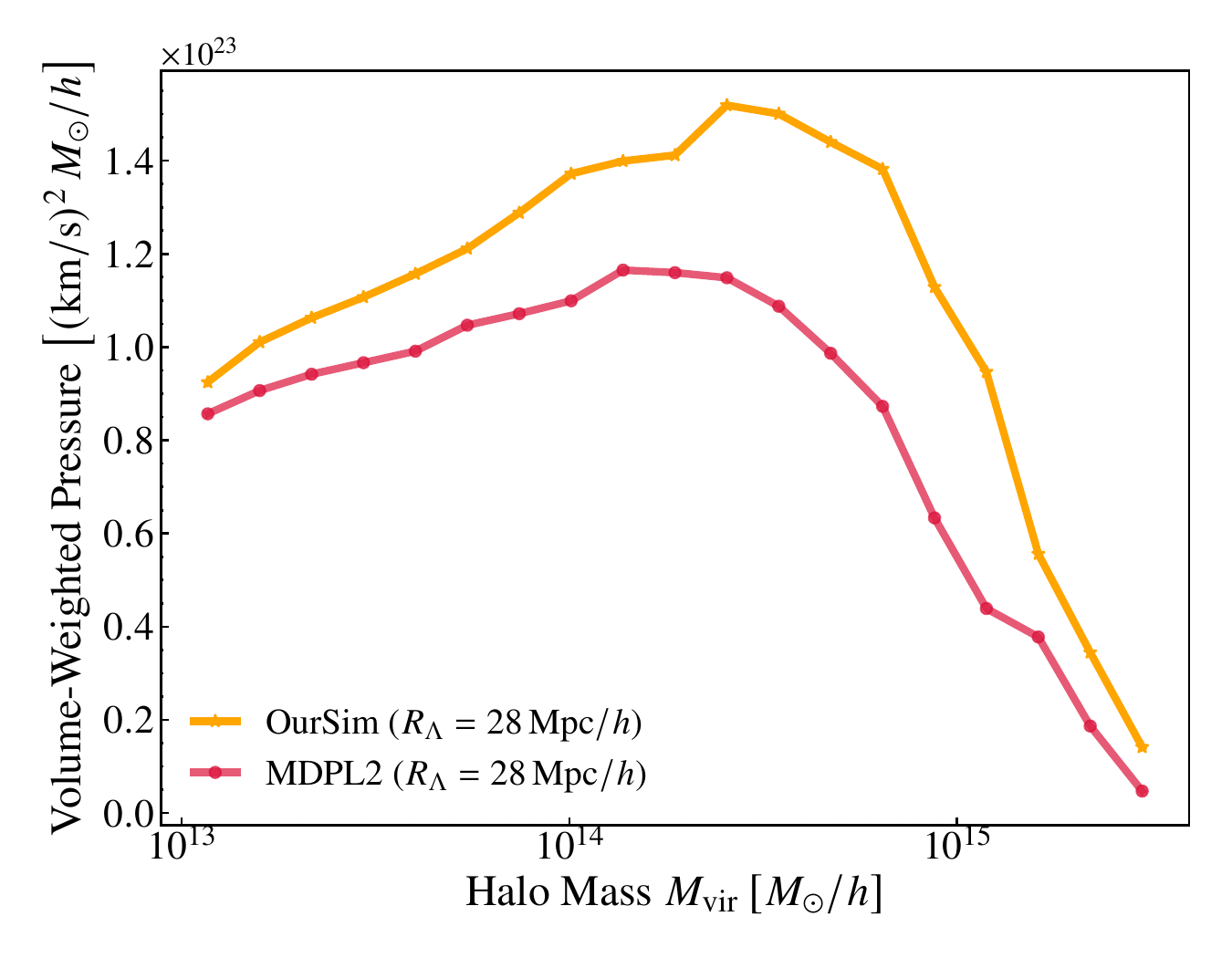}
    \end{subfigure}
    \hfill
    \begin{subfigure}{0.49\textwidth}
        \centering
        \includegraphics[width=\linewidth]{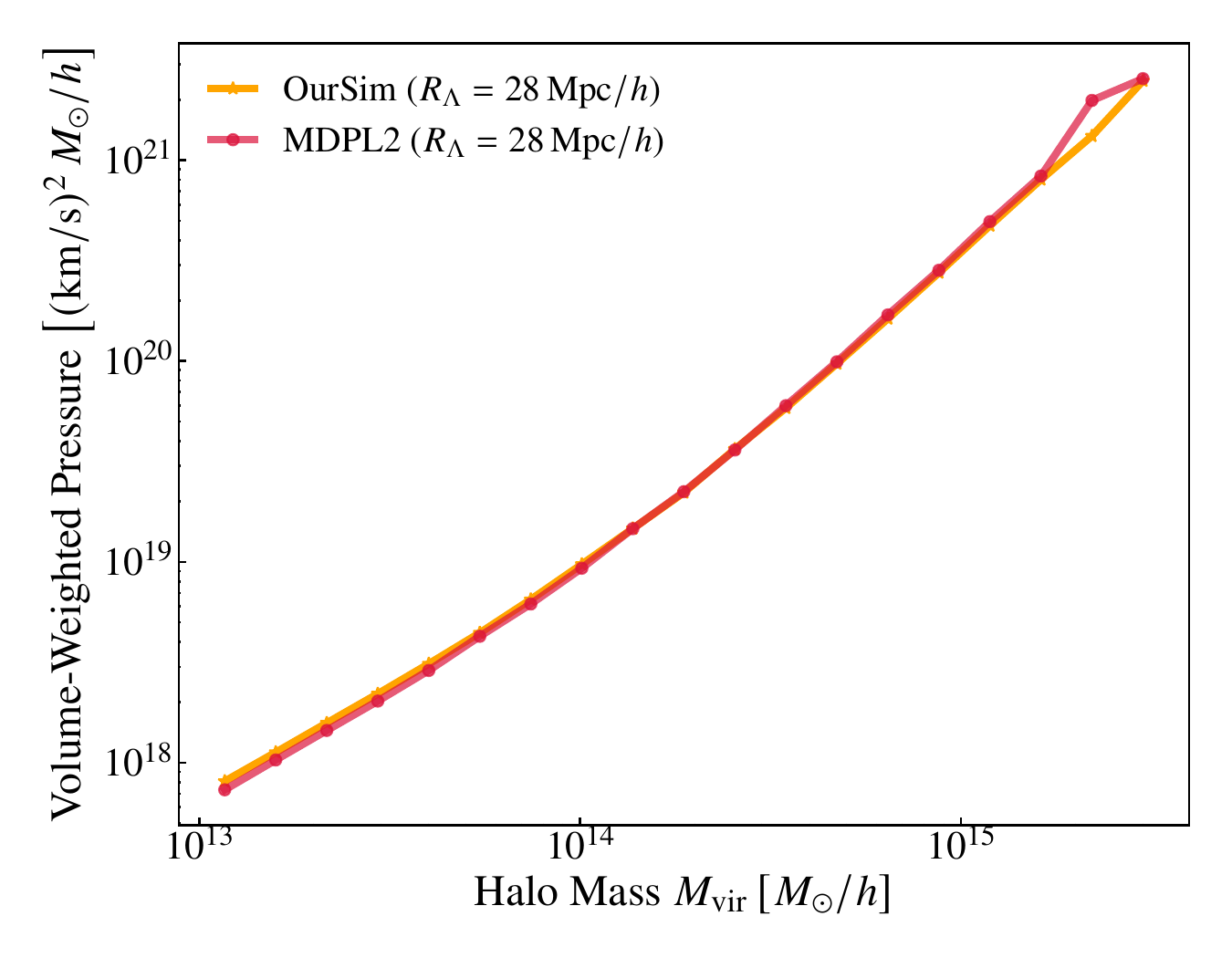}
    \end{subfigure}
    \caption{We compare the box-summed Volume-weighted Pressure Contributions from our simulations and \textsc{MDPL2} (\textit{left}), and the per-halo average \(\langle V^h P^h \rangle\) in each mass bin (\textit{right}). As expected, the analysis on our simulations gives larger counterterm magnitudes, but the average per-halo pressure contribution is the same, leading us to conclude that the difference in the summed results is due to the different halo mass functions, and not a systematic difference in the pressure of halos of a given mass.}
    \label{fig:mdpl2compare2}
\end{figure}

\subsection{Comparison with MDPL2 Simulation Analysis}\label{app:mdpl2}

In this brief section, we compare the results of our halo-based analysis on our simulations and the \textsc{MDPL2} simulation \cite{Klypin_2016}. As mentioned in Section~\ref{sec:counter1}, the cosmologies used by \textsc{MDPL2} and our simulations are not the same, leading to different Halo Mass Functions and hence, different overall results. Figure~\ref{fig:mdpl2compare1} shows a comparison between the Halo Mass Functions, before and after removing halos which fall inside the analysis radius of a larger halo. Clearly our simulations contain more halos, leading to larger halo-based estimates. As seen in Figure~\ref{fig:mdpl2compare2}, the volume-weighted pressure contributions are larger in the analysis on our simulations, but there is strong agreement between the average volume weighted pressure in each mass bin. This implies that the stress tensor calculation is consistent between the two analyses, and that the enhanced total pressure contribution in our simulations is primarily driven by the increased abundance of halos rather than a systematic difference in the pressure associated with halos of a given mass. 

\begin{figure}[!t]

    \centering
    \begin{subfigure}{0.49\textwidth}
        \centering
        \includegraphics[width=\linewidth]{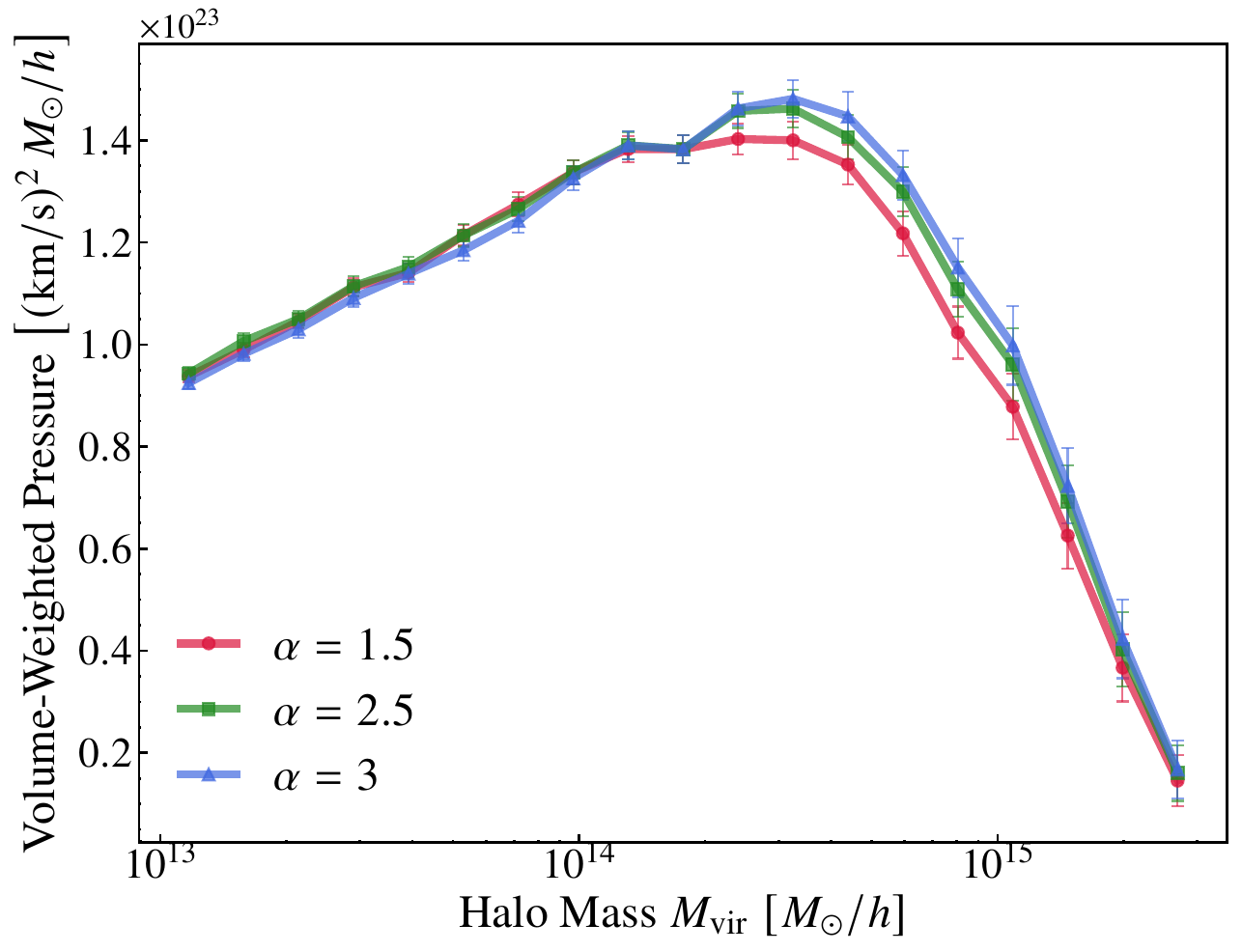}
    \end{subfigure}
    \hfill
    \begin{subfigure}{0.49\textwidth}
        \centering
        \includegraphics[width=\linewidth]{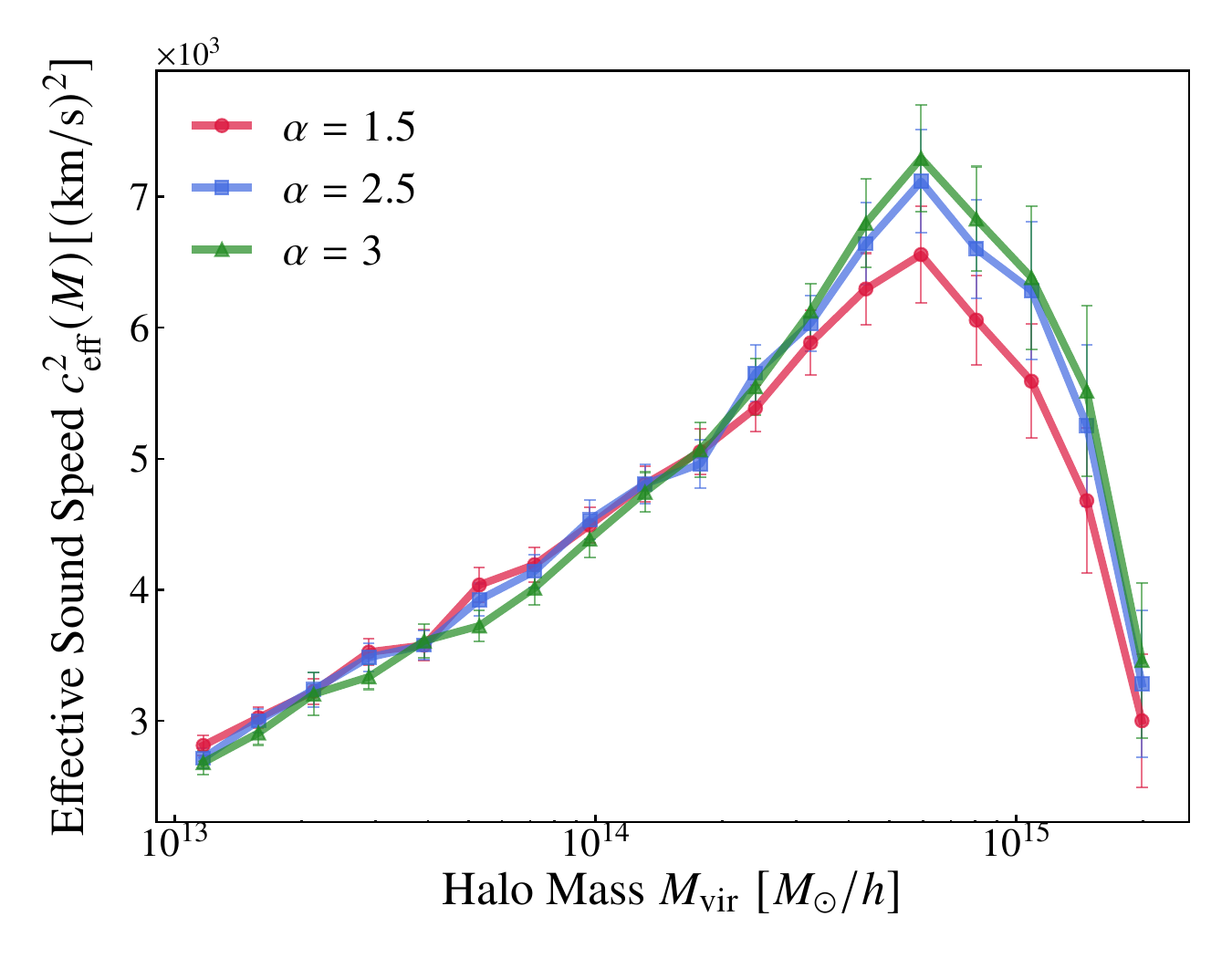}
    \end{subfigure}
    \begin{subfigure}{0.49\textwidth}
        \centering
        \includegraphics[width=\linewidth]{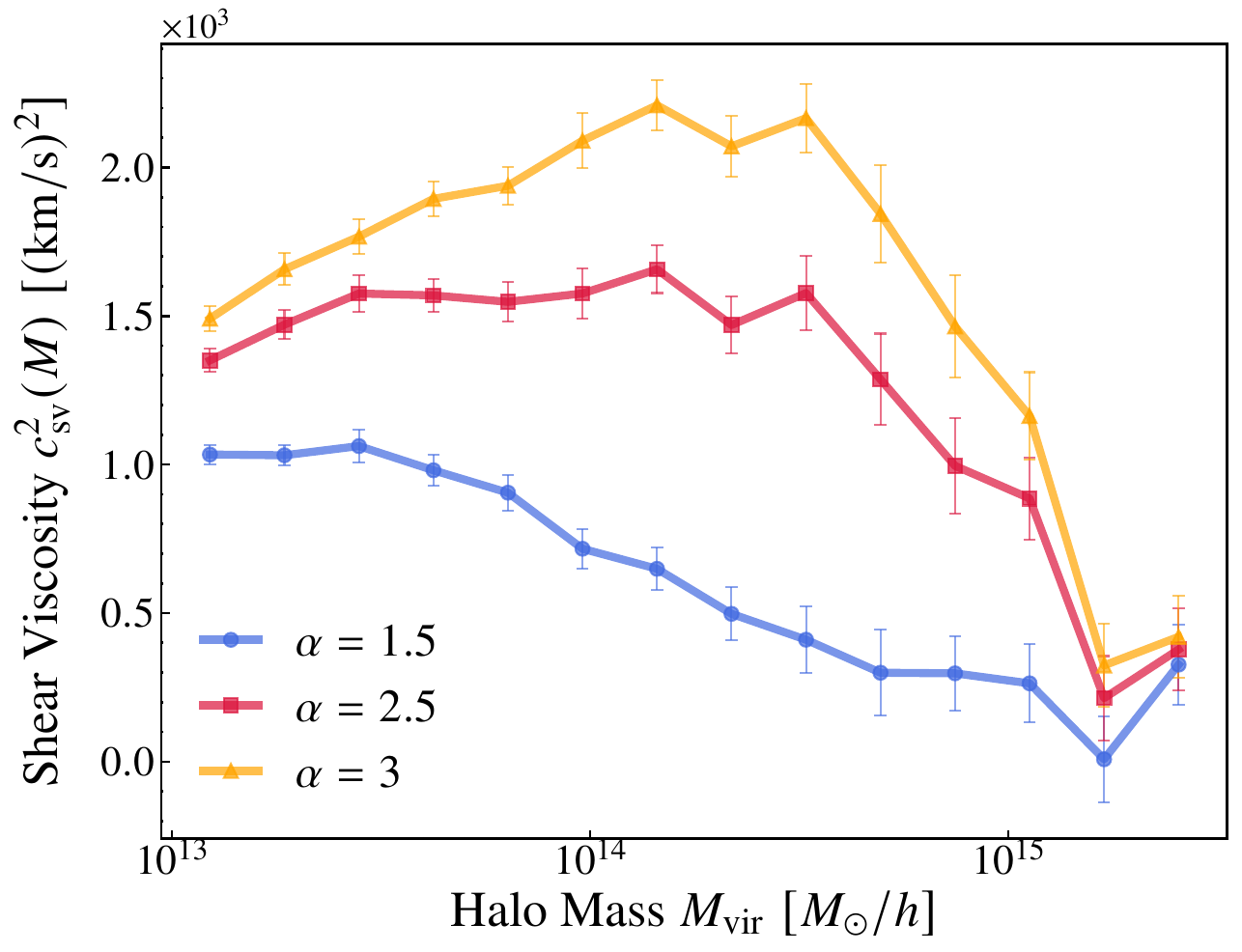}
    \end{subfigure}

    \caption{Comparing \(\bar{P}\), \(c_{\mathrm{eff}}^2\) and \(c_{sv}^2\) contributions for varying the radius cut \(\alpha\). We note that the results remain stable for \(\bar{P}\) and \(c_{\mathrm{eff}}^2\), but show larger variations for \(c_{sv}^2\). This analysis is performed at \(R_{\Lambda} = 28\,\mathrm{Mpc}/h\) See the text for a discussion.}

    \label{fig:varycut}

\end{figure}

\section{Dependence on the Radius Cut \(\alpha\)}\label{app:cut}

In the previous sections, we fixed the radius cut \(\alpha = 3\). This section will be focused on investigating the sensitivity of our results to the choice of \(\alpha\). We will compare the results for \(\alpha \in \{1.5, 2.5, 3\}\). As seen in Figure~\ref{fig:varycut}, the background pressure and effective sound speed contributions do not change drastically as we vary \(\alpha\). As expected, the contribution increases as we increase \(\alpha\), and we note that contributions show systematic increase near the peak of the contribution in both cases. These results validate our expectation that halo accretion is the underlying physical process. 

Figure~\ref{fig:varycut} also shows that the shear viscosity contributions show much larger changes than in the other counterterms. By splitting by the Anisotropic Flux Tensor, we have shown that the shear viscosity is sensitive to the accretion geometry around the halo. The computation of the Anisotropic Flux Tensor involves the shell between \(R_\mathrm{vir}\) and \(3 \times R_\mathrm{vir}\) around each halo. Thus, this level of sensitivity of the shear viscosity to the \(\alpha\) is reasonable. This strong dependence also suggests that a significant fraction of the shear viscosity contribution originates from the filamentary structure surrounding the halo, which becomes increasingly captured as \(\alpha\) grows, consistent with the alignment between the anisotropic stress and the feeding filament illustrated in Figure~\ref{fig:haloalign}.

\section{The Anisotropic Flux Tensor}\label{app:aniso}

In this section we describe the computation of the total inflow rate and Anisotropic Flux Tensor, which are essential to the arguments provided in Section~\ref{sec:extract}. We construct these quantities to characterise how the environment of a halo affects its contributions to the counterterms. We focus on the shell between \(R_\mathrm{vir}\) and \(3 \times R_\mathrm{vir}\) for a given halo.

\begin{figure}[!t]
    \centering
    \includegraphics[width=0.9\textwidth]{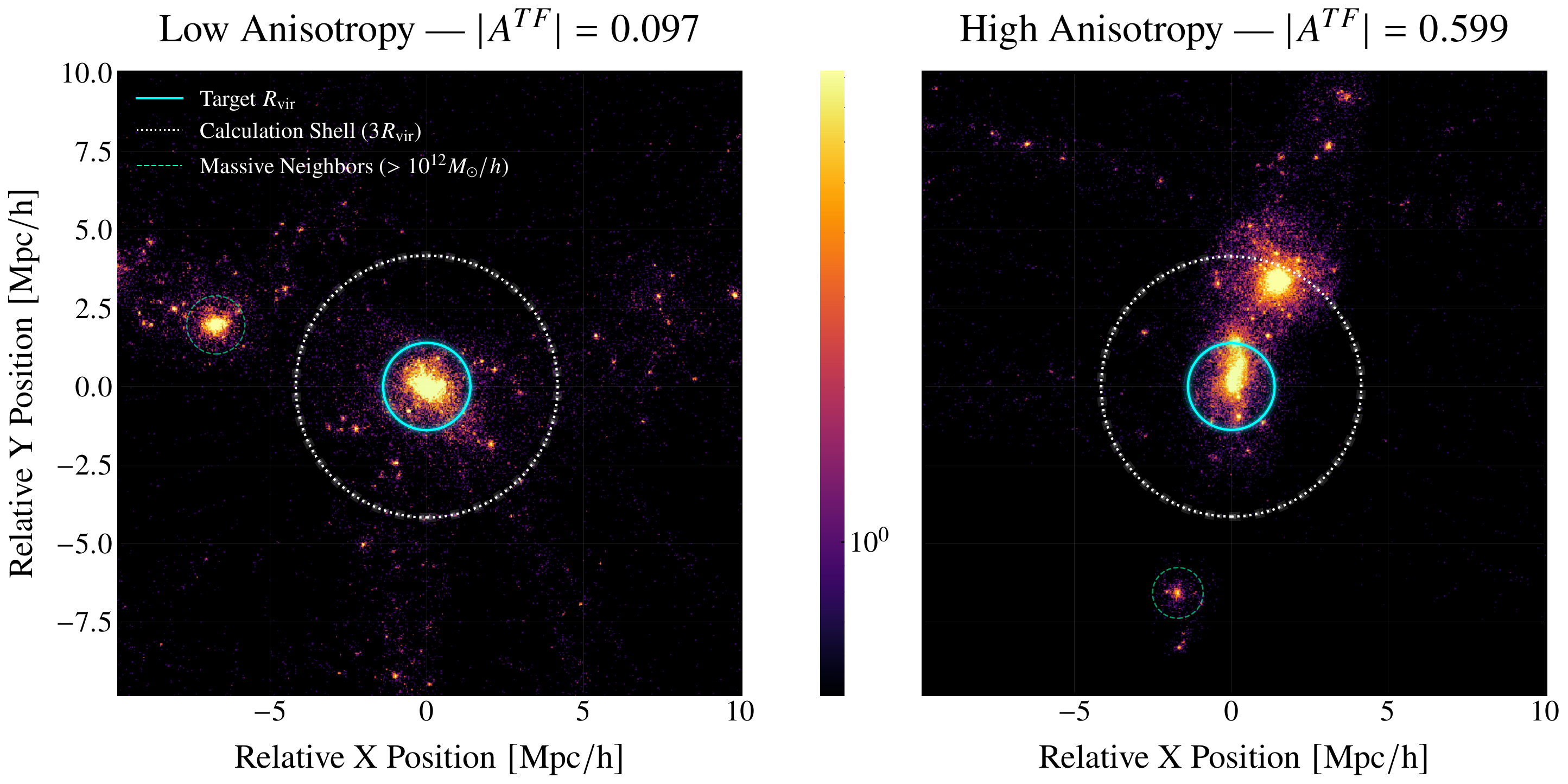}
    \caption{Comparison between two Halo Environments with different \(|A^{TF}|\). The halo environment low anisotropy image shows no distinct features, whereas in the high anisotropy case, the halo is undergoing a merger event.}
    \label{fig:haloenv}
\end{figure}

We include all infalling particles in this shell, by isolating particles with radial velocity \(v_r < 0\), with respect to the centre of the halo. Now to compute the total inflow rate, we sum over the infalling radial momenta: 
\begin{equation}
    \mathrm{Total\,Inflow\,Rate} = \sum_{p} m_p (- v_r) = I\,.
\end{equation}
Thus the total inflow rate acts as a proxy for instantaneous mass accretion. Now we define the Flux Tensor as: 
\begin{equation}
     A^{ij} = \sum_{p} \left(\frac{m_p (- v_r)}{I}\right) \hat n^i \hspace{0.1cm}\hat n^j\,,
\end{equation}
where \(\hat n^i\) denotes the ith component of the unit vector pointing in the direction of the particle from the halo centre. By normalising by total inflow rate, the diagonal elements of the Flux Tensor are a measurement of the magnitude of isotropy in the geometry of the infalling particles. We are interested in the anisotropy of the infall, leading us to the construction of the Anisotropic Flux Tensor: 
\begin{equation}
    \hat A^{ij} = A^{ij} - \frac{1}{3}Tr(A^{ij}) \hspace{0.1cm}\delta^{ij}\,.
\end{equation}
Thus we use the median of the magnitude of the Anisotropic Flux Tensor, \(|A^{TF}| = \sqrt{\hat A^{ij} \hat A_{ij}}\), to split the halo contributions. This is a measurement of the magnitude of anisotropy in the geometry of the infalling particles. Figure~\ref{fig:haloenv} shows a comparison between the environment of two halos with very different \(|A^{TF}|\) values. In this example, matter is infalling in both cases, but clearly the high anisotropy case is a merger event leading to a large \(|A^{TF}|\). We complete this section with a simple consistency check, by showing that the off-diagonal elements of the Flux Tensor vanish in the case of purely isotropic infall, in the continuum limit. Say the magnitude of the isotropic infall is \(F_0\), then we have: 
\begin{equation}
    A^{ij} = \frac{\int F_0 \hspace{0.1cm}\hat n^i \hspace{0.1cm}\hat n^j \hspace{0.1cm} d\Omega}{\int F_0\hspace{0.1cm} d\Omega}\,.
\end{equation}
And we choose a coordinate system such that \(\hat n^x = \mathrm{sin}\theta\hspace{0.1cm}\mathrm{cos}\phi \), \(\hat n^y = \mathrm{sin}\theta\hspace{0.1cm}\mathrm{sin}\phi \) and \(\hat n^z = \mathrm{cos}\theta\). Thus we get: 
\begin{equation}
    A^{ij} = \frac{1}{4 \pi} \int \hat n^i \hspace{0.1cm}\hat n^j \hspace{0.1cm} d\Omega\,.
\end{equation}
From here it's trivial to show that all the off-diagonal elements vanish and by spherical symmetry \( A^{ij} = \delta^{ij}/3\). Thus, for the case of completely isotropic infall, the Anisotropic Flux Tensor vanishes, allowing us to use \(|A^{TF}|\) as a measurement of the anisotropy in the infall. 

\section{Comparison of Full Box Methods}\label{app:fullbox}

In Section~\ref{sec:fullbox} we used a real space random sphere approach to smooth the effective stress tensor and performed all the correlations in real space, as averages over these spheres. In this brief section we compare the results of this methodology with the standard Fourier space regression approach to extract the counterterms. We continue to use the real space Top Hat filter for consistency, but all of the smoothing operations are performed in Fourier space. The construction of all the relevant fields and the effective stress tensor remains the same as the procedure in Section~\ref{sec:fullbox}. Thus, the methodology remains the same for the background pressure computation, as it is simply one-third of the trace of the stress tensor. 

Rather than extracting a single value of the counterterm from an individual Fourier mode, we perform an ordinary least-squares (OLS) regression: 
\begin{equation}
    c^2_{\mathrm{eff}} =
\frac{1}{\bar{\rho}}
\frac{\displaystyle\sum_{\mathbf{q}} P_{\tau\delta}(\mathbf{q})}
{\displaystyle\sum_{\mathbf{q}} P_{\delta\delta}(\mathbf{q})}
\qquad
c^2_{\mathrm{sv}} =
-\frac{4Ha}{3\bar{\rho}}
\frac{\displaystyle\sum_{\mathbf{q}} P_{\tau\sigma}(\mathbf{q})}
{\displaystyle\sum_{\mathbf{q}} P_{\sigma\sigma}(\mathbf{q})}\,.
\end{equation}
Here, \(\mathbf{q}\) denotes an individual Fourier wavevector, and the sums run over every such mode in the simulation box.

The difference arises in the methodology of the \(c^2_{\mathrm{eff}}\) and \(c^2_{\mathrm{sv}}\) extractions. They are computed as: 
\begin{equation}
    c^2_{\mathrm{eff}}(k) = \frac{1}{\bar{\rho}} \frac{P_{\tau \delta}(k)}{P_{\delta \delta}(k)}, \hspace{0.3 cm}c^2_{\mathrm{sv}}(k) = - \frac{4Ha}{3\bar{\rho}} \frac{P_{\tau \sigma}(k)}{P_{\sigma \sigma}(k)}\,,
\end{equation}
where \(P_{\tau\delta}(k) = \langle Tr[\tau^{ij}(k)] \delta(k)\rangle/3\), \(P_{\tau\sigma}(k) = \langle \tau^{ij}(k) \sigma^{ij}(k)\rangle\) and so on.

\begin{table}[!t]
\centering
\begin{tabular}{|c|c|c|c|}
\hline
\(R_{\Lambda}\) (in \(\mathrm{Mpc}/h\)) & RS & FS & \% Difference \\
\hline
16 & $2.85\times10^{4}$ & $2.85\times10^{4}$ & $-0.14\%$ \\
20 & $2.84\times10^{4}$ & $2.85\times10^{4}$ & $-0.38\%$ \\
24 & $2.85\times10^{4}$ & $2.85\times10^{4}$ & $-0.01\%$ \\
28 & $2.87\times10^{4}$ & $2.87\times10^{4}$ & $0.10\%$ \\
\hline
\end{tabular}
\caption{\label{tab:pressure_comparison}Comparing the Real Space Random Sphere (RS) and Fourier Space OLS Regression (FS) full box methods to compute \(\bar{P}/\bar{\rho}\), in units of \((\mathrm{km/s})^2\), for different EFT cutoff scales \(R_{\Lambda}\) (in \(\mathrm{Mpc}/h\)). The two methods agree to well within a percent.}
\end{table}

\begin{table}[!t]
\centering
\begin{tabular}{|c|c|c|c|}
\hline
\(R_{\Lambda}\) (in \(\mathrm{Mpc}/h\)) & RS & FS & \% Difference \\
\hline
16 & $1.07\times10^{5}$ & $1.07\times10^{5}$ & $-0.55\%$ \\
20 & $1.03\times10^{5}$ & $1.03\times10^{5}$ & $-0.19\%$ \\
24 & $1.02\times10^{5}$ & $1.02\times10^{5}$ & $0.07\%$ \\
28 & $1.01\times10^{5}$ & $1.01\times10^{5}$ & $-0.18\%$ \\
\hline
\end{tabular}
\caption{\label{tab:sound_speed_comparison}Comparing the Real Space Random Sphere (RS) and Fourier Space OLS Regression (FS) full box methods to compute \(c_{\mathrm{eff}}^2\), in units of \((\mathrm{km/s})^2\), for different EFT cutoff scales \(R_{\Lambda}\) (in \(\mathrm{Mpc}/h\)). The two methods agree to well within a percent.}
\end{table}

\begin{table}[!t]
\centering
\begin{tabular}{|c|c|c|c|}
\hline
\(R_{\Lambda}\) (in \(\mathrm{Mpc}/h\)) & RS & FS & \% Difference \\
\hline
16 & $1.62\times10^{4}$ & $1.69\times10^{4}$ & $-4.14\%$ \\
20 & $2.21\times10^{4}$ & $2.32\times10^{4}$ & $-5.03\%$ \\
24 & $2.78\times10^{4}$ & $2.94\times10^{4}$ & $-5.43\%$ \\
28 & $3.46\times10^{4}$ & $3.58\times10^{4}$ & $-3.51\%$ \\
\hline
\end{tabular}
\caption{\label{tab:shear_viscosity_comparison}Comparing the Real Space Random Sphere (RS) and Fourier Space OLS Regression (FS) full box methods to compute \(c_{sv}^2\), in units of \((\mathrm{km/s})^2\), for different EFT cutoff scales \(R_{\Lambda}\) (in \(\mathrm{Mpc}/h\)). Unlike \(\bar{P}/\bar{\rho}\) and \(c_{\mathrm{eff}}^2\), the two methods differ at the \(5\%\) level, reflecting the noisier shear viscosity measurement.}
\end{table}

In Tables~\ref{tab:pressure_comparison}--\ref{tab:shear_viscosity_comparison}, we compare the results through the real space Random Sphere approach (RS) and the Fourier space OLS regression approach (FS). The background pressure and effective sound speed agree to within a percent, whereas the shear viscosity shows deviations at percent level. We attribute this to the excess noise in the shear viscosity measurement, and use the RS approach in the main text to remain consistent with the halo-based calculation. 

\section{Layzer-Irvine Cosmic Energy Equation Test}\label{app:cee}

In this brief section we test the robustness of our full box calculation, following the methodology in \cite{nascimento2025semi} by checking against the Layzer-Irvine cosmic energy equation: 
\begin{equation}
    \frac{d}{d\eta}(\kappa + u) + \mathcal{H}(2\kappa + u) = 0\,,
\end{equation}
where \(\kappa = \langle K^i_i\rangle/2\bar{\rho}\) and \(u = \langle P^i_i\rangle/\bar{\rho}\). We analyse 21 snapshots, equally spaced in scale factor, \(a = 1/3\) to \(a = 1\), from an N-body simulation with a box size of \(1000 \hspace{0.1 cm} \mathrm{Mpc}/h\) and \(1024^3\) particles. We perform the test by comparing the computation of \(\kappa\) in two independent ways. \(\kappa_{\mathrm{sim}}\) is computed directly from the full box stress tensor computation. On the other hand, \(\kappa_{\mathrm{pred}}\) is computed as an integral over \(u\), which itself is computed from the full box calculation:
\begin{equation}
    \kappa_{\mathrm{pred}}(a) = -u(a) + \frac{1}{a^2} \int_0^a da'\, a' u(a')\,.
\end{equation}

\begin{figure}[!t]
    \centering
    \includegraphics[width=0.7\textwidth]{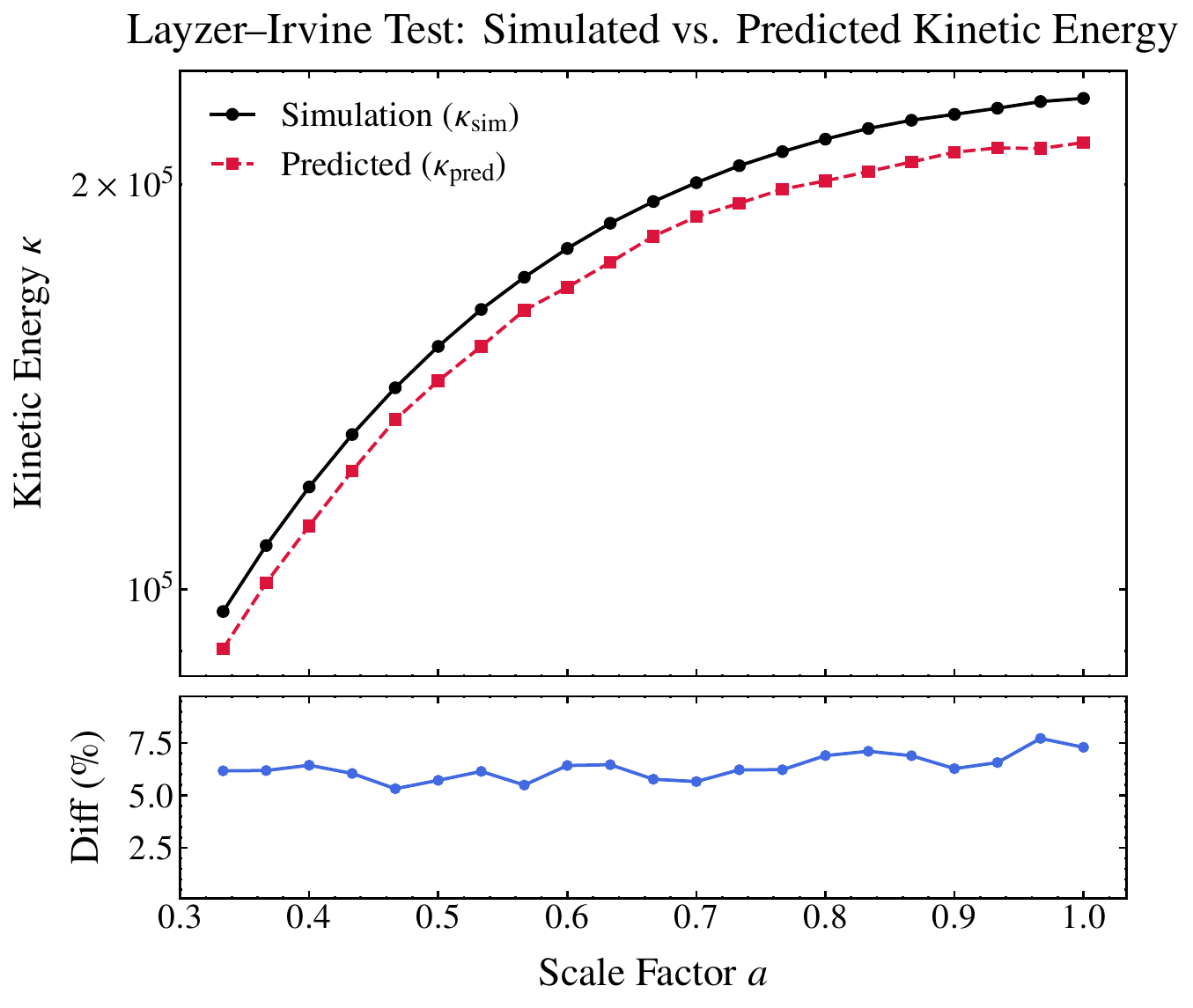}
    \caption{Comparing the simulated kinetic term \(\kappa_{\mathrm{sim}}\), computed directly from the full box stress tensor, to the predicted \(\kappa_{\mathrm{pred}}\), obtained by integrating the Layzer-Irvine cosmic energy equation. We note that \(\kappa_{\mathrm{sim}}\) systematically overestimates \(\kappa_{\mathrm{pred}}\) by \(\sim 7\%\) across redshifts. This is consistent with a similar overestimate found in previous work, and is likely explained by known systematics discussed in the text.}
    \label{fig:cee}
\end{figure}

The integral extends to \(a = 0\), but our snapshots begin at \(a = 1/3\). As noted in \cite{nascimento2025semi}, the integral is expected to be dominated by the upper limit. Thus we approximate the integral at early times assuming the early-time, matter-domination scaling \(u(a) \sim a\). This allows us to write:
\begin{equation}
     \int_0^{a_{\mathrm{min}}} da'\, a' u(a') = \frac{u_{\mathrm{min}}}{a_{\mathrm{min}}} \int_0^{a_{\mathrm{min}}} da' \, (a')^2 = u_{\mathrm{min}} \hspace{0.1cm} a_{\mathrm{min}}^2 /3\,.
\end{equation}

Figure~\ref{fig:cee} shows the results of this analysis. We notice that \(\kappa_{\mathrm{sim}}\) systematically overestimates \(\kappa_{\mathrm{pred}}\) by \(\sim 7\%\), similar to what was found in \cite{nascimento2025semi}, where this residual (with \(\kappa_{\mathrm{sim}}\) running above the theoretical prediction) was tentatively attributed to noise from empty or low-density grid cells, which require dividing by a near-zero density when constructing the velocity field. Since our velocity field is likewise obtained on a grid, this same effect could plausibly contribute to the residual we observe here. Up to this offset, the test provides an independent consistency check of our full box computation.

\bibliographystyle{JHEP}
\bibliography{mybib}

\end{document}